\documentclass[11pt]{article}
\pdfoutput=1
\usepackage{jheppub}

\usepackage{slashed}
\usepackage{color, verbatim}
\usepackage{latexsym}
\usepackage{epsfig}
\usepackage{amsmath,accents}

\usepackage{amssymb}
\usepackage{graphicx}
\usepackage{bm}
\usepackage{stackrel}

\usepackage[font={small}]{caption}

\usepackage{hyperref}
\usepackage{epstopdf}
\definecolor{myred}{rgb}{0.7, 0, 0}
\definecolor{myblue}{rgb}{0, 0, 0.7}
\definecolor{mygreen}{rgb}{0.04, 0.7, 0.07}
\hypersetup{colorlinks,citecolor=myred,linkcolor=myblue,urlcolor=mygreen,linktocpage=true}

\newcommand{\be}{\begin{equation}}
\newcommand{\ee}{\end{equation}}
\newcommand{\bea}{\begin{eqnarray}}
\newcommand{\eea}{\end{eqnarray}}

\newcommand{\g}{\mathfrak g}

\newcommand{\nn}{\nonumber}

\newcommand{\s}{\sigma}

\def\a{\alpha}
\def\b{\beta}
\def\g{\gamma}

\def\d{\delta}
\def\D{\Delta}

\def\m{\mu}
\def\n{\nu}
\def\l{\lambda}
\def\L{\Lambda}
\def\r{\rho}
\def\s{\sigma}

\newcommand{\eq}[1]{Eq.~(\ref{#1})}
\newcommand{\fig}[1]{Fig.~\ref{#1}}

\newcommand{\sect}[1]{Section~\ref{#1}}

\begin{document}

\thispagestyle{empty}

\begin{center}

\begin{center}

\vspace{.5cm}

{\Large
UV-complete and stable Quintom Dark Energy models in the light of DESI DR2
\vspace{0.3cm}
}\\

\end{center}

\vspace{1.cm}

\textbf{
Fotis Koutroulis$^\S$
}
\\

\vspace{1cm}

{\textit{Theoretical Physics Division, Institute of High Energy Physics, Chinese Academy of Sciences,\\ 19B Yuquan Road, Shijingshan District, Beijing 100049, China}
}

{\em 
China Center of Advanced Science and Technology, Beijing 100190, China}

\end{center}

\vspace{0.8cm}

\centerline{\bf Abstract}
\vspace{2 mm}

\begin{quote}\small
We propose that Quintom dark energy models, the simplest scenario that predicts the crossing of cosmological boundary, can be embedded in a UV-motivated five-dimensional anisotropic orbifold-lattice construction, the so called Non-Perturbative Gauge Higgs Unification (NPGHU) model. The latter contains purely a 5d $SU(2)$ gauge field on the bulk which, in the continuum, projects on our 4d boundary a complex scalar and a $U(1)$ gauge field. These are identified as the dynamical Dark Energy sector of our universe while Standard Model and Dark Matter are completely 4d localized. The bulk-originated fields come (at late-times) with dim-6 higher derivative operators which inherit the spectrum with both physical and phantom, scalar and gauge, degrees of freedom. We show that the 4d effective dark energy action is that of a modified Quintom model and that the associated background equation of state ($w_q$) can naturally be of a Quintom-B type. We expose the crucial contribution of the massive gauge ghost on $w_q$ which allows representative Quintom-B trajectories to remain compatible with DESI DR2 BAO distance ratios, with substantially reduced tuning compared to standard Quintom lore. We further propagate the model-derived Quintom-B backgrounds into late-time observables, showing for several benchmarks their impact on $H(z)$, BAO distance ratios compared with DESI DR2 and the smooth-growth observable $f\sigma_8(z)$.
The NPGHU model has a plethora of interesting features. Among others it is by definition free from ghost instabilities, forbids the presence of any potential term, as well as, it possesses a finite low-energy cut-off ($\L$) which determines the full localization of the brane and renders Lorentz invariance approximate. These characteristics are inherited to the effective Quintom action and we explicitly show their role for the consistency of the model investigating its behavior under both linear perturbations (classical instability) and vacuum decay (quantum instability). Interestingly, for the most natural realization of the NPGHU model which suggests that today we live in the vicinity of the cut-off (setting $\L \approx {\cal O}(10) H_0$), we find that the model remains free of immediate linear and vacuum-decay instabilities within the controlled effective regime, despite the presence of IR-emergent phantom degrees of freedom. 
In total, NPGHU model provides a natural framework such that the Quintom models can acquire a higher-dimensional origin within a controlled effective regime, where the phantom behavior is IR-emergent, while retaining benchmark-level predictivity for the late-time background evolution.

\end{quote}

$^\S$Email: \href{mailto:fkoutroulis@ihep.ac.cn}{fkoutroulis@ihep.ac.cn}

\vfill

\newpage

\tableofcontents

\newpage
\section{Introduction}
Almost 25 years have past since the first robust experimental evidences regarding the accelerated expansion of our universe. The associated experiments were conducted by two groups through the observations of the luminosity distances of high-redshift supernovae \cite{SupernovaSearchTeam:1998fmf, SupernovaCosmologyProject:1998vns}.
Following this discovery a bunch of new observations coming for instance from the Cosmic Microwave Background \cite{Planck:2018vyg} or the Large-Scale Structure (LSS) \cite{Chuang:2013hya} independently confirmed that the universe is not static but rather is expanding with an accelerated pace.
In order to describe this behavior the concept of Dark Energy (DE) was introduced which initially was expressed in the standard cosmological scenario ($\Lambda$CDM) via the cosmological constant $\Lambda$.
Unfortunately such a description leads to various issues such as fine-tuning and coincidence problems in $\Lambda$CDM cosmology as was presented first by Weinberg \cite{Weinberg:2000yb}.
In addition, the recently released measurements of Baryon Acoustic Oscillations (BAO) by the Dark Energy Spectroscopic Instrument (DESI) \cite{DESI:2024mwx, DESI:2025zgx} provide strong evidence, at the level of $\lesssim 5\sigma$ (with some recent analysis on the systematics \cite{Capozziello:2025qmh} challenging this number) for deviations from $\L$CDM paradigm towards a Dynamical Dark Energy (DDE). 
Due to the previous drawback, in the last decades a big effort has been put on developing cosmological models that go beyond the standard cosmological scenario exploiting DDE scenarios. One still viable subset of these cases refers to the scalar-field models which include quintessence \cite{Ratra:1987rm, Wetterich:1987fm, Gomez-Valent:2025mfl}, phantom \cite{Caldwell:1999ew} and K-essence \cite{Armendariz-Picon:2000ulo, Nojiri:2026uvn} models or emergent dark energy cases \cite{Najafi:2026kxs} among others. An alternative direction that has more recently been developed to explain the DDE behavior, is based on the possible existence of non-trivial dark matter-dark energy interactions which leads to energy exchange between the two sectors \cite{Wang:2016lxa, Giare:2024smz, Khoury:2025txd, Petri:2025swg, Nojiri:2025low, LaPenna:2026avs}.
The main common characteristic among these scenarios is that they predict a time-evolving equation-of-state (EoS) $w$, spanning values from $w > -1$ (quintessence) to $w < -1$ (phantom).
However, they usually fail to reproduce a good fit to data compared to the one obtained using a phenomenological parametrization that allows for crossing the cosmological boundary, such as the Chevallier-Polarski-Linder (CPL) \cite{Chevallier:2000qy, Linder:2002et} dark energy parametrization.
On the other hand, there is a class of the above models of particular interest and simplicity known as quintom models \cite{Feng:2004ad, Zhao:2005vj, Guo:2006pc, Cai:2006dm} (see also \cite{Cai:2009zp, Ren:2026jyw} for a thorough review and \cite{Carloni:2024zpl, Giare:2024gpk, Giare:2024oil, Yin:2024hba, Jiang:2024xnu, Pang:2025lvh} for recent theoretical studies). Among other features, quintom scenarios entertain the possibility that $w$ could cross $-1$. In that sense they are enabling the description of a broader range of cosmological evolution incorporating naturally the phantom crossing.
Notice that models of coupled DM-DE may incorporate the crossing of the dark energy equation of state through the phantom divide (seen by DESI) as well, however they face some strong pressure from constraints on long-range DM forces \cite{Chen:2025ywv}.
Despite quintom simplicity and success these models suffer from several theoretical and phenomenological problems. Among them the most crucial are the following: i) the existence of a No-Go theorem, at least in the single-scalar realizations, ii) the lack of fundamental justification regarding the presence of non-physical (ghost) degree of freedom (dof) in the spectrum at late-times, iii) the lack of a consistent way to cure the pathologies originating from the latter and iv) the need of extensive fine-tuning in order to be aligned to the recent data since usually they predict a Quintom-A behavior for the DE equation of state in opposition to the observations. Finally, this class of models has been characterized as purely phenomenological without a more fundamental origin.

\begin{figure}[t]
\centering
\includegraphics[width=8cm]{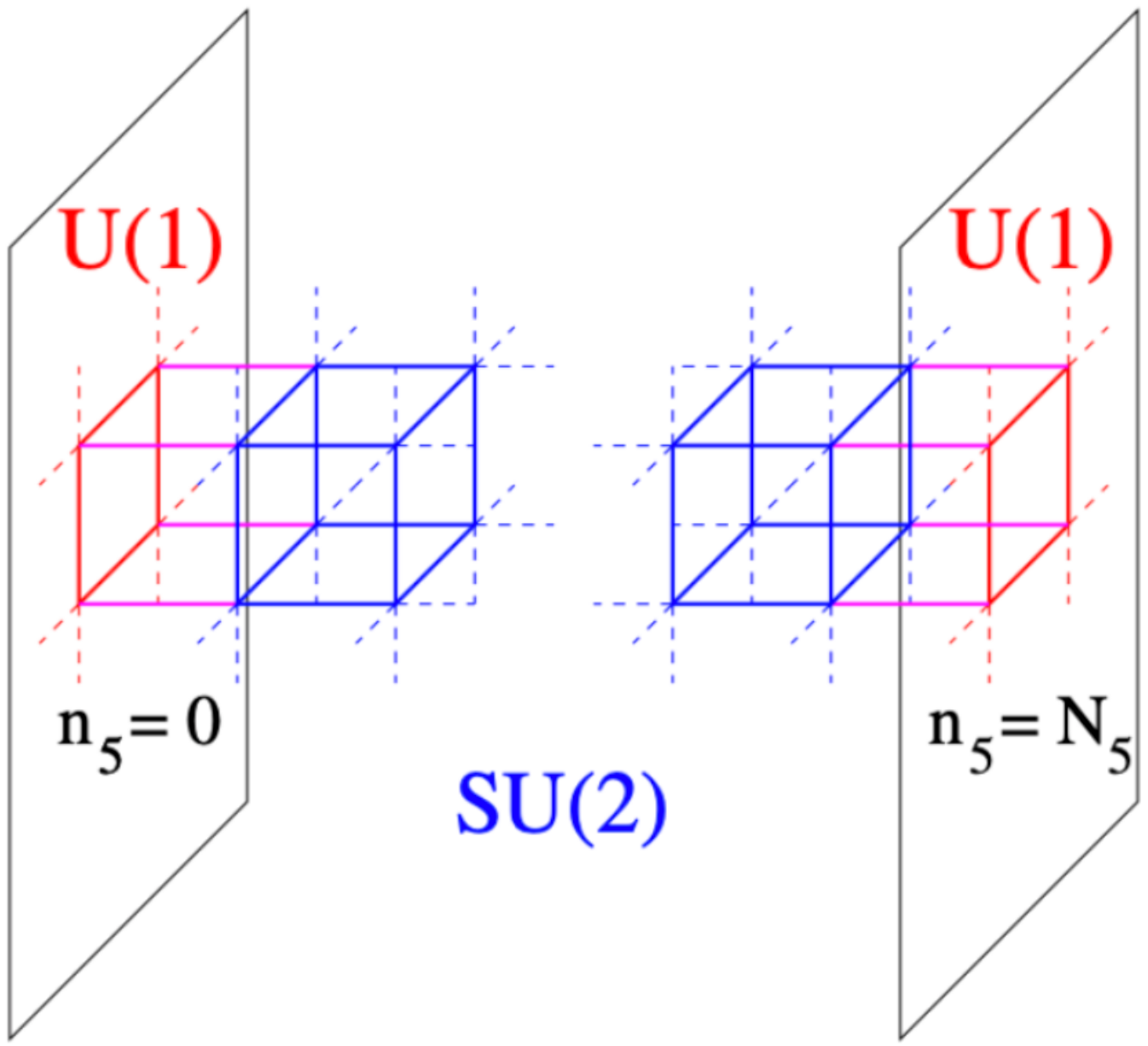} 
\caption{\it The 5d anisotropic orbifold lattice spacetime, constracted for the first time in \cite{Irges:2006hg, Knechtli:2007ea}, where we live on the 4d $U(1)$ brane at the fixed point $n_5 = 0$. At each point on the extra dimension, $n_5 = \{1, ..., N_5 -1 \}$, lies a 4d brane with an SU(2) gauge field on it.
}
\label{Higgsphase}
\end{figure}

In this work we develop, to the best of our knowledge, the first NPGHU-based higher-dimensional completion of the original quintom models in which the main theoretical problems listed above are regulated within a controlled effective regime and which can naturally reproduce Quintom-B-like trajectories similar to those preferred by CPL-based DESI-era reconstructions or, in other words, it can predict the Quintom-B behavior of DE with ameliorated fine-tuning.
The UV completion that we develop here is based on the Non-Perturbative Gauge-Higgs Unification (NPGHU) model, an anisotropic orbifold lattice construction originally studied by \cite{Knechtli:2007ea, Irges:2006hg, Alberti:2015pha} as an alternative solution to Higgs hierarchy problem (with characteristically negligible parameter fine-tuning).
However, here we follow a different path and propose that the NPGHU model is an attractive UV completion of the Standard Model (SM) which could naturally provide a consistent quitnom-like spectrum that accounts for a DDE candidate.   
In particular, according to NPGHU scenario the spacetime is fundamentally discrete and 5-dimensional such that we live on the 4d boundary of this model while on the bulk exists purely a 5d non-Abelian gauge field (\fig{Higgsphase}). Moreover, in the simplest realization that we consider here we assume that both the SM and the Dark Matter (DM) are completely localized on the 4d boundary and they do not interact with the DE sector (except gravitational interactions). Then we entertain the scenario that the components of the 5d gauge field, that survive on the boundary due to the orbifold conditions, play (at the naive continuum limit) the role of a dynamical dark energy sector.
In this regard the total 4d effective action is given by
\be\label{Stot4deff}
S_{\rm tot} = S_{\rm DE} + S_{\rm DM} + S_{\rm SM} 
\ee
where $S_{\rm DE}$ is the 4d effective dark energy part.
NPGHU model possesses several non-trivial characteristics which we thoroughly develop and explain in \sect{BoQLM}. Among them is the (proven both non-perturbatively and on the continuum) presence of a finite, low energy, cut-off scale $\L$ on which the 5d originated dof exhibit a 1st order phase transition (\fig{Higgsphase2}). Due to the latter, bulk and boundary are completely decoupled (full localization) for energies above $\L$ distinguishing two different effective actions
\be\label{SDEbsL0}
S_{\rm DE} \equiv 
\left\{
\begin{alignedat}{3} 
 & S^{\rm orb}_{\m > \L} & ~~{\rm for } ~~ \m \ge \L 
& &  &  \\
  & S^{\rm orb}_{\m < \L} & ~~{\rm for } ~~ \m < \L 
\end{alignedat}
\right.  
\ee
Above $\L$ (Coulomb phase) the action involves a massless Scalar QED (leading order in lattice-spacing expansion) 
\be
S^{\rm orb}_{\m > \L} \equiv \int d^4x \sqrt{-g} \left [  - \frac{1}{4} F_{\m\n}F^{\m\n}  + |D_\m \phi|^2 \right] \nn
\ee
and describes "early-times" dark energy.
Below $\L$ (Higgs phase) the effective action which is relevant for the late-time dark energy is parameterized by a massless Abelian Lee-Wick-like model (next-to-leading order in lattice-spacing expansion) without polynomial terms \cite{Irges:2018gra, Irges:2020nap}  
\bea\label{SDE30}
S^{\rm orb}_{\m < \L} &\equiv& S^{\rm orb}_{\m > \L} + \int d^4x \sqrt{-g} \left [ \frac{c_\a}{2\L^2} \left( \nabla^\m F_{\m\n}\right)\left( \nabla_\m F^{\m\n}\right) - \frac{c_6}{\L^2} |D^\m D_\m \phi|^2  + {\cal L}_{{\rm R},\chi} + {\cal L}_{{\rm R}, B} \right ]
\eea
The scalar $\chi$ and gauge field $B_\m$ are introduced by fixing the reparameterization freedom of the action and play an important role in the consistency of the late-time model as we explicitly demonstrate in \sect{BoQLM}.
The dim-4 operators and dim-6 derivative operators involved in $S^{\rm orb}_{\m < \L}$ are sufficient to capture the physics near the phase transition \cite{Irges:2020nap}.\\
Here the focus is on \eq{SDE30} which lies below but still in the vicinity of the cut-off towards the Higgs phase.
In this case the spectrum is essentially described by a modified and extended quintom-like model since it includes both scalar and gauge phantoms but the lattice origin of the model forbids the presence of any scalar potential.
As the simplest example we consider the small lattice anisotropy limit and demand (phenomenologically) that today we live near the vicinity of the phase transition such that localization is strong but not exact (the bulk is not fully decoupled from the boundary).
Therefore, the starting point of our analysis in this work involves the 4d effective late-time dark energy action
\be\label{SDEgs10}
S^{\g \ll 1}_{{\rm DE}} \approx \int d^4x \sqrt{-g} \left[ - \frac{1}{4} F_{\m\n,1}F^{\m\n,1} +  \frac{1}{4} F_{\m\n,2}F^{\m\n,2} -\frac{1}{2} m_{A_2}^2 A_{\m,2}^2 + |\partial_\m \phi_1|^2 - |\partial_\m \phi_2|^2 + m_{\phi_2}^2  |\phi_2|^2 \right] + S_\chi + S_{B_\m} 
\ee
We then calculate the background energy and pressure densities and solve the associated equation of motion (eom) to obtain the full time-evolution of the background EoS. The latter depends essentially on four initial conditions coming from the fields and their derivatives. After an extensive numerical scan we find that a Quintom-B scenario can be naturally realized with substantially reduced tuning within the scanned region.
Then we put our model under the stress of cosmological perturbations and instabilities in both classical and quantum level. Regarding the former, in this first attempt to UV complete the quintom models, we focus only on the linear level and determine the conditions under which the spectrum remains stable in the presence of non-zero interactions among the fields.
Quantum instabilities, on the other hand, are parameterized by the spontaneous vacuum decay to phantom and physical fields due to the universal gravitational interaction even in the absence of other couplings among the sectors. These are the main reason that render quintom-like models severely problematic. However, in our case we find that even though a vacuum decay is possible, it will actually never complete within the lifetime of the universe. There are two mechanisms that help regulate the dangerous instabilities within the effective regime, both coming purely from the lattice origin of our UV completion. The first is due to the existence of the finite $\L$ which suggests that Lorentz invariance is emergent and not exact in the IR, providing naturally a preferred lattice frame (the CMB frame). $\L$ then constrains the otherwise infinite boosts (phantom phase space) on the calculation of the vacuum decay rate and we find an upper bound ($\L_{\rm max}$) below which the lifetime of the latter is much higher than the age of universe.
In particular our scenario naturally suggests that the cut-off, where the initial conditions of the fields are defined, could be given by any value $\L \lesssim 100 H_{\rm m,0}$ which is higher but close to the Hubble parameter today, $H_0 \equiv \Omega_{\rm m, 0 }^{-1/2} H_{\rm m,0} \approx 1.5\times 10^{-33} ~\rm eV$. $H_{\rm m,0}$ refers to the matter part of the Hubble parameter today. Here we developed our analysis by choosing to coincide the initial field conditions with the dawn of the localization phase transition and the deep matter dominated era of the universe getting $\L \approx 10 H_{\rm m,0}$.
We show that such choice is many orders of magnitude lower than $\L_{\rm max}$ rendering the vacuum sufficiently long-lived within the effective regime without introducing severely catastrophic gravity modifications at length scales that contradicts the experiments.
An instability stressed to scales defined by this cut-off, $\l_{\rm cr} \approx 0.1 H^{-1}_{\rm m, 0} $, 
may affect BAO and modify super-cluster growth potentially giving a (mild) scale-dependent growth on very large scales, 
however this analysis requires a full late-time phenomenological study which is outside the scope of the current work.
The second reason which keeps safe the IR behavior of the model is that, moving towards to the far future, localization becomes more and more weak and higher than dim-6 derivative operators are now entering to the spectrum obtaining gradually the full lattice plaquettes. The model does not admit a 4d description any more and is dominated by the full lattice construction, whose gauge-invariant formulation is expected to control the purely four-dimensional ghost description beyond the validity of the truncated effective theory. In that sense, the lattice 5d origin of our model will regulate the possibility of the 4d vacuum decay at some point in the very far future.   

Let us close the introduction by mentioning a set of previous and recent attempts with seemingly similar characteristics with the current work which, however, have neither an essential overlap with our construction nor they address all the issues that a quintom-like model may develop, in contrast to our case.  
A recent, interesting, approach which combines a broken-$U(1)$ generalized Proca vector field with a scalar field as dark energy dof, that could give the phantom crossing, was presented in \cite{Tsujikawa:2026xqm}. Even though in similar spirit, this work has no overlap with the quintom UV-completion that we propose here.   
On the other hand, UV-completions involving hidden sectors which admit breaking of the Lorentz-invariance and/or crossing of the phantom divide are found in \cite{Holdom:2004yx, Elizalde:2004mq, Rubakov2006PhantomWU, Libanov:2007mq}. However these constructions, even in the absence of instabilities, can at best produce a Quintom-A phantom crossing in disagreement with the current experimental data.
Finally, an attempt to use a brane-world scenario on a 5-dimensional spacetime which allows for a phantom phase was developed in \cite{Dvali:2000hr, Mishra:2025goj}. However, the former (under infinite fifth dimension) was focused on the emergence of 4d Newtonian gravity on a brane in 5d flat space and for a minimal setup suffers from several phenomenological subtleties. Among them is the presence of ghost instabilities which cannot be (naturally) cured, in contrast to our case. The latter, although interesting and successful on the phantom crossing, depends strongly on the presence of the extra dimension at the effective Hubble rate as well as does not provide a clear motivation regarding the origin of the scalar potential on the brane. Therefore, to our knowledge, the present construction differs from these works in its NPGHU origin, its higher-derivative boundary action and its finite-cutoff mechanism for regulating the phantom sector. 

The structure of the paper is organized as follows. In \sect{BoQLM} we give a very detailed analysis of our set up, we explain the main characteristics of the NPGHU model and establish the physical picture that we consider here. Separating the spectrum in two regimes, above and below the cut-off, we define the associated action for the dark energy sector (explicitly explaining \eq{Stot4deff}) and we justify which dof give a consistent late-time action.
Then, in \sect{TEoSwqOTEQM} we explicitly calculate the energy and pressure densities along with the eom for the fields of the late-time action in order to determine the evolution of the background EoS. We give one Quintom-B and one quintessence benchmark example before we perform an extensive numerical scan over the free parameters of the EoS, evaluating the full range of its evolution during the low redshifts. To connect the NPGHU-induced quintom dynamics with observations, we supplement the equation-of-state analysis with a benchmark-level phenomenological study of $H(z)$, BAO distances, DESI DR2 distance ratios and the smooth linear-growth observable $f\sigma_8(z)$. The presence of phantom fields may render the model unstable either in classical or quantum level which is the subject of \sect{IFPI}. There, we determine the conditions under which our effective quintom model avoids immediate linear perturbative instabilities and remains sufficiently long-lived against gravity-induced vacuum decay. Lastly, we present our conclusions in \sect{Conclusions} while we show the connection of the conformal-time with the e-fold derivatives and we explain the choice of the initial conditions in App. \ref{dBasis} and App. \ref{CTIFV} respectively.

\section{The setup of the UV-complete quintom-model}\label{BoQLM}
In this section we initiate a brief but rather compact description about the origin of our model and then we present its main characteristics. After obtaining the late-time effective quintom action we develop the needed technology for the evaluation of the EoS. The latter is defined for our case as $w_q$ in order to be distinguished from the notation $w$ used by $\L$CDM.\\
As we have already advertised our starting point involves the extension of spacetime to a 5-dimensional space which, on top of that, is fundamentally discreetized leading to an extradimensional lattice.
The extra dimension is a circle with radius $R$ and we construct an orbifold lattice by projecting the circle by the discrete group $\mathbb{Z}_2$. The latter identifies the upper with the lower semicircle and results in a discretized interval with two fixed points, at $n_5 = 0$ and $n_5=N_5$. On each of these fixed points lives a 4d slice and the produced geometry is that of $\mathbb{R}^4 \times S_1/\mathbb{Z}_2$.
To be even more general we consider that this orbifold lattice is anisotropic in the fifth dimension meaning that the 4d lattice spacing ($a_4$) and its 5d counterpart ($a_5$) do not evolve in the same way leading to the anisotropy factor $\gamma = a_4/a_5$. Finally any field content introduced on the bulk of this 5d anisotropic lattice obeys specific orbifold boundary conditions. The scenario that we consider here is the simplest, non-trivial, possible case  of a pure $SU(2)$ gauge field leaving on the bulk of the lattice which gives rise to the context of Non-Perturbative Gauge-Higgs Unification (NPGHU).
Such a construction was originally studied purely on the lattice \cite{Knechtli:2007ea, Irges:2006hg, Alberti:2015pha} and is depicted on \fig{Higgsphase}. According to that picture our world lives\footnote{Note that there is a $\mathbb{Z}_2$-symmetry along the midpoint of the orbifold lattice therefore we can choose to localize the SM either on the $n_5 =0$ or $N_5$ without loss of generality.} on the 4d $U(1)$ brane at the fixed point $n_5 = 0$. At each point ("node") in the bulk, $n_5 = \{1, ..., N_5 -1 \}$, lies a 4d brane with an SU(2) gauge field ($A^\mathbf{a}_\m$) on it. On the other hand, its continuum counterpart was developed in \cite{Irges:2018gra, Irges:2020nap}. In both cases was proven that this model could account for a novel Higgs mechanism with naturally low cut-off proposing an alternative solution to the Higgs-hierarchy problem.
 
Here, however, we will consider the above 5d anisotropic orbifold lattice as the full spacetime where on the bulk lives a 5d $SU(2)$ gauge field, $A^{\mathbf{A}}_M$ with $\mathbf{A} = 1,2,3$ and a dimensionful coupling $g_5$, while on the 4d boundary the full SM and the DM are localized. Of course we may assume that all or part of the SM and the DM live on the 5d bulk and analyze the way that they are projected on the 4d boundary, but this is a more involved scenario and we leave it for a future work. Based on the previous argumentation, we study the possibility that the imprints of this bulk gauge field on our 4d spacetime play the role of a dynamical dark energy. 
\begin{figure}[t]
\centering
\includegraphics[width=9cm]{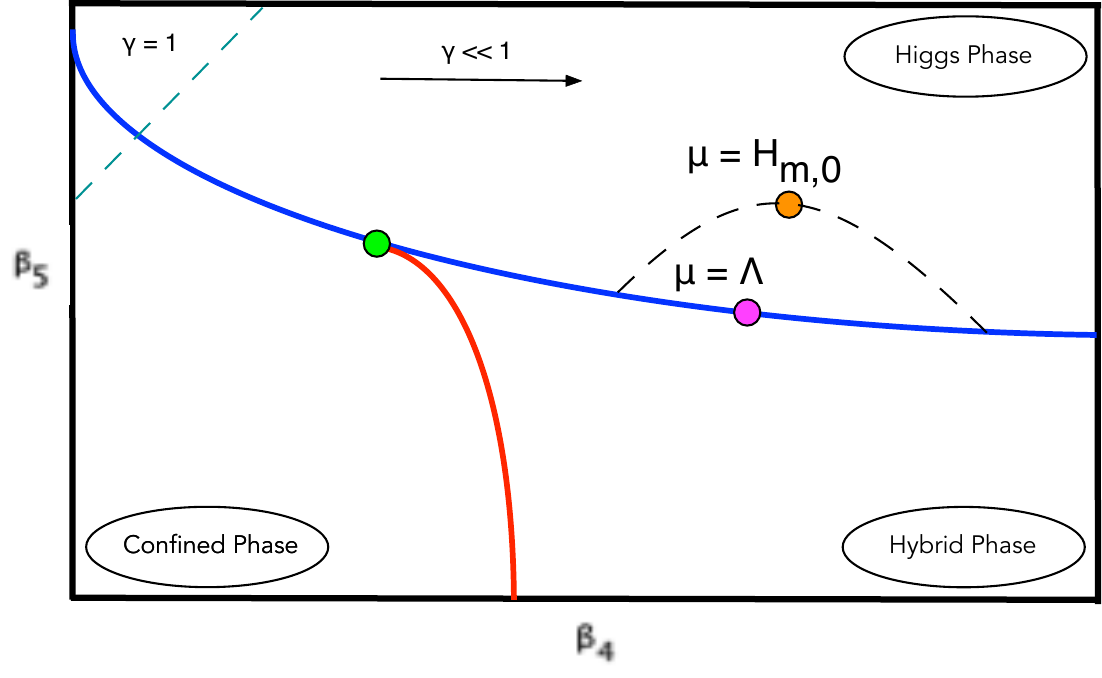}
\caption{\it 
A 2d cartoon picture of the phase diagram of our model. There is the Confined, the Hybrid and the Higgs phase, all of them separated by a 1st order phase transition. Each point on the blue-line corresponds to a finite cut-off. In this work we are interested only in the Higgs-Hybrid transition part which lies after the triple point (green dot) with $\g \ll 1$. There the purple-dot represents the energy scale of the localization 1st order phase transition, at the finite cut-off $\m = \L$, while the orange-dot denotes the energy scale of our universe today ($\m = H_{\rm m, 0}$). The black dashed line suggests that today the universe lives in the vicinity of the phase transition in the Higgs phase but not far from $\L$.
}
\label{Higgsphase2}
\end{figure} 
In particular, due to the non-trivial boundary orbifold conditions the projection of $A^{\mathbf{A}}_M$ on the boundary brane involves only its $U(1)$ subgroup, realized by the $A^3_\m \equiv A_\m$ part, as well as its fifth components $A_5^1$ and $ A_5^2$ which are combined to a complex scalar field $\phi \equiv A_5^1 + i A_5^2$. In that sense these extra degrees of freedom (dof), $A_\m$ and $\phi$, will accompany the visible and dark matter ones as prospect dynamical dark energy components. 
\\
Let us now emphasize on the characteristics of this construction which make it a distinguished UV completion of SM. Since the 4d effective DE action originates from the lattice, it will be constrained by the fundamental UV cut-off
\be\label{L45UV}
\L_{4,\rm UV} \sim \frac{\pi}{a_4} ~ \left(\L_{5, \rm UV } \sim \frac{\pi}{a_5} \right)
\ee
defining a very high energy scale as $(a_4, a_5) \to 0$. However, it is proven both non- and perturbatively that the NPGHU model naturally possesses also a finite, much lower, cutoff related to the localization of the 4d brane given by 
\be\label{LFa4a5}
\L \propto F(a_4^{-1},a_5^{-1}) 
\ee
as it is explained in \cite{Irges:2020nap}.
When the energy scale reaches this $\L$, the system exhibits a bulk-driven 1st order phase transition without affecting the dof that are purely localized on the boundary. The associated phase diagram of the model is schematically depicted on \fig{Higgsphase2} showing the existence of three phases, the Confined, the Hybrid and the Higgs phase. $\b_4$ and $\b_5$ are the dimensionless lattice couplings multiplying the plaquettes in the lattice action and $H_{\rm m,0}$ represents the Hubble rate today. The interesting reader may see \cite{Irges:2018gra, Irges:2020nap} for more details. Let us mention though that the NPGHU model was originally described purely on the lattice and was used as a probe solution to the Higgs hierarchy problem. It was shown non-perturbatively \cite{Irges:2006hg, Alberti:2015pha} that one obtains the SM value $\rho_{\rm SM} = m_H/m_W \approx 1.38$ only along the blue line and only in the vicinity of the localization transition where the model remains non-perturbatively 4d. Therefore NPGHU suggests that any viable 4d description in this framework exists only near the phase transition justifying our assumption that the Universe lies in this regime today.\\
The order of $\L$, which depends strongly on the level of the anisotropy and the lattice characteristics, as well as the importance of its existence is explicitly analyzed in \sect{PSGPI}. However, let us clarify that NPGHU non-perturbatively suggests that the mass ratio of the associated dof near the transition is of ${\cal O}(1)$ and is related with the lattice order parameter (the Polyakov loop). The latter defines the relevant scale and dictates when the localization happens. Here the Polyakov loop is associated with the $H_{\rm m,0}$ and the related localization cut-off is $\L \sim N H_{\rm m,0}$ with $N$ a numerical factor the value of which depends on the position on the blue line of \fig{Higgsphase2} and on how fast the anisotropy runs. The latter varies slowly and in order to avoid an unreasonable breaking of scale invariance via the Polyakov loop, a choice of $N\equiv {\cal O}(10)$ puts us near the continuum limit of the NPGHU model (close to the very right part of \fig{Higgsphase2}) reducing the fine tuning.
For any purpose of this work we will consider $\g \ll 1$ which lies after the triple point (green dot), therefore the only relevant for the boundary-observer transition will be the Higgs-Hybrid one. The latter involves the blue-line part on \fig{Higgsphase2} which includes the purple-dot showing the energy scale of the localization transition, $\m = \L$.
The so called Higgs phase exists only on the 4d boundary while the Hybrid phase characterizes the 4d bulk slices. Each point on the blue-line corresponds to a UV cut-off from the point of view of both $A_\m, \phi$ and $A^\mathbf{a}_\m$ \cite{Irges:2020nap}. This means that the running parameters (couplings and masses) in both sides increase with the same energy scale (the standard renormalization scale $\m$) and reach their maximum value on the same $\L$ simultaneously. More schematically, in \fig{Higgsphase} the Higgs-Hybrid phase transition is realized between the brane which lives on $n_5 = 0$ and the brane which lives on $n_5 = 1$.\\ 
Lattice simulations \cite{Irges:2006hg, Knechtli:2007ea, Irges:2012mp, Alberti:2015pha} have proven that as one approaches the vicinity of any point on the blue line from the Higgs phase, the model exhibits dimensional reduction via localization. The latter means that the closer one lies to the cut-off, then the more the 4d boundary decouples from the brane at $n_5 = 1$. When one hits exactly on the cut-off, the whole 5d space becomes fully layered and the boundary is completely decoupled from the bulk.
At that point the bulk-driven 1st order phase transition takes place and from the Higgs phase ($\m \lesssim \L$) the system moves to a Coulomb phase ($\m \gtrsim \L$). The appearance of effective mass scales on the boundary is due to its non-zero coupling with the bulk. On the other hand, as $\mu$ approaches the deep IR and the model moves towards the interior of the Higgs phase (top-right corner of \fig{Higgsphase2}) localization is lost, the interpretation of the spectrum as 4-dimensional is wrong and the full 5d lattice is obtained.
On the (naive) continuum limit the previous description is parameterized for the 4d boundary as follows: The Coulomb phase at $\m \gtrsim \L$ (no bulk-boundary interaction) is realized by a massless Scalar QED (SQED) with only dim-4 operators \cite{Irges:2018gra}. The Higgs phase at $\m \lesssim \L$ is described by a massless Lee-Wick SQED (LWSQED) model \cite{Irges:2020nap}. Notice that an interesting feature of the lattice origin of our 4d model is that polynomial terms are forbidden and there are no potential interactions in neither of the two phases.

Based on the above arguments, here we propose that our universe lives on the 4d boundary of this NPGHU model and today lies not very far from the vicinity of the cut-off (orange-dot on \fig{Higgsphase2}) towards the Higgs phase ($ \m \lesssim \L$). For the rest we neglect the details on how to produce the continuum actions from the lattice while the interested reader may see \cite{Irges:2018gra, Irges:2020nap} and references therein. 
Then, for scales $\m > \L$ the boundary is fully 4d-localized, in Coulomb phase, without any explicit knowledge about the bulk and the associated total action reads 
\be\label{StmbL}
S^{\rm 4d}_{\rm tot} = S^{\rm orb}_{\m > \L} + S_{\rm DM} + S_{\rm SM} \equiv \int d^4x \sqrt{-g} \left [  - \frac{1}{4} F_{\m\n}F^{\m\n}  + |D_\m \phi|^2 \right] + S_{\rm DM} + S_{\rm SM}
\ee
On the contrary, in the vicinity of the purple-dot on \fig{Higgsphase2}, for $\m < \L$ localization is still there but now the effects from the bulk-boundary interaction should be taken into account. A non-trivial (but incomplete as we explain below) analysis suggests that the total boundary action will now read    
\bea\label{StmsL}
S^{\rm 4d}_{\rm tot} &=& S^{\rm orb}_{\m < \L} + S_{\rm DM} + S_{\rm SM} \nn\\
 &\equiv& S^{\rm orb}_{\m > \L} + \int d^4x \sqrt{-g} \left [ \frac{c_\a}{2\L^2} \left( \nabla^\m F_{\m\n}\right)\left( \nabla_\m F^{\m\n}\right) 
- \frac{c_6}{\L^2} |D^\m D_\m \phi|^2 + \cdots \right] + S_{\rm DM} + S_{\rm SM}
\eea 
where $\phi$ is a complex scalar, $A_\m$ is a photon-like gauge field and as usual $F_{\m\n} = \nabla_\m A_\n - \nabla_\n A_\m$. The couplings $c_\a$ and $c_6$ as well as the cutoff have been introduced due to the higher derivative operators and they are complicated functions of the lattice couplings and spacings (see (2.2) of \cite{Irges:2020nap}). 
The covariant derivative is given by 
\be
D_\m = \nabla_\m - i g_4 A_\m
\ee
where $g_4$ is the dimensionless Abelian-gauge 4d coupling which encodes the information of the extradimensional and anisotropic origin of this model since it is defined as
\be\label{g4gg}
g_4^2 = \frac{g_5^2}{a_4} \g = g^2 \g
\ee
Everything in the above is adjusted to take into account that the model lives on a Friedmann-Robertson-Walker (FRW) universe with a background metric included in $\sqrt{-g}$ and given by
\be\label{FRW}
ds^2 = a(\tau)^{2b} d\tau^2 -  a(\tau)^2 \delta_{ij} dx^i dx^j
\ee
where $a(\tau)$ is the normalized with its today value, $a_0 $, scale factor and for $b = 1$ we have that $d\tau$ corresponds to the conformal time while if $b=0$ then $d\tau \to dt$, the cosmic time. Here we use the mostly negative metric signature $(+---)$.

Let us clarify momentarily that the presence of phantom (ghost) dof is due to the appearance of the higher derivative operators, the origin of which is not by hand but rather can be understood directly from the lattice origin of the model. Schematically, the anisotropic orbifold Wilson action contains plaquettes purely along the four-dimensional slices and plaquettes involving the fifth direction (see App. A of \cite{Irges:2020nap}), which after expanding in the lattice spacings becomes
\bea
S_{\rm lat}^{\rm orb} &=& \frac{\beta_4}{2N}\sum_{n_\mu}\sum_{\mu<\nu}{\rm tr}\!\left[1-U_{\mu\nu}^{b}(n_\mu)\right] +\frac{\beta_5}{2N}\sum_{n_\mu}\sum_{\mu}{\rm tr}\!\left[1-U_{\mu5}^{h}(n_\mu,0)\right] \nn\\
 &\equiv & \sum_{n_\mu}a_4^4\sum_{\mu<\nu}\left[\frac14F_{\mu\nu}^{3}F^{3\mu\nu} +\frac{1}{16\,\mu^2(a_4)}\left(\hat\Delta_\mu F^{3\mu\nu}\right)\left(\hat\Delta^\rho F^3_{\rho\nu}\right)+\cdots\right] \nn\\
  && + \sum_{n_\mu}a_4^4\sum_{\mu}\left[|\hat D_\mu\phi|^2 +\frac{1}{4\,\mu^2(a_4)}|\hat D_\mu\hat D^\mu\phi|^2+\cdots\right]
\eea
where the orbifold projection leaves on the boundary the Abelian gauge component $A_\mu\equiv A_\mu^3$ and the scalar field $\phi\equiv A_5^1+iA_5^2 $ and and $\mu(a_4)$ is the intrinsic scale associated with the remnant lattice spacing and the finite localization cut-off as we have already explained. 
Therefore, already at the lattice-spacing expansion level,
\be
S_{\rm lat}^{\rm orb} \supset \sum_{n_\mu}a_4^4\left[ \frac{1}{16\,\mu^2(a_4)}(\hat\Delta_\mu F^{3\mu\nu})(\hat\Delta^\rho F^3_{\rho\nu}) +\frac{1}{4\,\mu^2(a_4)}|\hat D_\mu\hat D^\mu\phi|^2 \right]
\ee
Considering the (naive) continuum limit we obtain \eq{StmsL}.
In that sense, the higher-derivative terms are the leading corrections inherited from the orbifold plaquette expansion near the localization transition. Once these terms are retained, the auxiliary-field form of the theory necessarily contains the phantom scalar and gauge poles displayed below (\eq{Rghosts}), rather than an independently imposed quintom sector.

Now, the $\cdots$ in action \eq{StmsL} indicate higher than dim-6 derivative operators, so before we move on let us clarify the physical picture that we imply here which justifies the inclusion of only dim-6 operators in our late-time effective action.
In order to obtain $S^{\rm orb}_{\m > \L}$ or $S^{\rm orb}_{\m < \L}$ one should start from the lattice plaquette action of our model in \fig{Higgsphase} and consider the (naive) continuum limit. The latter inherits in principle the action (after tedious algebra) with an infinite tower of higher (dimension-$d$) derivative operators, $O_d$, such as
\be
O_6 = \left \{ \frac{|D^\m D_\m \phi|^2}{\L^2}, \frac{\left( \nabla^\m F_{\m\n}\right)^2}{\L^2} \right\}, ~ O_8 = \left\{ \frac{|D^\rho D^\m D_\m \phi|^2}{\L^8}, \frac{\left( \nabla^\rho\nabla^\m F_{\m\n}\right)^2}{\L^8} \right\},~ \cdots  
\ee
However, one can truncate the series according to the level of localization even though its detailed dynamical nature is not understood up to date beyond lattice simulations. 
However to maintain consistency, the non-perturbative four-dimensional characteristics of NPGHU model (depicted on the associated phase diagram) should be matched by its continuum counterpart. In \cite{Irges:2020nap} (\cite{Irges:2018gra}) has shown that in the vicinity of the phase transition (exactly on the cut-off) the full non-perturbative picture of the NPGHU phase diagram is already successfully reproduced by dim-6 (dim-4) operators.
The reproduction was done by matching the RG flows between the Hybrid and Higgs phase under the presence of the dim-6 operators while the inclusion of higher than six dimensional operators had a mild effect on both runnings near the vicinity of the transition. On the other hand, these operators become necessary to reproduce the rest of the phase diagram in the deep Higgs phase, meaning in the deep IR. Therefore the inclusion of only dim-6 operators in the regime of the localization transition is an implicit characteristic which originates by demanding that the NPGHU model and its continuum counterpart remain consistent. Of course further investigation about the correlation between truncation and localization is needed but it is outside the scope of the current work.
For the rest therefore we keep that the closer we are to the phase transition, the more the boundary decouples from the bulk and a more truncated 4d effective action can reproduce the physics. Exactly on the phase transition, the characteristics of our model can be reproduced by an action truncated to dim-4.\\
Justifying the presence of only dim-6 operators, however, is not the end of the story regarding the consistency of this action. As it was first shown in \cite{Irges:2019bzb} (for a similar recent analysis\footnote{The work of \cite{Cembranos:2025dor} is concentrated on the field theoretical counterpart of the well-known Pais-Uhlenbeck oscillator instability problem, however the authors' solution to this problem is applicable here and matches the one proposed in \cite{Irges:2019bzb}.} see also \cite{Cembranos:2025dor}) the basis of \eq{StmsL} is not complete since under the most general and gauge-invariant field reparameterization \cite{Irges:2020nap}, one should take into account also a non-trivial shift in the measure of the associated path-integral. The former then inherits the action with one more dof for each phantom field, in the same spirit with the BRST symmetry, the so called R-ghosts.
Then, the complete version of $S^{\rm orb}_{\m < \L}$ reads 
\be\label{StmsL2}
S^{\rm orb}_{\m < \L} \equiv S^{\rm orb}_{\m > \L} + \int d^4x \sqrt{-g} \left [ \frac{c_\a}{2\L^2} \left( \nabla^\m F_{\m\n}\right)\left( \nabla_\m F^{\m\n}\right) - \frac{c_6}{\L^2} |D^\m D_\m \phi|^2  + {\cal L}_{{\rm R},\chi} + {\cal L}_{{\rm R}, B} \right ]
\ee
where
\be\label{Rghosts}
{\cal L}_{{\rm R},\chi} =  |\partial_\m \chi|^2 + \frac{\L^2}{ c_6}  |\chi|^2 - \l_\chi |\chi|^2  |\phi|^2 ~~ {\rm and} ~~ {\cal L}_{{\rm R}, B} =  - \frac{1}{4} B_{\m\n}B^{\m\n} -\frac{1}{2} \frac{\L^2}{2 c_\a} B_{\m}^2
\ee
correspond to the Lagrangian of the scalar and gauge R-ghosts, $\chi$ and $B_\m$ respectively.
Notice that both have the correct sign for the kinetic term but wrong mass-term sign meaning that they are tachyonic dof. The magnitude of $\l_\chi$ is a complicated function of the lattice characteristics which are out of our control so here we will constraint it using phenomenological arguments.\\
As it was explained in \cite{Irges:2019bzb}, the presence of the R-ghosts is catalytic for the validity of the theory since they cure the Ostrogradsky instabilities which initially appeared due to the problematic, extra, pole of the propagators.
It was shown that one needs to impose specific pole-cancellation conditions at both tree- and loop-level in order to obtain a consistent spectrum with the former forcing the R-ghost masses to be equal to the phantom ones, as it will be clarified momentarily. An important feature of the above analysis is that the pole cancellation is not completely equivalent with diagrammatic cancellations, therefore in principle both the R-ghosts and the phantoms participate in the spectrum.
However, as we explain later on for the background evolution of the DE and the associated EoS, both R-ghosts will be negligible since their evolution will always remain subdominant compared to the rest of the spectrum.

The problematic extra poles mentioned above appear due to the presence of the Higher Derivative Operators (HDO) in the spectrum. In the following we make the presence of these Ostrogradsky ghosts, named effectively as phantom fields, clear and facilitate our analysis by decoupling the extra dof. This is done here in the usual manner by introducing auxiliary\footnote{One should be careful to use the covariant derivatives when necessary and to avoid integration by parts. Under this steps and since the field redefinitions are happening on Lagrangian level, the final result should be independent of the background geometry.} scalar and gauge fields and making the change of variables 
\be\label{phitophi12AtoA12}
\phi = \phi_1 - \phi_2 ~~ {\rm and} ~~ A_\m = A_{\m,1} - A_{\m,2}
\ee
respectively \cite{Chivukula:2010kx}. Then the action of the boundary-projected fields at $\m \lesssim \L$ becomes 
\bea\label{SDE3}
S^{\rm orb}_{\m < \L} &\equiv& \int d^4x \sqrt{-g} \Bigg [ - \frac{1}{4} F_{\m\n,1}F^{\m\n,1} +  \frac{1}{4} F_{\m\n,2}F^{\m\n,2} -\frac{1}{2} m_{A_2}^2 A_{\m,2}^2 + |\partial_\m \phi_1|^2 - |\partial_\m \phi_2|^2  \nn\\
&& + m_{\phi_2}^2 |\phi_2|^2 - i g_4 \left( A_{\m, 1} - A_{\m, 2} \right) \left( \phi_1 \partial^\m \bar \phi_1 - \phi_2 \partial^\m \bar \phi_2 + {\rm h.c.} \right) \nn\\
   &&- g_4^2 \left( A_{\m, 1} - A_{\m, 2} \right)^2 \left( |\phi_1|^2 - |\phi_2|^2 \right) + {\cal L}_{{\rm R},\chi} + {\cal L}_{{\rm R}, B} \Bigg ]
\eea
with
\be\label{Rghosts2}
{\cal L}_{{\rm R},\chi} =  |\partial_\m \chi|^2 + m_{\phi_2}^2  |\chi|^2 - \l_\chi |\chi|^2  |\phi_1 - \phi_2|^2 ~~ {\rm and} ~~ {\cal L}_{{\rm R}, B} =  - \frac{1}{4} B_{\m\n}B^{\m\n} -\frac{1}{2} m_{A_2}^2 B_{\m}^2
\ee
where note that we have exchanged the couplings $c_\a, c_6$ and the cut-off with the gauge and scalar phantom (and R-ghost) mass through $m_{A_2}^2 = \L^2/c_\a$ and $m_{\phi_2}^2 = \L^2/c_6$ respectively.\\
In summary and based on the above discussions, here we consider that today the universe sits somewhere near the vicinity of the phase transition towards the Higgs phase. Localization is still strong but the bulk is not completely decoupled from the boundary and only dim-6 operators are needed to capture the physics. The cut-off then is assumed to be $\L \gtrsim H_{\rm m,0}$ as we will also justify later on.
Given the previous arguments our starting point is the total action 
\be
S_{\rm tot} = S_{\rm DE} + S_{\rm DM} + S_{\rm SM} 
\ee
where the effective dark energy part will read
\be\label{SDEbsL}
S_{\rm DE} \equiv 
\left\{
\begin{alignedat}{3} 
 & S^{\rm orb}_{\m > \L} & ~~{\rm for } ~~ \m \ge \L ~~\rm{\eq{StmbL}}\\
& &  &  \\
  & S^{\rm orb}_{\m < \L} & ~~{\rm for } ~~ \m < \L ~~\rm{\eq{SDE3}}
\end{alignedat}
\right. 
\ee
The main goal of this work is to evaluate the dark energy EoS which in our case will be given by the effective action $S^{\rm orb}_{\m < \L}$ at relatively late times. Notice that $S^{\rm orb}_{\m < \L}$ is an effective quintom model and we will study whether it could provide a viable Quintom-B scenario free of immediate instabilities within the controlled effective regime (which is usually extremely hard to achieve) or not.
In what follows and since the various sectors interact only gravitationally among each other, we will forget about $S_{\rm SM}$ and keep as a barotropic fluid the DM's contribution.
In addition, we will forget about the 5d origin of this model exploiting only implicitly this information via the couplings, the anisotropy and the localization effect. 

\section{The EoS $w_q$ of the 4d effective quintom-model}\label{TEoSwqOTEQM}
In the previous section we motivated the possibility that the nature of dark energy originates from the components of 5d non-Abelian gauge field projected on the 4d boundary of a 5d anisotropic orbifold lattice.
Of course one should always move hand-by-hand with the experimental data which are presenting some very convincing hints towards a DDE. Such arguments were very recently presented by the latest measurements of BAO by DESI \cite{DESI:2024mwx, DESI:2025zgx}.
Since our goal is to entertain the possibility that our model can produce late-time histories compatible with the DESI-motivated preference for dynamical-dark-energy rendering NPGHU as a viable candidate for DE's origin, here we develop the necessary technology in order to pin down the evolution of $w_q$ and to compare it with the observed one ($w_{\rm obs}$). Keep in mind that recent DESI-era reconstructions suggest a preference for quintom-like behavior \cite{Yang:2024kdo} and more precisely the Quintom-B type evolution since $w_{\rm obs}$ starts in a phantom phase ($< -1$) at early times and crosses the cosmological boundary $-1$ to enter a quintessence phase ($w > -1$) at late times.
In that sense, we essentially need to check whether $w_q$ of the NPGHU-induced effective quintom model \eq{SDE3} could be of Quintom-B type or not.
The needed ingredients for the calculation of the EoS are the background energy and pressure densities, $\rho_{\rm DE}$ and  $p_{\rm DE}$ respectively.
These quantities are determined by the Stress-Energy (SE) tensor $T_{\m\n,\rm DE}$ given by 
\be\label{SETmn}
T_{\m\n,\rm DE} = \frac{2}{\sqrt{-g}} \frac{\d {\cal L}_{\rm DE}}{\d g^{\m\n}} 
\ee
as $\rho_{\rm DE} = g^{00}\langle T_{00,\rm DE} \rangle$ and $p_{\rm DE} = -g^{ij}\langle T_{ij,\rm DE} \rangle/3$ where we have considered the isotropic-averaging of the SE parts to ensure the isotropy of the FRW spacetime. In the following, even though implied, we drop this notation for simplicity.
The associated equation of state then is given by
\be
w_q = \frac{p_{\rm DE}}{\rho_{\rm DE}}
\ee
As we have already mentioned, we facilitate our analysis regarding the running of $w_q$ with time by considering that the anisotropy in the 5d orbifold lattice is small, $\g \ll 1$. The latter just makes the calculations simpler since sets $ g_4^2 \ll g_4 \ll 1$ in \eq{SDEbsL} such that scalar-gauge interactions are very feeble and we neglect them at leading order in the small-anisotropy expansion.
In principle there is a range of smallness for $\g$ and one may just consider a value such that $g_4 \gg g_4^2$ neglecting only the quadratic interactions. This choice will lead to a different EoS which will clearly have a more complicated form without, however, giving any deeper insight.
For the current work the presence of interactions will only introduce some complications without altering drastically our results.

However, in order to further justify our choice to consider negligible interactions here, recall that as we have already mentioned in text below \eq{LFa4a5}, NPGHU model was proven to provide a reasonable 4d description only near the localization transition and after the triple point of the phase diagram. This case is naturally lying in the regime where the fifth dimension is much larger than the fourth one, or in other words $a_5 \gg a_4$, which in turn suggests a small anisotropy limit since $\g = a_4/a_5 \ll 1$. Then from \eq{g4gg} we see that $g_4 = g_5/\sqrt{a_5}$ so for a large $a_5$ and a perturbative $g_5$ we already get $g_4 < 1 $ and $g_4^2 \ll 1$. This statement should be reconsidered when one goes to regions above the triple point in \fig{Higgsphase2}, however this will correspond to high moments in the running of the anisotropy which is very slow as it was shown in \cite{Irges:2020nap} so it may affect only the very high energy scales. 
Finally, under $g_4 <1$ we briefly show at the end of \sect{TEoSEQM} that even if we keep the interaction terms, then any shift in the background evolution remains negligible compared to the retained free kinetic and mass contributions.

\subsection{The role of the 5d-lattice originated dof}\label{TEoSEQM}
Based on the previous argumentation and the chosen hierarchy of couplings, we neglect both the linear and quadratic in $A_{\m, (1,2)}$ interactions and the effective quintom action of \eq{SDE3} is simplified to  
\be\label{SDEgs1}
S^{\g \ll 1}_{{\rm DE}} \approx \int d^4x \sqrt{-g} \left[ - \frac{1}{4} F_{\m\n,1}F^{\m\n,1} +  \frac{1}{4} F_{\m\n,2}F^{\m\n,2} -\frac{1}{2} m_{A_2}^2 A_{\m,2}^2 + |\partial_\m \phi_1|^2 - |\partial_\m \phi_2|^2 + m_{\phi_2}^2  |\phi_2|^2 \right] + S_\chi + S_{B_\m}
\ee
which provides us with the following SE tensor 
\bea\label{TDEgs1}
T^{\g \ll 1}_{\m\n,\rm DE} &=&  - g^{\a\b} F_{\m\a,1}F_{\n\b,1} +  g^{\a\b} F_{\m\a,2}F_{\n\b,2} -m_{A_2}^2 A_{\m,2}A_{\n,2} + 2 \partial_\m \phi_1 \partial_\n \bar \phi_1 - 2\partial_\m \phi_2 \partial_\n \bar \phi_2  \nn\\
&& -  g_{\m\n}\Bigg [ - \frac{1}{4} F_{\a\b,1}F^{\a\b,1} +  \frac{1}{4} F_{\a\b,2}F^{\a\b,2} -\frac{1}{2} m_{A_2}^2 A_{\a,2}^2 + |\partial_\a \phi_1|^2 - |\partial_\a \phi_2|^2 + m_{\phi_2}^2 |\phi_2|^2 \Bigg ] + T_{\m\n, \chi} + T_{\m\n, B_\m} \nn
\eea
Notice that we keep the contribution of the R-ghosts implicit since their effect on the background EoS is negligible as we will show momentarily. Moreover, it is straightforward to obtain the R-ghost SE combining \eq{Rghosts} with \eq{SETmn}. 
For the rest of this analysis we exploit the spatially flat, homogeneous and isotropic FRW background, working with conformal time ($b = 1$ in \eq{FRW}), 
\be\label{gmnb}
\bar g_{\m\n} = a(\tau)^2 \left( d\tau^2 - dx^2 \right)
\ee
which allows us to consider only the background field values, such as $\phi_{(1,2)}(\tau, \vec x) \equiv  \varphi_{(1,2)}(\tau)$, $A_\m (\tau, \vec x) \equiv A_\m (\tau)$, $\chi(\tau, \vec x) \equiv  \chi_0(\tau)$ and $B_\m (\tau, \vec x) \equiv B_\m (\tau)$.

In that terms and before we move on, we now justify our argument about the negligible effect of the R-ghosts on the background EoS. Here we choose to focus only on the scalar part of \eq{SDEgs1} in order to make the calculation transparent, however we have checked that similar results hold also for the gauge sector of our model.
The associated action, along with \eq{Rghosts2}, then reads
\be\label{Sscalar}
S_{\rm scalar} = \int d^4x \sqrt{-g} \left[ |\partial_\m \phi_1|^2 - |\partial_\m \phi_2|^2 + |\partial_\m \chi|^2 +  m_{\phi_2}^2  |\phi_2|^2  +  m_{\phi_2}^2  |\chi|^2 - \l_\chi |\chi|^2  |\phi_1 - \phi_2|^2 \right]
\ee
and using
\be
\frac{1}{\sqrt{-g}} \partial_\m \left( \sqrt{-g} \frac{\partial {\cal L_{\rm scalar}}}{\partial (\partial_\m  \bar \Phi)} \right) = \frac{ \partial {\cal L_{\rm scalar}}}{\partial \bar \Phi}
\ee
with $\Phi = \{\varphi_{(1,2)}, \chi_0 \} $ and $\bar \Phi = \{\varphi^*_{(1,2)}, \chi^*_0 \} $, we obtain the system of the background scalar eom
\bea\label{phi1phi2chi0}
\ddot\varphi_{1,n} + \left[2 + \frac{\dot{\cal H}}{{\cal H}} \right] \dot \varphi_{1,n} + e^{2N}  \l_\chi \frac{|\varphi_{2,\a}|^2}{{\cal H}^2} |\chi_{0,n}|^2 (\varphi_{1,n} - \varphi_{2,n}) &=& 0 \nn\\
 \ddot\varphi_{2,n} + \left[2 + \frac{\dot{\cal H}}{{\cal H}} \right] \dot\varphi_{2,n} + e^{2N}\frac{ m^2_{\phi_2}}{{\cal H}^2} \varphi_{2,n} + e^{2N}  \l_\chi \frac{|\varphi_{2,\a}|^2}{{\cal H}^2} |\chi_{0,n}|^2 (\varphi_{1,n} - \varphi_{2,n}) &=& 0 \nn\\
 \ddot\chi_{0,n} + \left[2 + \frac{\dot{\cal H}}{{\cal H}} \right] \dot\chi_{0,n} - e^{2N}\frac{ m^2_{\phi_2}}{{\cal H}^2} \chi_{0,n} + e^{2N}  \l_\chi \frac{|\varphi_{2,\a}|^2}{{\cal H}^2} \chi_{0,n} |(\varphi_{1,n} - \varphi_{2,n})|^2 &=& 0
\eea
In the above, as we will see more explicitly in the next section, in order to facilitate our analysis we choose to work with dimensionless quantities according to the following algorithm: First we shift from conformal time derivatives to efolds, $N = \ln a/a_0$, via $d\tau = dN/{\cal H}$. The notation $\cdot$ represents derivative with respect to $N$ as we explain in App. \ref{dBasis}.
\begin{figure}[t]
\centering
\includegraphics[width=8.5cm]{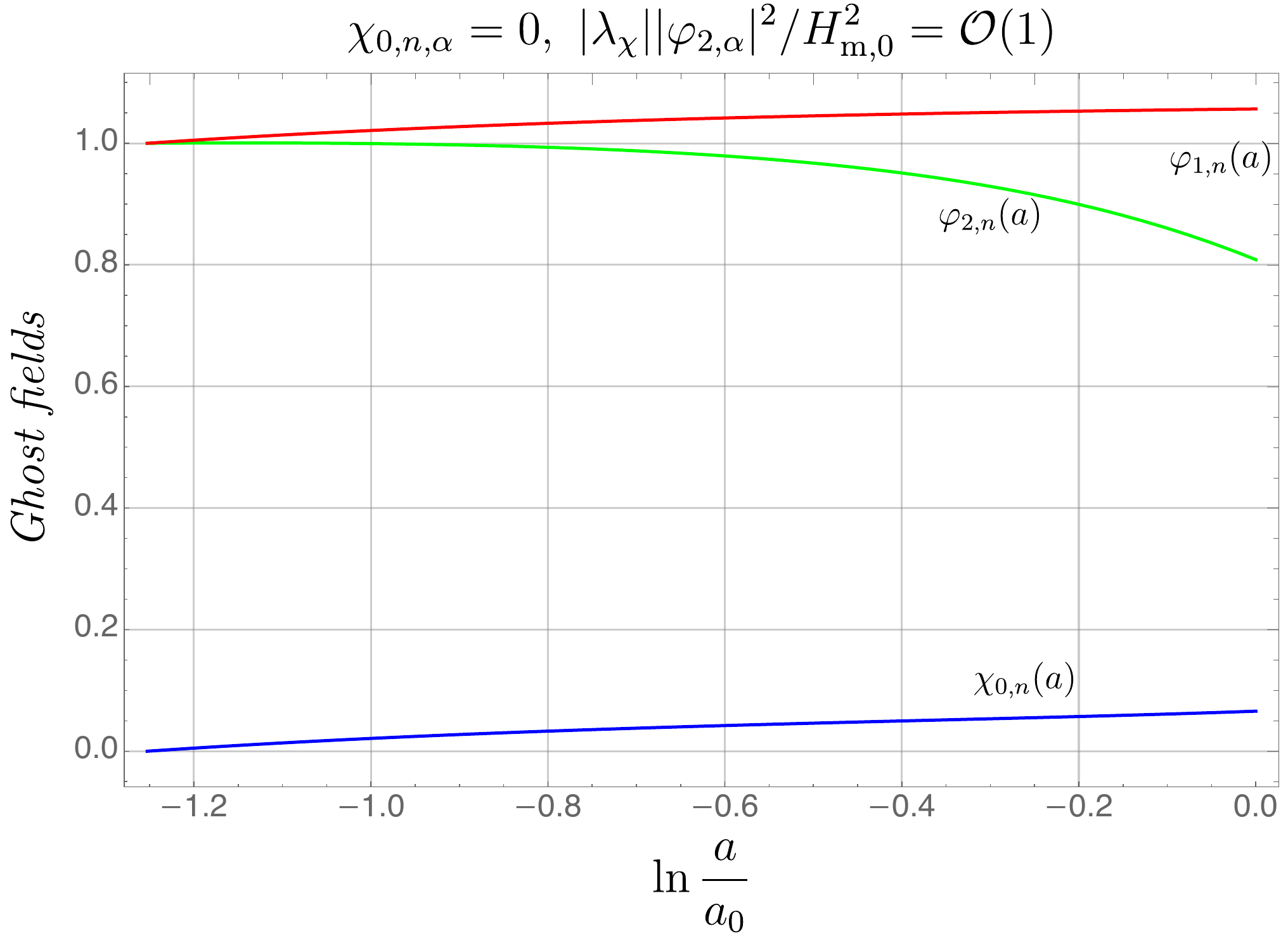} \quad
\includegraphics[width=8.5cm]{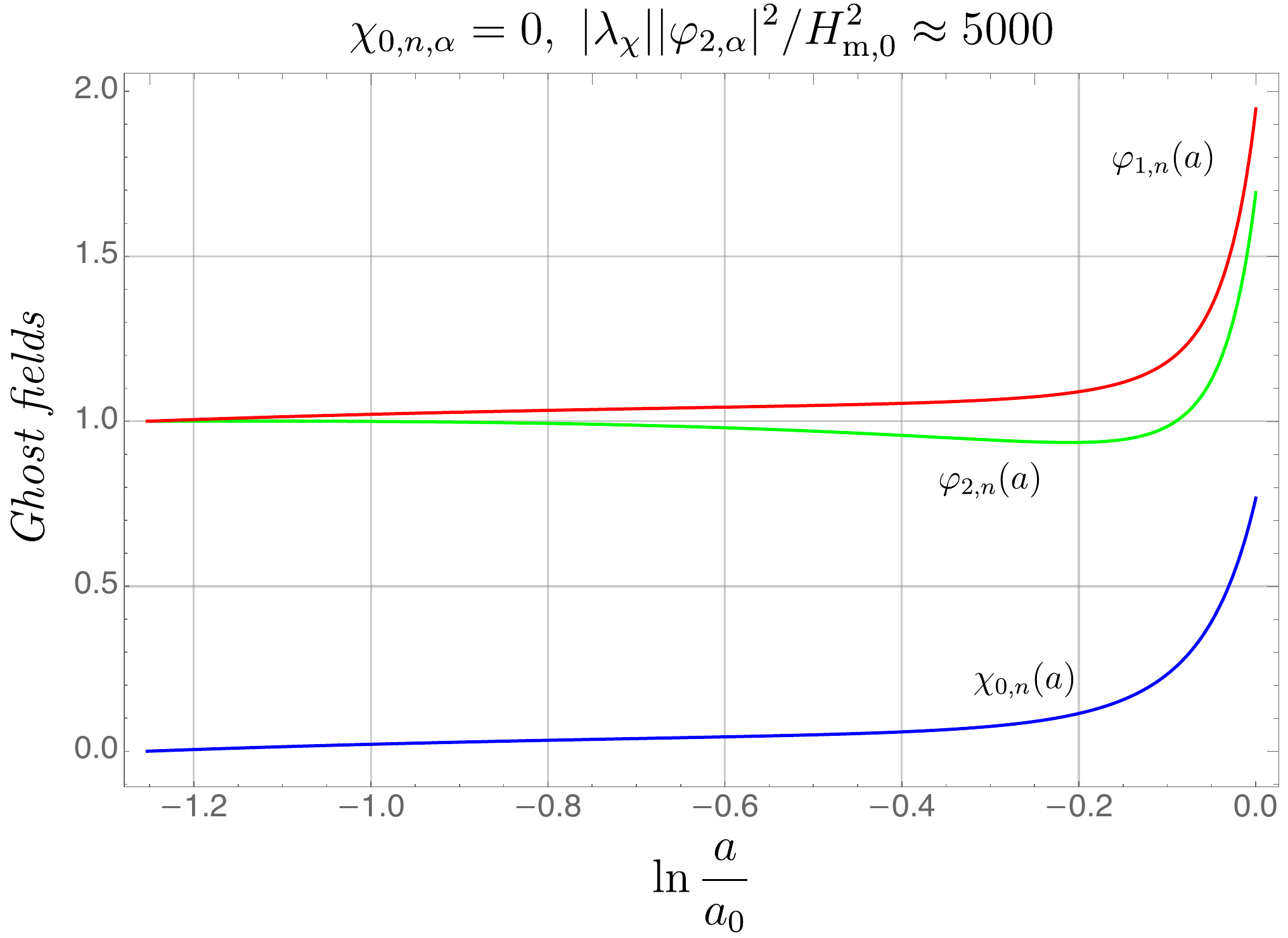} 
\caption{Left panel:{\it The evolution of the R-ghost $\chi_{0,n}$ remains always subdominant compared to the phantom $\varphi_{2,n}$ for $|\l_\chi||\varphi_{2,\a}|^2/H^2_{\rm m,0} =  {\cal O}(1)$ and $\chi_{0,n, \a} = 0$.} Right panel: {\it Given $\l_\chi < 0$, increasing the quartic coupling we reach a number above which $\chi_{0,n}$ surpasses the phantom at the very late redshift and dominates the EoS. }}
\label{chi-phantom}
\end{figure} 
The background fields, collectively denoted as $\Phi$, admit their initial value at $\m \equiv \L$ which corresponds to $N = N_\a$ and gives $\Phi_\a \equiv \Phi(N_\a)$. Then we choose to normalize all the fields with the initial value $|\varphi_2(N_\a)| \equiv |\varphi_{2,\a}| $ defining the normalized fields as $\Phi_n = \Phi/|\varphi_{2,\a}|$. ${\cal H} = H_{\rm m,0} F(N)$ denotes the comoving Hubble rate given by Friedmann’s first equation as we analytically describe in \eq{fFe} in the next section. We solve the above system imposing democratic initial conditions for the normalized $\varphi_{(1,2)}$ such that $\varphi_{1,n}(N_\a) = \varphi_{2,n}(N_\a) = 1$. On the other hand, for the R-ghost there is no such freedom since we are forced to choose $|\chi_0(N_\a)| = 0$ exactly on $\m = \L$ (or in other words on $N = N_\a$) as we explain in App. \ref{CTIFV}.

The question to be addressed here is whether $\chi_0$ is subdominant to $\varphi_2$ (and $\varphi_1$) during the late times such that it will not affect the EoS or not.
The simplest way to answer this is by comparing the evolution of $\varphi_{2,n}$ and $\varphi_{1,n}$ with $\chi_{0,n}$ in order to see which one dominates during the low redshifts.
Note that the vanishing initial value for $\chi_0$ is crucial to keep its evolution suppressed compared to the phantom, however it is not sufficient. One should also take into account the order of the coupling $\l_\chi $ compared to $|\varphi_{2,\a}|^2/H^2_{\rm m,0}$.
In order to see the effect of these arguments we proceed by solving the system of \eq{phi1phi2chi0} numerically, since the non-trivial coupling among the fields constrains our analytical power. Assuming initially that $|\l_\chi| |\varphi_{2,\a}|^2/H^2_{\rm m,0} =  {\cal O}(1)$, for either $\l_\chi > 0$ or $<0$, we show the subdominant evolution of the R-ghost compared to the phantom for all the redshift regime until today on the left panel of \fig{chi-phantom}. Then, on the right panel we present the same evolution but now considering $|\l_\chi| |\varphi_{2,\a}|^2/H^2_{\rm m,0} \approx 5000 $ keeping the initial conditions fixed and choosing $\l_\chi < 0$ since for $\l_\chi > 0$ the R-ghost remains always subdominant.
In total we find that 
the contribution of the R-ghost to the background EoS remains negligible compared to both $\varphi_{2,n}$ and $\varphi_{1,n}$, as long as $|\chi_0(N_\a)| = 0$ and
\be\label{lchibound1}
|\l_\chi| \lesssim 5000 \frac{H^2_{\rm m,0} }{|\varphi_{2,\a}|^2} \Rightarrow \frac{H_{\rm m,0} }{\sqrt{|\l_\chi|} |\varphi_{2,\a}|} \gtrsim 0.042
\ee
for $\l_\chi < 0$.
As we will justify in \sect{PSGPI}, assuming that $|\varphi_{2,\a}|^2 \equiv {\cal H}^2(N_\a) \approx \left(10 H_{\rm m,0} \right)^2$
\be\label{lchibound2}
|\l_\chi| \lesssim 50 
\ee
which is a well suited value for the perturbative analysis that we consider here.\\
In summary, even though the presence of the R-ghosts is important to cure the pole instability of the phantom fields, their background evolution is not strong enough to influence the EoS when it is contrasted to $\varphi_{(1,2)}$ (and $A_{\m, (1,2)}$). Therefore, for the rest of the investigation in this section we will neglect the presence of $\chi$ and $B_\m$, simplifying a lot our analysis. However, R-ghosts will be reintroduced in the next section where we will consider perturbations around the background and the stability of our  model will be put under tension.

\subsubsection*{The evolution of the background EoS under the physical and phantom fields}
Back to the calculation of the background equation of state for our effective quintom model and based on the previous discussions, the relevant for us SE parts (after neglecting the R-ghosts) will read
\be\label{Tgs100}
T^{\g \ll 1}_{00,\rm DE} \approx  |\varphi'_1|^2 - |\varphi'_2|^2 + \frac{a^2}{2} \left[ \mathbf{E}^2_1 + \mathbf{B}^2_1 \right] - \frac{a^2}{2} \left[ \mathbf{E}^2_2 + \mathbf{B}^2_2 \right] - \frac{1}{2} m_{A_2}^2 \left[ A_{0,2}^2 + A_{i,2}^2 \right] - a^2m_{\phi_2}^2 |\varphi_2|^2 
\ee
for the $00$ component and 
\be\label{Tgs1ij}
T^{\g \ll 1}_{ij,\rm DE} \approx  \left( |\varphi'_1|^2 - |\varphi'_2|^2 + \frac{a^2}{6} \left[ \mathbf{E}^2_1 + \mathbf{B}^2_1 \right] - \frac{a^2}{6} \left[ \mathbf{E}^2_2 + \mathbf{B}^2_2 \right] - \frac{1}{6} m_{A_2}^2 \left[ 3A_{0,2}^2 - A_{i,2}^2 \right] + a^2m_{\phi_2}^2 |\varphi_2|^2 \right) \d_{ij}
\ee
for the $ij$ isotropic averaged component respectively. The primes denote derivative with respect to conformal time.
For the previous calculation we used the following relations (in accordance to \cite{Durrer:2024ibi}) 
\bea\label{EiBk}
F_{0i} &=& a^2 E_i ~~ {\rm and} ~~ F^{0i} = -\frac{1}{a^2} E^i \nn\\
F_{ij} &=&  a^2 \varepsilon_{ijk} B^k ~~ {\rm and} ~~  F^{ij} = \frac{1}{a^4} F_{ij} 
\eea
where the above define the background electric and magnetic fields as measured by a comoving observer on FRW spacetime.
Since the gauge ghost is massive, then $A_{i,2}$ has in principle 3 polarization modes (2 transverse and 1 longitudinal). However, as we explain in App. \ref{CTIFV}, consistency of the model before and after the localization phase transition is ensured when the ghost longitudinal modes are initially not populated and their evolution is absent or negligibly small.
Therefore, in the relations above and also for the rest of this work the gauge phantom is essentially realized only by the transverse modes, $A_{i,2} \equiv A^T_{i,2}$.

Based on the relations \eq{Tgs100} and \eq{Tgs1ij} we now calculate $\rho_{\rm DE}$ and $p_{\rm DE}$ in conformal time via
\bea\label{rhoDEsg}
\rho_{\rm DE} &\approx& \frac{|\varphi'_1|^2}{a^2} - \frac{|\varphi'_2|^2}{a^2} + \frac{1}{2} \left[ \mathbf{E}^2_1 + \mathbf{B}^2_1 \right] - \frac{1}{2} \left[ \mathbf{E}^2_2 + \mathbf{B}^2_2 \right] - \frac{1}{2 a^2} m_{A_2}^2 \left[ A_{0,2}^2 + A_{i,2}^2 \right] - m_{\phi_2}^2 |\varphi_2|^2 
\nn\\
 \rho_{\rm DE} &\approx&  \frac{|\varphi'_1|^2}{a^2} - \frac{|\varphi'_2|^2}{a^2} + \frac{1}{2} \left[ \mathbf{E}^2_1 + \mathbf{B}^2_1 \right] - \frac{1}{2} \left[ \mathbf{E}^2_2 + \mathbf{B}^2_2 \right] + V_{\rho} 
\eea
and
\bea\label{pDEsg}
p_{\rm DE} &\approx& \frac{|\varphi'_1|^2}{a^2} - \frac{|\varphi'_2|^2}{a^2} + \frac{1}{6} \left[ \mathbf{E}^2_1 + \mathbf{B}^2_1 \right] - \frac{1}{6} \left[ \mathbf{E}^2_2 + \mathbf{B}^2_2 \right] - \frac{1}{6 a^2} m_{A_2}^2 \left[ 3 A_{0,2}^2 - A_{i,2}^2 \right] + m_{\phi_2}^2 |\varphi_2|^2 
\nn\\
p_{\rm DE} &\approx& \frac{|\varphi'_1|^2}{a^2} - \frac{|\varphi'_2|^2}{a^2} + \frac{1}{6} \left[ \mathbf{E}^2_1 + \mathbf{B}^2_1 \right] - \frac{1}{6} \left[ \mathbf{E}^2_2 + \mathbf{B}^2_2 \right] + V_p
\eea
respectively, where have defined for simplicity the energy and pressure potential terms
\be
V_{\rho} = - m_{\phi_2}^2 |\varphi_2|^2 - \frac{1}{2 a^2} m_{A_2}^2 \left[ A_{0,2}^2 + A_{i,2}^2 \right]
\ee
and
\be
V_p = m_{\phi_2}^2 |\varphi_2|^2 - \frac{1}{6 a^2} m_{A_2}^2 \left[ 3A_{0,2}^2 - A_{i,2}^2 \right]
\ee
and note that $V_{\rho} \ne - V_p$ in our case.\\
In that sense we can obtain a relatively compact form regarding the EoS give by
\be\label{wqDEgs1}
w_q = \frac{|\varphi'_1|^2 - |\varphi'_2|^2 + \frac{a^2}{6} \left[ \mathbf{E}^2_1 + \mathbf{B}^2_1 \right] - \frac{a^2}{6} \left[ \mathbf{E}^2_2 + \mathbf{B}^2_2 \right] + a^2 V_p}{|\varphi'_1|^2 - |\varphi'_2|^2 + \frac{a^2}{2} \left[ \mathbf{E}^2_1 + \mathbf{B}^2_1 \right] - \frac{a^2}{2} \left[ \mathbf{E}^2_2 + \mathbf{B}^2_2 \right] + a^2 V_{\rho}}
\ee
or after some massaging by
\be\label{wqDEgs1b}
w_q = - 1 + 2 \frac{|\varphi'_1|^2 - |\varphi'_2|^2 +  \frac{a^2}{3} \left[ \mathbf{E}^2_1 + \mathbf{B}^2_1 \right] - \frac{a^2}{3} \left[ \mathbf{E}^2_2 + \mathbf{B}^2_2 \right] - \frac{1}{6} m_{A_2}^2 \left[ 3A_{0,2}^2 + A_{i,2}^2 \right] }{|\varphi'_1|^2 - |\varphi'_2|^2  + \frac{a^2}{2} \left[ \mathbf{E}^2_1 + \mathbf{B}^2_1 \right] - \frac{a^2}{2} \left[ \mathbf{E}^2_2 + \mathbf{B}^2_2 \right] - a^2 m_{\phi_2}^2 |\varphi_2|^2 - \frac{1}{2} m_{A_2}^2 \left[ A_{0,2}^2 + A_{i,2}^2 \right] }
\ee
The above expression shows that the EoS in our case is that of an extended quintom model and despite its simple form it is not straightforward to determine the exact way that it evolves with time. The latter depends on the relative size and sign between the ghosts masses $m_A^2 $ and $ m_{\phi_2}^2$ as well as on the exact evolution of the physical and phantom scalar and gauge dof.
In principle one should consider a complicated system of eom to be solved simultaneously. However, in the small anisotropy limit that we study here the field eom are decoupled and the system with the Einstein eom is much simpler to be solved.\\
Before we enter to these details note that $w_q$ could be simplified even further since for a homogeneous and isotropic background one may exploit the temporal gauge $A_\m = \left(0, -A_i \right)$ along with the Coulomb gauge $\nabla^i A_i = 0$, following \cite{Durrer:2024ibi}. These suggest from \eq{EiBk} that the background components become 
\be
E_i = - \frac{1}{a^2} A'_i ~~ {\rm and} ~~ B_i = 0
\ee
according to which the background EoS becomes 
\be\label{wqDEgs1c}
w_q = - 1 + 2 \frac{|\varphi'_1|^2 - |\varphi'_2|^2 + \frac{1}{3 a^2}(A'_{i,1})^2  - \frac{1}{3 a^2}(A'_{i,2})^2 - \frac{1}{6} m_{A_2}^2  A_{i,2}^2 }{|\varphi'_1|^2  - |\varphi'_2|^2 + \frac{1}{2 a^2}(A'_{i,1})^2  - \frac{1}{2 a^2}(A'_{i,2})^2 - a^2 m_{\phi_2}^2 |\varphi_2|^2 - \frac{1}{2} m_{A_2}^2  A_{i,2}^2 }
\ee
Finally and based on the above, let us show the usefulness of the above expression by considering an indicative example. For that we consider the time scale of the universe today, $\tau_0$, where the evolution of both scalar and gauge fields is negligible compared to the mass terms, $|\varphi'_k|, A'_{i,k} \ll a^2 m_{\phi_2}^2 |\varphi_2|^2 , m_{A_2}^2 A_{i,2}^2$, with $k = 1,2$. 
According to this, $w_q$ admits the approximate and asymptotic form 
\be\label{wqap}
w_q(\tau_0) \approx -1 + \frac{- \frac{m_{A_2}^2}{3}  A_{i,2}^2 }{ - a(\tau)^2 m_{\phi_2}^2 |\varphi_2|^2 - \frac{1}{2} m_{A_2}^2 A_{i,2}^2 } \Bigg|_{\tau = \tau_0}\equiv  -1 + \frac{1}{3}\frac{ 1 }{  \frac{1}{2} + \frac{ m_{\phi_2}^2 }{m_{A_2}^2} \frac{ |\varphi_2(\tau_0)|^2 }{ A_{i,2,}(\tau_0)^2} }
\ee
and justifies our expectation that the exact phase of the DDE today will be determined by the relative sign and magnitude of the ghosts masses (or equivalently of the HDO couplings in lattice language) in combination with the relative magnitude of the scalar and gauge ghost fields.\\
Before closing this section let us emphasize that even if we have kept the interaction terms of \eq{SDE3} explicit, then the shift in the energy density ($\d\rho_{\rm DE}$) would have been in our framework 
\be
\d\rho_{\rm DE} = -\frac{g_4^2}{a^2} \left(|\varphi_1|^2-|\varphi_2|^2\right) \left(A_{i,1}-A_{i,2}\right)^2 \nn
\ee
and the shift of the pressure density ($\d p_{\rm DE}$) after isotropic averaging would have been
\be
\d p_{\rm DE} = \frac{g_4^2}{3a^2} \left(|\varphi_1|^2-|\varphi_2|^2\right) \left(A_{i,1}-A_{i,2}\right)^2 \nn
\ee
Then the associated shifted EoS becomes
\be
w_{q, \rm shift}= \frac{p_{\rm DE} +\d p_{\rm DE}}{\rho_{\rm DE}+\d\rho_{\rm DE}} \approx w_q \nn
\ee
since $g_4^2 \ll 1$.

\subsection{Full background evolution of $w_q$ in the small anisotropy limit $\g \ll 1$}\label{FBSGL}
Based on the previous section and as we have already discussed, for a concrete and complete analysis of the EoS we need to solve the system of eom generated by the dark energy effective action 
\be\label{SDEgs1b}
S^{\g \ll 1}_{{\rm DE}} \approx \int d^4x \sqrt{-g} \left[ - \frac{1}{4} F_{\m\n,1}F^{\m\n,1} +  \frac{1}{4} F_{\m\n,2}F^{\m\n,2} -\frac{1}{2} m_{A_2}^2 A_{\m,2}^2 + |\partial_\m \phi_1|^2 - |\partial_\m \phi_2|^2 + m_{\phi_2}^2  |\phi_2|^2 \right]
\ee
and the gravity part. Regarding the former we have to find for each field dof the associated eom using again 
\be
\frac{1}{\sqrt{-g}} \partial_\m \left( \sqrt{-g} \frac{\partial {\cal L_{\rm DE}}}{\partial (\partial_\m \bar \Phi)} \right) = \frac{ \partial {\cal L_{\rm DE}}}{\partial \bar \Phi}
\ee
but now only for $\Phi = \{\varphi_{1,2}, A_{i, 1,2} \} $. 
Following the algorithm that we developed below \eq{phi1phi2chi0} to obtain dimensionless and normalized quantities, the above translates to the following scalar-gauge system of eom
\bea\label{eomdot}
\ddot\varphi_{1,n} + \left[2 + \frac{\dot{\cal H}}{{\cal H}} \right] \dot \varphi_{1,n} &=& 0 ~~{\rm and} ~~ \ddot\varphi_{2,n} + \left[2 + \frac{\dot{\cal H}}{{\cal H}} \right] \dot\varphi_{2,n} + e^{2N}\frac{ m^2_{\phi_2}}{{\cal H}^2} \varphi_{2,n} = 0\nn\\
\ddot A_{i,1, n} + \left[4 + \frac{\dot{\cal H}}{{\cal H}} \right] \dot A_{i,1, n} &=& 0 ~~{\rm and} ~~ \ddot A_{i,2, n} +\left[4 + \frac{\dot{\cal H}}{{\cal H}} \right] \dot A_{i,2, n} + \frac{m_{A_2}^2}{{\cal H}^2} A_{i,2, n} = 0
\eea
with identical equations for the scalar fields complex conjugate. As we have already mentioned, ${\cal H}$ denotes the comoving Hubble rate given by Friedmann’s first equation
\be\label{fFe}
{\cal H}^2 = (a H)^2  = \frac{1}{3 M_{\rm Pl}^2} a^2 \left( \rho_{\rm DE} + \rho_{\rm m} \right) \equiv {\cal H}_{\rm DE}^2 + {\cal H}^2_{\rm m} 
\ee
where $H$ is the physical Hubble rate.
Note that the total system that we should evolve includes both the dark energy fields and the dark and baryonic matter (parameterized on an FRW spacetime by a fluid). The latter is expressed here via $\rho_{\rm m}$, the energy density of a barotropic fluid, with EoS $p_{\rm m} = ({\rm f} - 1) \rho_{\rm m} =  w_{\rm m} \rho_{\rm m}$. The associated continuity equation in the dot representation becomes
\be\label{rhomdot}
\dot \rho_{\rm m} + 3 \rho_{\rm m}  = 0\Rightarrow \rho_{\rm m} = \rho_{\rm m, 0} e^{-3N}
\ee
where $\rho_{\rm m,0 } \equiv \rho_{\rm m}(a_0)$ is the matter's total energy density today.
In principle $0 \le {\rm f} \le 2$ but for our purposes we consider $\rm f=1$, which corresponds to DM (and baryonic matter), suggesting that 
\be\label{cHm}
{\cal H}^2_{\rm m} = a^2 \frac{\rho_{\rm m, 0}}{3 M_{\rm Pl}^2} e^{-3N} \equiv  H^2_{\rm m,0} e^{-N}
\ee
with $M_{\rm Pl} \approx 2.4 \times 10^{18}~ {\rm GeV}$ the reduced Planck mass.\\
Regarding the Hubble rate for the dark energy we use the obtained energy density of \eq{rhoDEsg} which, due to our gauge and normalization choices, is given in the dot basis by
\be\label{cHDEf}
{\cal H}_{\rm DE}^2  \equiv H^2_{\rm m,0} \frac{|\varphi_{2,\a}|^2}{3 M_{\rm Pl}^2} \left[ \left( | \dot \varphi_{1,n}|^2 - | \dot \varphi_{2,n}|^2 + \frac{e^{-2N}}{2} (\dot A_{i,1, n})^2  - \frac{e^{-2N}}{2} (\dot A_{i,2, n})^2 \right)e^{-N} - \frac{m_{A_2}^2}{2 H^2_{\rm m,0}}  A_{i,2, n}^2 - e^{2N} \frac{m_{\phi_2}^2}{H^2_{\rm m,0}} |\varphi_{2,n}|^2  \right]
\ee
as we show in App. \ref{dBasis}.
In that sense, the total Hubble rate is rewritten\footnote{Note that this justifies our choice for ${\cal H}$ defined below \eq{phi1phi2chi0}.} as
\be\label{cHtn}
{\cal H}^2 = H^2_{\rm m, 0} \left[ \frac{|\varphi_{2,\a}|^2}{3 M_{\rm Pl}^2} \widetilde{\cal H}_{{\rm DE}}^2 + e^{-N} \right]
\ee
where $\widetilde{\cal H}_{{\rm DE}}^2$ is the normalized and dimensionless DE Hubble rate
\be
\widetilde{\cal H}_{{\rm DE}}^2 = \left( | \dot \varphi_{1,n}|^2 - | \dot \varphi_{2,n}|^2 + \frac{e^{-2N}}{2} (\dot A_{i,1, n})^2  - \frac{e^{-2N}}{2} (\dot A_{i,2, n})^2 \right)e^{-N} - \frac{m_{A_2}^2}{2 H^2_{\rm m,0}}  A_{i,2, n}^2 - e^{2N} \frac{m_{\phi_2}^2}{H^2_{\rm m,0}} |\varphi_{2,n}|^2 
\ee
From the above analysis is clear that even though the scalar-gauge system is decoupled, the eom remain correlated since the Hubble rate is not just constant and evolves with time as a function of all the fields. Solving the system is highly non-trivial however let us first to present an analytic estimation under some legit approximation and later on to support it with a numerical analysis.\\
For both cases, based on the previous relations, we define a set of free parameters which are: the initial values for the normalized scalar and gauge fields, $\Phi(N_\a)/|\varphi_{2,\a}| \equiv \Phi_{n,\a} $ along with their associated derivatives, $\dot \Phi(N = N_\a)/|\varphi_{2,\a}| \equiv \dot \Phi_{n,\a} $, the relevant magnitude and signs of $m_{\phi_2}^2/H_{\rm m,0}^2 $ and $ m_{A_2}^2/H_{\rm m,0}^2$ as well as the ratio $|\varphi_{2,\a}|^2/M_{\rm Pl}^2$. For the latter we have seen that as long as $|\varphi_{2,\a}|^2 < M_{\rm Pl}^2$ our results are untouched therefore we fix it to a $|\varphi_{2,\a}|^2/M_{\rm Pl}^2 \ll 1 $ value that will be motivated in the following sections. The non-trivial part in our system of equations is hidden in the running of ${\cal H}^2$ as a function of e-folds according to \eq{cHtn}. So in the following we assume that its running is dominated initially by the barotropic fluid and only at the very late times the DE takes control suggesting ${\cal H}^2 \approx H_{\rm m,0}^2 e^{-N}$ for the most of the evolution. Based on that we can solve \eq{eomdot} analytically obtaining
\bea\label{f1A1sol}
\varphi_{1,n}(N) &\equiv&  k_2 - \frac{2}{3} k_1 e^{-3/2 N} 
\nn\\
A_{1,n}(N) &\equiv&  k_4 - \frac{2}{7} k_3 e^{-7/2 N} 
\eea
for the physical fields and 
\bea\label{f2A2sol}
\varphi_{2,n}(N) &\equiv&  \left(\frac{m_{\phi_2}^2}{H_{\rm m,0}^2} \right)^{1/4} \frac{ 1  }{ \sqrt{3} \theta_\varphi(N) } \left[ 2 k_5  \cos  \left(\theta_\varphi(N) \right) + k_6  \sin  \left(\theta_\varphi(N)\right)   \right]  \nn\\
A_{2,n}(N) &\equiv& 128 \left(\frac{m_{A_2}^2}{H_{\rm m,0}^2} \right)^{7/4} \theta^{-7}_A(N)   \Bigg\{ k_7 \Bigg[ \theta_A(N) \left( 1 -  \frac{\theta^2_A(N)}{15} \right) \sin  \left(\theta_A(N)\right) + \left( 1 - \frac{6\theta^2_A(N)}{15} \right) \cos  \left(\theta_A(N)\right)  \Bigg] \nn\\
&&+ 12 k_8 \Bigg[ \theta_A(N) \left( \frac{\theta^2_A(N)}{15} -1 \right) \cos  \left(\theta_A(N)\right) + \left( 1 - \frac{6\theta^2_A(N)}{15} \right)\sin  \left(\theta_A(N)\right) \Bigg]  \Bigg\}
\eea
for the phantoms where we have defined the angles
\be\label{fiAangle}
\theta_\varphi(N) = \frac{2}{3} \sqrt{e^{3 N} \frac{ m_{\phi_2}^2 }{H_{\rm m,0}^2} } ~~{\rm and}~~ \theta_A(N) = 2 \sqrt{e^N \frac{ m_{A_2}^2}{H_{\rm m,0}^2} }
\ee
for the scalar and the gauge ghost respectively. Note that we have abandoned the $A_i$ notation for transparency, since only the transverse part of $A_1$ and $A_2$ contributes.
\begin{figure}[t]
\centering
\includegraphics[width=8.5cm]{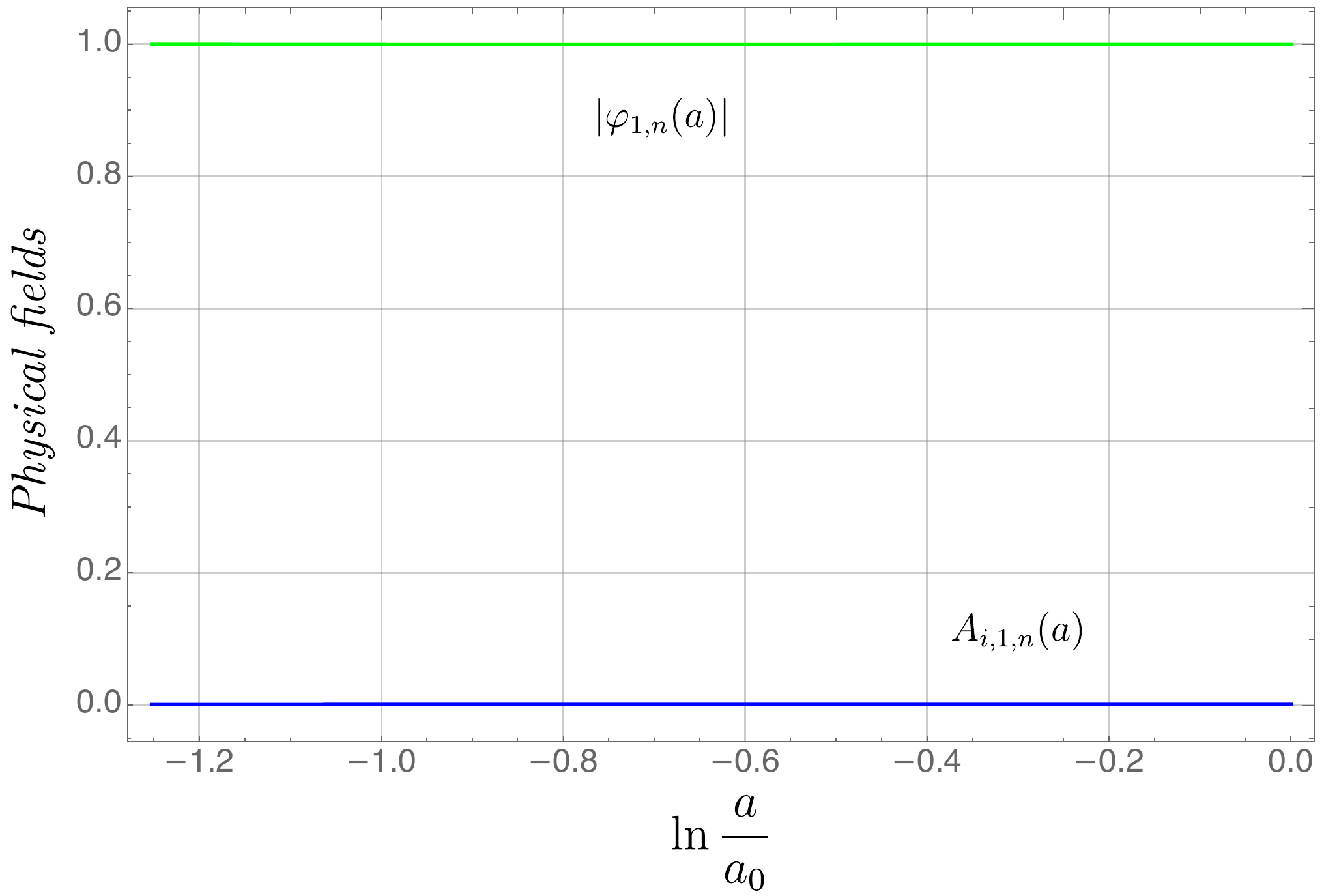} \quad
\includegraphics[width=8.5cm]{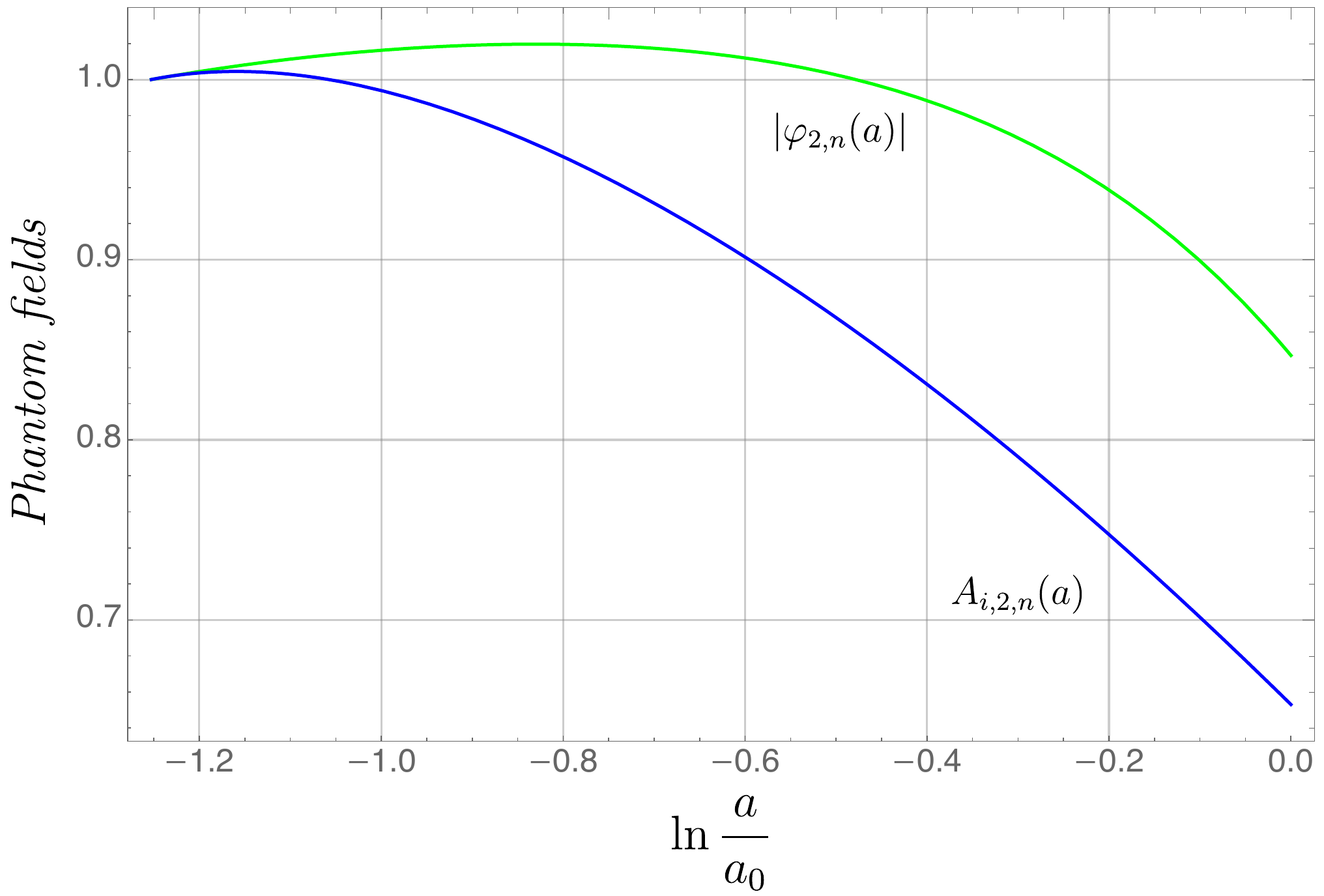} 
\caption{\it The evolution of the scalar and gauge physical and phantom fields as a function of the $N = \ln a/a_0$ under the mass-ratio $ m_{A_2}^2/H_{\rm m,0}^2 = m_{\phi_2}^2/H_{\rm m,0}^2  = 3$ and the democratic initial conditions $|\varphi_{1,n, \a}| \approx {A}_{2,n,\a} \approx |\varphi_{2,n, \a}| = 1 $. As we have explained below \eq{freeparameters} the evolution of $A_{i,1,n}$ is negligible compared with the others. }
\label{vf12nA12n}
\end{figure} 
A set of comments regarding the free parameters of the above system are in order: First note that $\theta_\varphi$ and $\theta_A$ could in principle be zero which corresponds to the limit that the ghost masses approaching zero. In that limit one gets 
\be
\varphi_{2,n}(N)\Big |_{m_{\phi_2}^2 \to 0} \approx - \frac{3\sqrt{3}}{2 }\left( \frac{m_{\phi_2}^2}{H_{\rm m,0}^2} \right)^{-1/4} k_6 e^{-3N/2} ~~{\rm and}~~ A_{2,n} (N) \Big |_{m_{A_2}^2 \to 0} \approx \left(\frac{m_{A_2}^2}{H_{\rm m,0}^2} \right)^{- 7/4}  k_7 e^{-7N/4}
\ee
which follows the same scaling with the solution of \eq{f1A1sol} for the massless physical fields, under the appropriate choice of $k_6$ and $k_7$, as expected. Now notice that the phantom eom are of the form 
\be
\ddot \Phi + A(N) \dot \Phi + M_{\Phi}^2(N) \Phi = 0 \nn  
\ee
resembling the standard case of inflation or axion misalignment suggesting that there would be an e-fold (time), $N_{\rm osc}$, where $M_{\Phi}(N_{\rm osc}) \gg A(N_{\rm osc}) $ indicating the era of fast field oscillations.
In our case the above analogy identifies to $A(N) \equiv 2 + \dot{\cal H}/{\cal H}$ and to $M_{\Phi}(N) \equiv (e^{2N} m_{\phi_2}^2/{\cal H}^2)^{1/2}$ for the scalar and $M_{\Phi}(N) \equiv ( m_{A_2}^2/{\cal H}^2)^{1/2}$ for the gauge ghost respectively.
In order to avoid the entrance of DE into that rapid oscillation period during the late-time instances of the universe, one should demand that around $N \approx 0$ holds 
\be
\sqrt{\frac{e^{2N} m_{\phi_2}^2}{{\cal H}^2}} \Bigg |_{N \approx 0} \lesssim {\cal O}(1) \left( 2 + \dot{\cal H}/{\cal H} \right) \Big |_{N \approx 0}  ~~{\rm and} ~~ \sqrt{\frac{m_{A_2}^2}{{\cal H}^2}} \Bigg |_{N \approx 0} \lesssim  {\cal O}(1) \left( 2 + \dot{\cal H}/{\cal H} \right)\Big |_{N \approx 0} \nn
\ee
and using the fact that ${\cal H}^2 \approx H_{\rm m,0}^2 e^{-N}$, the above suggest
\be\label{massbounds}
\left( \frac{m_{\phi_2}^2}{H_{\rm m,0}^2}, \frac{m_{A_2}^2}{H_{\rm m,0}^2}\right) \lesssim {\cal O}(1)
\ee
In that sense for the rest of our analysis and based on the previous relations we fix the masses squared to be positive. In addition, we choose the above mass ratios to be equal and also we fix $ m_{A_2}^2/H_{\rm m,0}^2 = m_{\phi_2}^2/H_{\rm m,0}^2 \approx 3 $ to facilitate our numerical analysis below without loss of generality since we have seen that we can always obtain the same behavior with any ${\cal O}(1)$ number. According to these we have eliminated two dof from our system of equations.
Finally, we are left with the unknown coefficients $k_i$, with $i = 1,\cdots,8$, which are determined by the initial conditions for the fields and their derivatives suggesting that 
\bea
k_{1,2} \equiv k_{1,2}(\varphi_{1,n,\a},\dot {\varphi}_{1,n,\a} ) ~~ &{\rm and}& ~~ k_{3,4} \equiv k_{3,4}(A_{1,n,\a},\dot {A}_{1,n,\a} ) \nn\\
k_{5,6} \equiv k_{5,6}(\dot {\varphi}_{2,n,\a}
)~~ &{\rm and}& ~~ k_{7,8} \equiv k_{7,8}(A_{2,n,\a},\dot {A}_{2,n,\a}, 
)
\eea
and of course the same set of parameters exists for the complex conjugate coefficients, $k_{1,2}^*$ and $ k_{5,6}^*$.\\
In total, it seems that we have seven free parameters (and the conjugates of three of them) however let us motivate how to fix some of them. From \eq{f1A1sol}, \eq{f2A2sol} and \eq{fiAangle} we see that the scaling of the fields with the scale factor suggests 
\be\label{freeparameters}
(\varphi_{1,n}, \dot \varphi_{1,n}, \varphi_{2,n}) \sim a^{-3/2}, ~ A_{2,n} \sim a^{-2} ~~{\rm and}~~ A_{1,n} \sim a^{-7/2}
\ee
so for the rest of this analysis we may get rid of $A_{1,n}$ and $\dot A_{1,n}$ along with their initial values (2 dof) since this field red-shifts much faster than the others.
Regarding the normalized scalars we consider without loss of generality the most democratic initial values (as it is explained in App. \ref{CTIFV}) fixing $\varphi_{1,n,\a} = 1$ (1 dof). Therefore we are left with the four initial conditions $(|\dot\varphi_{1,n, \a}|,|\dot \varphi_{2,n,\a}|, A_{2,n,\a},\dot {A}_{2,n,\a})$ which we consider as free parameters.
In \fig{vf12nA12n} we present the evolution of the physical and phantom fields where we keep fixed the mass-ratio parameter to $ m_{A_2}^2/H_{\rm m,0}^2 = m_{\phi_2}^2/H_{\rm m,0}^2  \approx 3$ while we choose as an illustrative example the most democratic initial conditions $|\varphi_{1,n, \a}| = {A}_{2,n,\a} = |\varphi_{2,n, \a}| = 1 $.

Under the above arguments and having settled down the freedom of our model, we follow the same spirit with App. \ref{dBasis} and we rewrite $w_q$ of \eq{wqDEgs1c} in the dotted and nomralized basis as
\be\label{wqDEgs1an}
w_q \approx - 1 + 2 \frac{|\dot \varphi_{1,n}|^2 - |\dot \varphi_{2,n}|^2 - e^{-2N}\frac{(\dot A_{2, n})^2}{3} - e^N\frac{m_{A_2}^2}{6 H_{\rm m,0}^2}  A_{2, n}^2 }{|\dot \varphi_{1,n}|^2 - |\dot \varphi_{2,n}|^2 - e^{-2N}\frac{(\dot A_{2, n})^2}{2} - e^{3N} \frac{ m^2_{\phi_2}}{H_{\rm m,0}^2} |\varphi_{2,n}|^2 - e^N\frac{m_{A_2}^2}{2 H_{\rm m,0}^2}  A_{2, n}^2 }
\ee
from which we may separate the evolution of the EoS into two regimes: a) initial time and b) today.
Regarding the former, $N_\a < 0 $, a naive estimation would be that the two mass terms are subdominant compared to the kinetic terms due to the e-folding suppression/enhancement, with the most dominant one that of the gauge ghost. In this regime $w_q \equiv w_{q,\rm \a}$ and after setting $N \equiv N_\a$ 
we collectively get 
\be
w_{q,\rm \a} \approx - 1 + 2 \frac{|\dot \varphi_{1,n, \a}|^2 - |\dot \varphi_{2,n, \a}|^2 - e^{-2N_\a}\frac{(\dot A_{2, n,\a})^2}{3} }{|\dot \varphi_{1,n, \a}|^2 - |\dot \varphi_{2,n, \a}|^2 - e^{-2N_\a}\frac{(\dot A_{2, n, \a})^2}{2} }
\ee
so in order that the EoS starts initially by a ghost phase around $N_\a$, with $w_{q,\a} <-1$, we should demand that 
$|\dot\varphi_{1,n, \a}|^2 >  |\dot\varphi_{2,n, \a}|^2$ and $ e^{-2N_\a}(\dot A_{2, n, \a})^2/3 <|\dot\varphi_{1,n, \a}|^2 - |\dot\varphi_{2,n, \a}|^2 < e^{-2N_\a} (\dot A_{2, n, \a})^2/ 2 $.
\begin{figure}[t]
\centering
\includegraphics[width=8.5 cm]{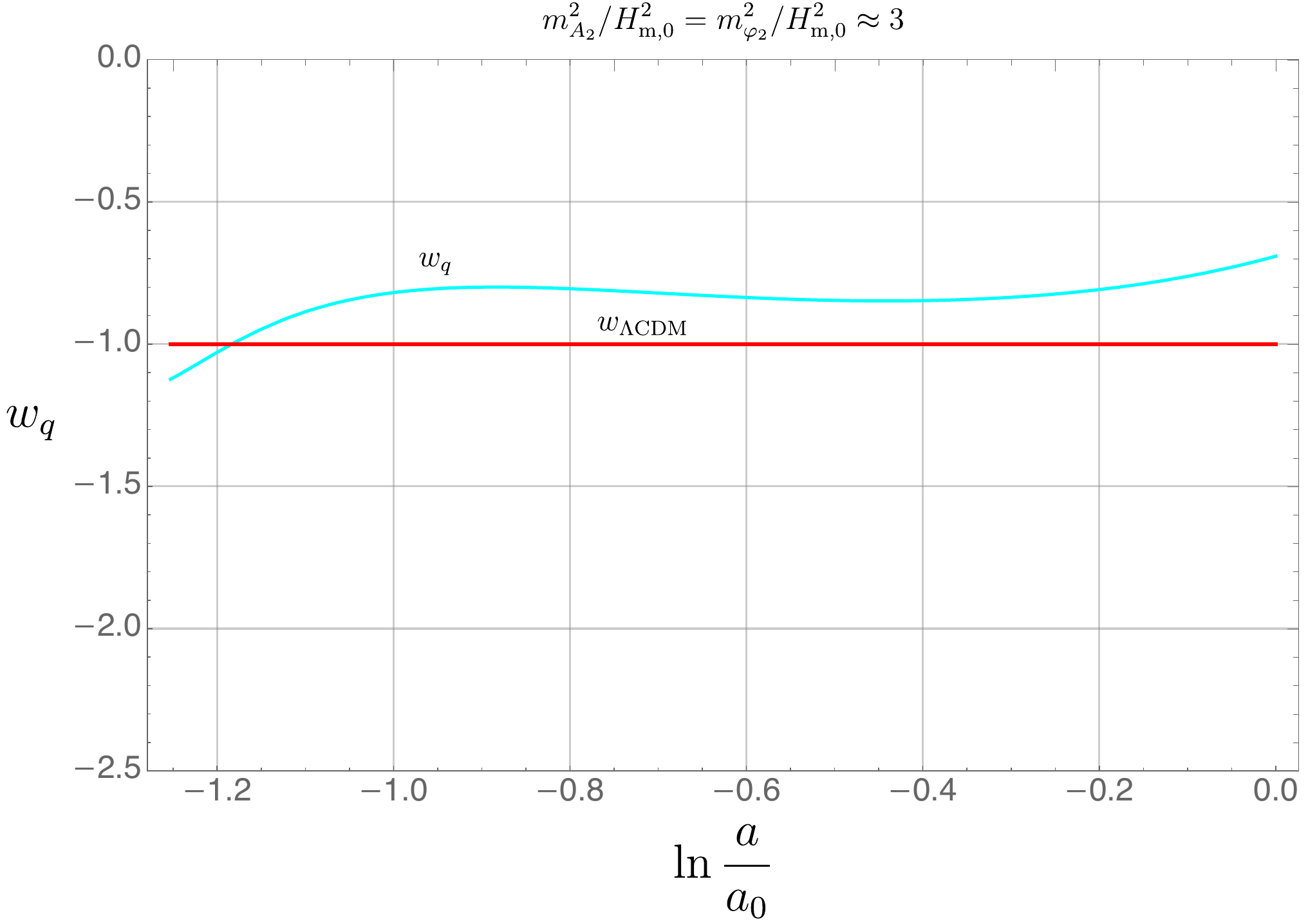} \quad
\includegraphics[width=8.5cm]{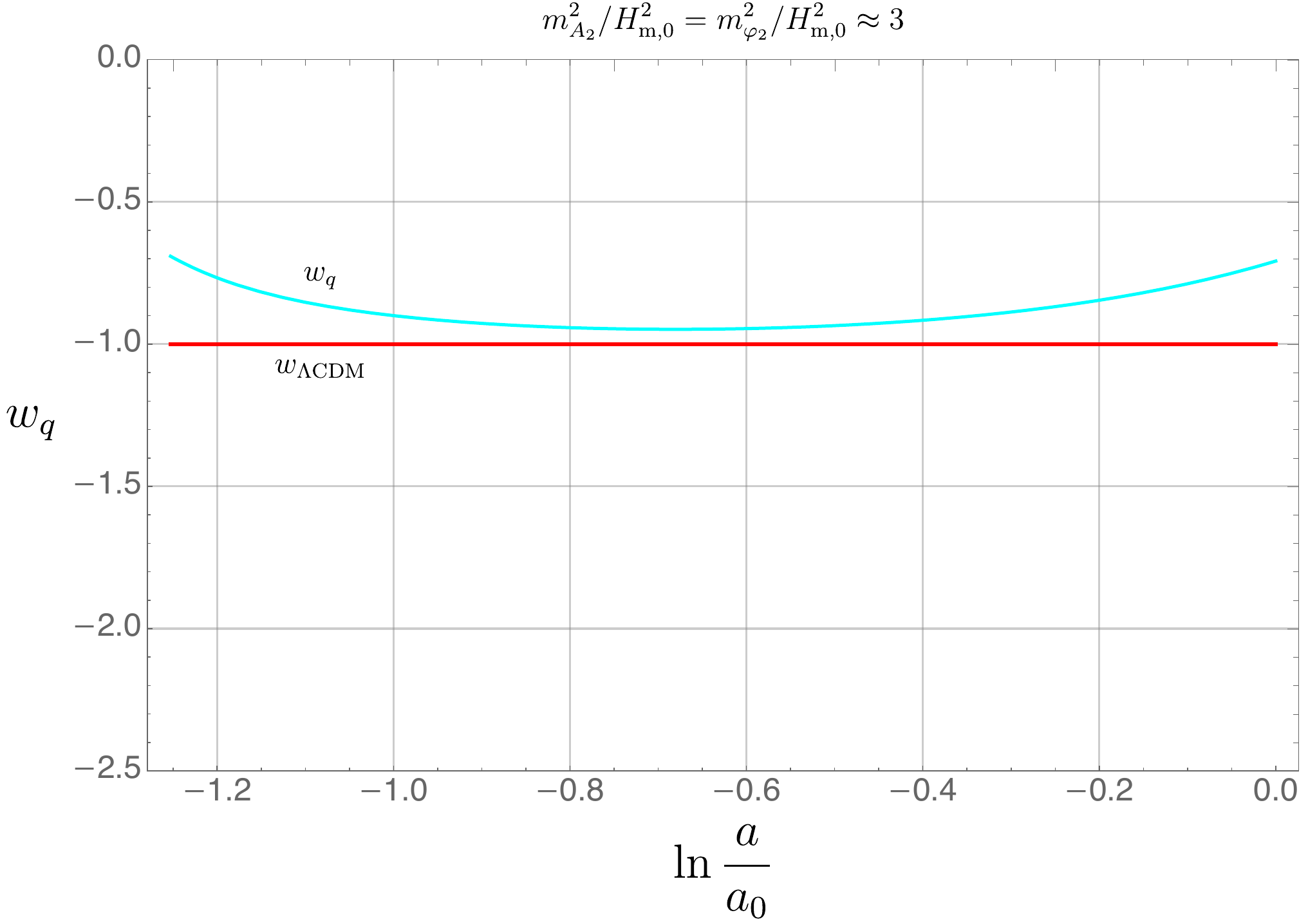} 
\caption{\it The evolution of $w_q$ (cyan line) as a function of the normalized scale factor for two benchmark initial conditions and fixed ghost field masses, $ m_{A_2}^2/H_{\rm m,0}^2 = m_{\phi_2}^2/H_{\rm m,0}^2  \approx 3$. The red line corresponds to the cosmological boundary $w_{\rm \L CDM} = -1$.
Left panel: A Quintom-B-like model is realized for the initial kinetic term hierarchy $|\dot\varphi_{1,n, \a}| > |\dot\varphi_{2,n, \a}|$ and $ e^{-2N_\a}(\dot A_{2, n, \a})^2/3 <|\dot\varphi_{1,n, \a}|^2 - |\dot\varphi_{2,n, \a}|^2 < e^{-2N_\a} (\dot A_{2, n, \a})^2/ 2 $. Right panel: A quintessence-like behavior is obtained for the inverse hierarchy, $|\dot\varphi_{1,n, \a}| < | \dot\varphi_{2,n, \a} |$.}
\label{wqmfmA1}
\end{figure} 
In the opposite hierarchy, $|\dot\varphi_{1,n, \a}|^2 <  |\dot\varphi_{2,n, \a}|^2$, we get $w_{q,\a} >-1$ and the DE lies initially in a quintessential phase. Note the counterintuitive scalar field hierarchy here, compared with the standard scalar quintom lore, due to the presence of the gauge ghost.
For our full numerical analysis, however, the mass terms are not neglected at $N_\a$ extending in that way the previous constrains.
Now, on the other hand, for the time regimes around today ($N \approx 0$) the mass terms dominate over the kinetic terms and we end up essentially with \eq{wqap} suggesting that $w_{q,0} \gtrsim -1$ since the ghost squared masses are always positive. In \fig{wqmfmA1} we present two benchmarks scenarios regarding the evolution of the dark energy EoS within the same redshift and mass ratios considered previously. Choosing $|\dot\varphi_{2,n, \a}| = 1/2 |\dot\varphi_{1,n, \a}|  = 10 \dot {A}_{2,n,\a} = 0.1 $ a Quintom-B model is realized (left panel) while for $|\dot\varphi_{1,n, \a}| = \dot {A}_{2,n,\a} = 0.1 |\dot\varphi_{2,n, \a}| = 0.01 $ we obtain a quintessence-like phase (right panel).
\begin{figure}[t]
\centering
\includegraphics[width=12cm]{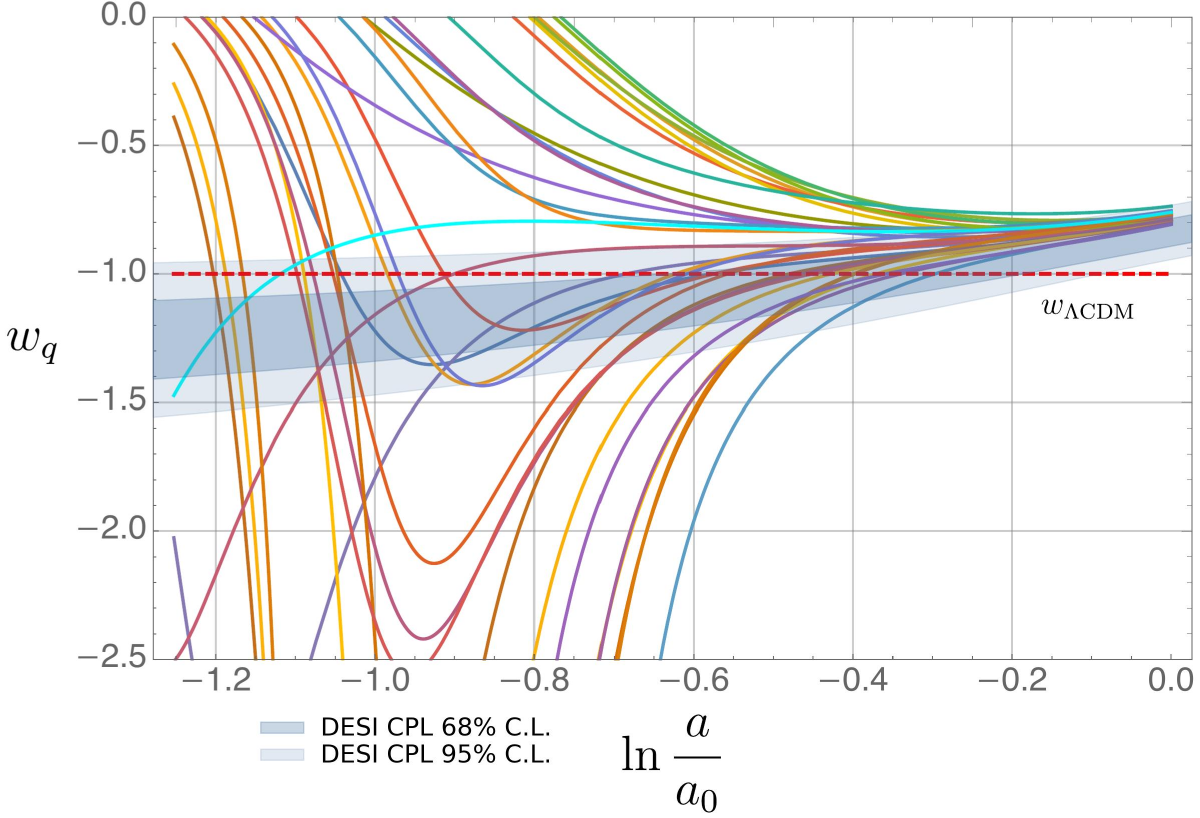} 
\caption{\it The EoS in \eq{wqDEgs1an} under the scanning of the initial conditions $(|\dot\varphi_{1,n, \a}|, |\dot \varphi_{2,n,\a}|, A_{2,n,\a},\dot {A}_{2,n,\a})$ which are the free parameters of the system. $m_{\phi_2}^2/H_{\rm m,0}^2$ and $m_{A_2}^2/H_{\rm m,0}^2$ are both fixed to some $ {\cal O}(1)$ value. 
The red dashed line corresponds to the cosmological boundary $w_{\rm \L CDM} = -1$. 
The shaded blue regions correspond to $68\%$ and $95\%$ confidence intervals (DESI DR2 BAO + CMB + Pantheon + SNa).
An interesting feature of the model is that the evolution at late-times always converges towards DESI intervals independently of the initial conditions. In other words either our model starts from a ghost or a quintessence phase, the EoS today always lies above the cosmological boundary.}
\label{wqScanned}
\end{figure} 
Of course, the above cases seem precisely selected to exploit the behavior that we have advertised so far regarding our effective quintom model. 
However, besides these two benchmark scenarios we have performed a full numerical analysis regarding the evolution of the DE equation of state under a parameter scanning for our initial conditions, $(|\dot\varphi_{1,n, \a}|,|\dot \varphi_{2,n,\a}|, A_{2,n,\a},\dot {A}_{2,n,\a})$, to support the previous claims. Based on that we find that almost half of the parameter space reproduces exactly the background that gives the best fit to the DESI + CMB + DESY5 data (within CPL parametrization of DE). In other words it produces a Quintom-B-type model accompanied with a crossing point around $N_c = \ln a_c/a_0 \approx -0.4$ or $z_c \approx 0.5 $. The other half lies within the quintessence-like phase. In addition, an interesting outcome of this analysis is that in both scenarios we always find a deviation from the cosmological boundary towards a quintessence phase at today's time scales. Such a behavior seems to be in agreement with the latest observations \cite{DESI:2024mwx, DESI:2025zgx} suggesting that at late-times the EoS is neither crossing the $w_{\rm \L CDM} = -1$ nor is aligned with that. These results are collectively presented in \fig{wqScanned}.

In conclusion, we show that our UV-complete effective quintom model can indeed accommodate the type of dynamical dark-energy evolution suggested by current DESI-era analyses. Due to its very restrictive and particular 5d-origin, the restrictive NPGHU origin reduces the freedom for arbitrary tuning of the initial conditions a fact that enhances its predictive power. In particular, our model provides a viable Quintom-B like scenario which is difficult to realize consistently in conventional four-dimensional quintom constructions. This is true even in the model building of higher derivative scalar theories unless one delicately designs (and fine-tunes) the terms of the Lagrangian to involve couplings with the kinetic term \cite{Cai:2025mas}. In addition, these models usually face ghost instabilities in classical (perturbations) and quantum level without being able to address both simultaneously. On the contrary, the NPGHU-induced effective quintom model contains mechanisms that regulate both the linear perturbative instability and the gravity-induced vacuum-decay problem as we demonstrate in the next section.
Finally, let us briefly comment on our choice to constraint the mass ratios within ${\cal O}(1)$ numbers. The bound of \eq{massbounds} suggest, under the definition of the ghost masses below \eq{SDE3}, essentially that 
\bea\label{mphi2Lambda}
(m_{\phi_2}^2, m_{A_2}^2) &\lesssim & H_{\rm m,0}^2  \Rightarrow \nn\\
\left(\frac{\L^2}{c_6}, \frac{\L^2}{c_\a} \right) &\lesssim &  H_{\rm m,0}^2 \approx \left(1.5 \times 10^{-33 } ~ \rm eV \right)^2 
\eea
where we have used that $H_{\rm m,0} \approx \sqrt{\Omega_{\rm m,0}} H_0 \approx 1.5 \times 10^{-33}~ \rm eV$.
Therefore the above bound gives an implicit constraint on the 5d lattice parameters from which our 4d effective model originates.
More precisely we may consider two extreme cases, one with a high scale cut-off (in the regime above $\rm MeV$ to be far from BBN) and huge couplings and one with extremely small cut-off and ${\cal O}(1)$ couplings. Both scenarios could be equally well justified within our 5d anisotropic lattice (especially in the regime of small anisotropy) both perturbatively \cite{Irges:2018gra} and from  Monte Carlo lattice simulations \cite{Alberti:2015pha}, admitting however different physical interpretation.\\
The former case suggests that the cut-off is bounded from below as $\L \gtrsim 10~\rm MeV$, such that full localization is achieved in high energies and before BBN, while today we leave relatively deep in the so called Higgs phase where localization is almost lost as we explained in \sect{BoQLM}.
In that sense one obtains the lower bound $(c_\a, c_6) \gtrsim 10^{79}$ where such quite huge value could, in principle, be justified in the regime of extremely small anisotropy.
The other scenario is that the cut-off is bounded from below as $\L \gtrsim H_{\rm m,0}$ leading to 
\be
(c_\a,c_6) \gtrsim {\cal O}(1) \nn
\ee
and essentially means that today we are relatively near the vicinity of the phase transition (orange-dot on \fig{Higgsphase2}) so that localization is still strong and the boundary is almost decoupled from the bulk.\\
The first case is essentially excluded phenomenologically since there are no cosmological observations which suggest deviations from the 4d description towards a 5d bulk (lost localization) today. 
Actually as we will show in the next section, the absence of ghost-instabilities even in the presence of interactions among the ghosts and physical fields provides us with an upper bound for the cut-off which is naturally well-within the physical picture of the second scenario described above. In particular we will justify why the choice $\L \lesssim 100 H_{\rm m,0}$ is consistent with the absence of ghost instabilities both above and below the cut-off, a non-trivial fact of our non-perturbative original construction, as well as why the former case above would also be deeply problematic for the evolution of the universe.

\subsection{Phenomenological predictions and observational viability}
Before we turn our attention to the stability analysis of the model let us close this section by supplementing the background EoS reconstruction with direct cosmological observables relevant for late-time data. This will make the analysis here even more robust. One should keep in mind that the Quintom DE models have quite recently been fully stressed under the cosmological observations coming from the latest DESI data \cite{Yang:2024kdo, Yang:2025mws, Yang:2025kgc}. These works show, using the Gaussian process regression to reconstruct the cosmological evolution, that quintom dynamics remain in consistent with the observational data motivating further the current attempt to UV complete this kind of models.\\
From the model prediction for $w_q$ we reconstruct the expansion rate, $H(z)$ (here the latter denotes the physical Hubble rate entering cosmological observables, not to be confused with the conformal Hubble rate used in the background eom ${\cal H} = aH$), the comoving and angular-diameter distances and the BAO combinations ($D_M(z)/r_d$), ($D_H(z)/r_d$) and ($D_V(z)/r_d$). For the scope of this work we assume that the present model modifies the late-time dark energy sector while leaving the pre-recombination thermal history unchanged\footnote{Note that the model is a late-time effective quintom action however above $\L$ the change of dark energy dof (only physical fields and no ghosts) will alter the early Universe evolution such that $r_d$ must be recomputed self-consistently. We leave this analysis for a future work.}, so the sound horizon ($r_d$) is evaluated using the standard early-universe sector. We further compute the linear growth factor and the observable $f\sigma_8(z)$ under the smooth-dark-energy approximation, thereby testing whether the viable background solutions show any immediate tension with the standard scale-independent growth prediction. A full computation of the matter power spectrum ($P(k,z)$) and the CMB angular spectra ($C_l$) would require implementing the NPGHU-induced effective perturbation sector in a Boltzmann solver which is therefore identified as a natural extension beyond the scope of the present revision. 
\begin{figure}[t]
\centering
\includegraphics[width=12cm]{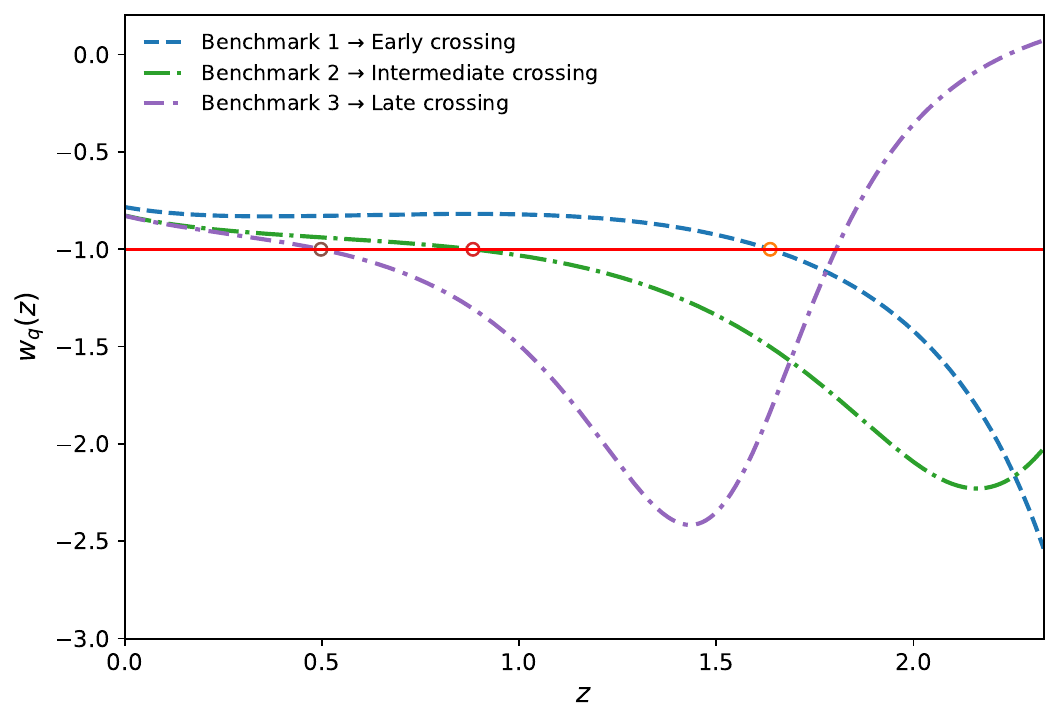} 
\caption{\it Evolution of the dark energy EoS $w_q(z)$ for three representative Quintom-B benchmarks of the effective quitnom model. The benchmarks are chosen such that the relevant crossing of the phantom divide, $w = -1 $, occurs at early, intermediate and late epochs respectively. The open circles indicate the relevant Quintom-B-like crossing for the three benchmarks at $z_{\rm cross} \approx 1.64$, $z_{\rm cross} \approx 0.88$ and $z_{\rm cross} \approx 0.5$ respectively. Note that Benchmark 3 had a Quintom-A crossing prior to the Quintom-B, relatively close to the localization phase transition, indicating the rich phenomenological implications of our effective quintom model.}
\label{wqz}
\end{figure} 
\begin{figure}[t]
\centering
\includegraphics[width=8.5cm]{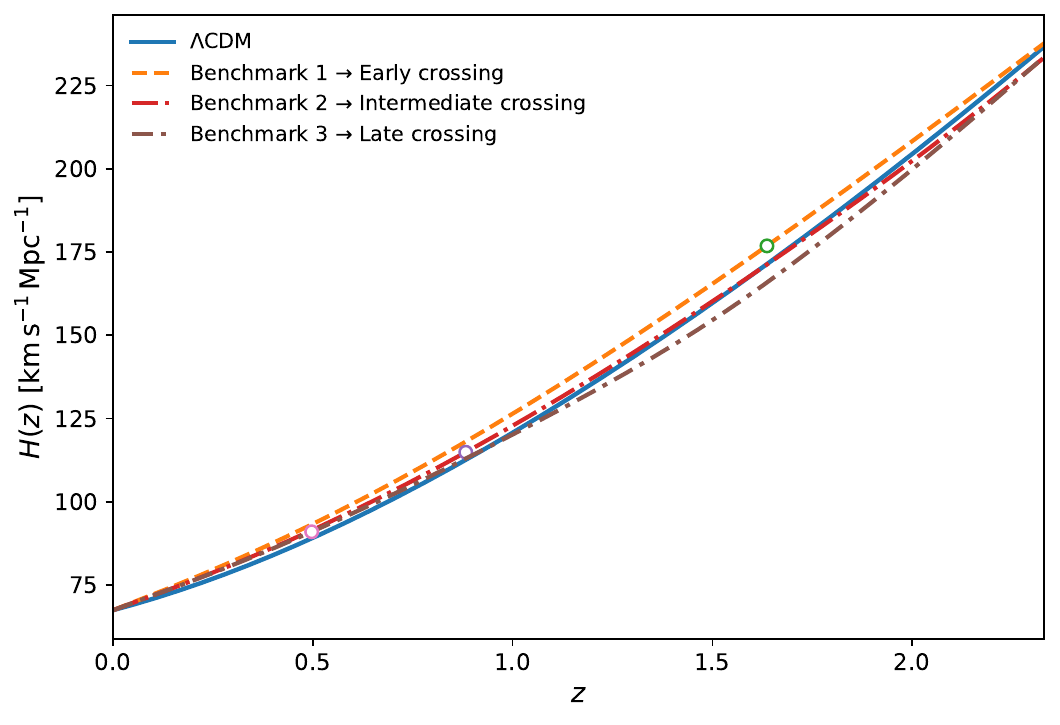} \quad
\includegraphics[width=8.5cm]{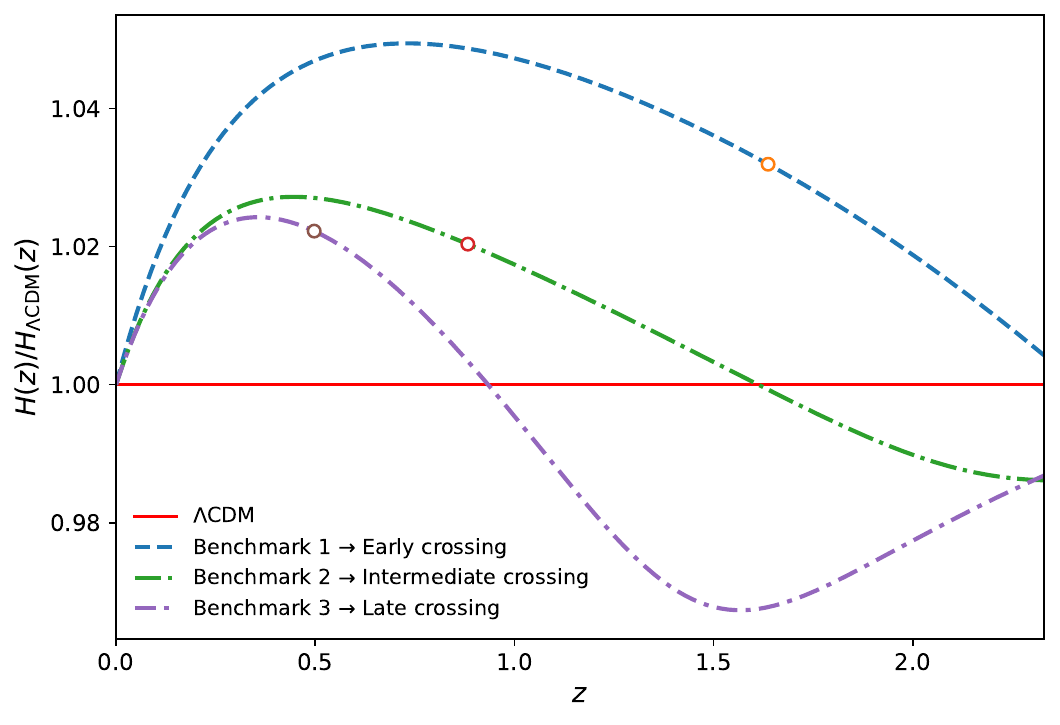} 
\caption{ Left panel: {\it $H(z)$ for the three Quintom-B benchmarks compared with the corresponding $\L$CDM expansion history for identical values of $H_0$, $\Omega_{\rm m,0}$ and the standard radiation sector. The benchmark curves are reconstructed directly from the model-derived equation of state $w_q(z)$, rather than from a phenomenological ansatz.} Right panel: {\it The ratio $H(z)/H_{\rm \L CDM} $ for the three Quintom-B benchmarks. This representation isolates and makes more explicit the departure from the standard $\L$CDM background expansion. In both panels open circles mark the redshift at which each benchmark undergoes the relevant phantom-divide crossing showing explicitly how the time of the Quintom-B transition correlates with the deviation from the $\L$CDM.}}
\label{HzHLCDM}
\end{figure} 

Having obtained the model prediction for the dark energy equation of state, $w_q(N)$, we can reconstruct the corresponding late-time expansion history without introducing an additional phenomenological parametrization. Then, the dark energy density follows from the continuity equation
\be
\dot{\rho}_{\rm DE} + 3(1+w_q) \rho_{\rm DE} = 0
\ee
in the dotted basis. However in order to facilitate our analysis let us move to the redshift basis using the relation $N=\ln(a/a_0)=-\ln(1+z)$. In addition, let us forget about the full parameter scan that we did in the previous section and focus on three characteristic Quintom-B benchmarks with (Benchmark 1) early-, (Benchmark 2) intermediate- and (Benchmark 3) late-crossing.
In that sense the result of the field evolution can be written as $w_q(z)$ and is depicted in \fig{wqz} for the Benchmark 1, 2 and 3 cases. Moreover, the solution of the energy density equation above gives
\be
\rho_{\rm DE}(z)=\rho_{\rm DE, 0} X_q(z) ~~{\rm with} ~~ X_q(z)= \exp\left[ 3\int_0^z \frac{1+w_q(z')}{1+z'} dz' \right]
\ee
and assuming spatial flatness, the normalized physical Hubble rate is therefore
\be
E^2(z)\equiv \frac{H^2(z)}{H_0^2} =  \Omega_{\rm m0}(1+z)^3 + \Omega_{r0}(1+z)^4 + \Omega_{q0}X_q(z),
\ee
with
\be
\Omega_{q0}=1-\Omega_{\rm m0}-\Omega_{r0}
\ee
such that the expansion history predicted by our effective quintom model is then
\be
H(z)=H_0E(z).
\ee
This construction uses the model-derived $w_q(z)$ rather than a CPL or other phenomenological dark energy parametrization. In this way, the comparison with late-time expansion data is directly tied to the microscopic NPGHU dynamics.
The evolution of physical Hubble rate and its comparison with the $\L$CDM case is shown in \fig{HzHLCDM} for the chosen Benchmarks.
In the present analysis we do not attempt to solve the Hubble tension since $H_0$ has been fixed rather than inferred from a joint likelihood. However, the model modifies the late-time expansion history relative to $\L$CDM, as shown by $H(z)/H_{\rm \L CDM}(z) \ne 1$. This indicates that the NPGHU dynamics can, implicitly, affect the background distances entering BAO and supernova constraints and therefore motivates a dedicated joint fit including $H_0$, BAO, SNe and CMB/early-Universe priors.
\begin{figure}[t]
\centering
\includegraphics[width=8.5cm]{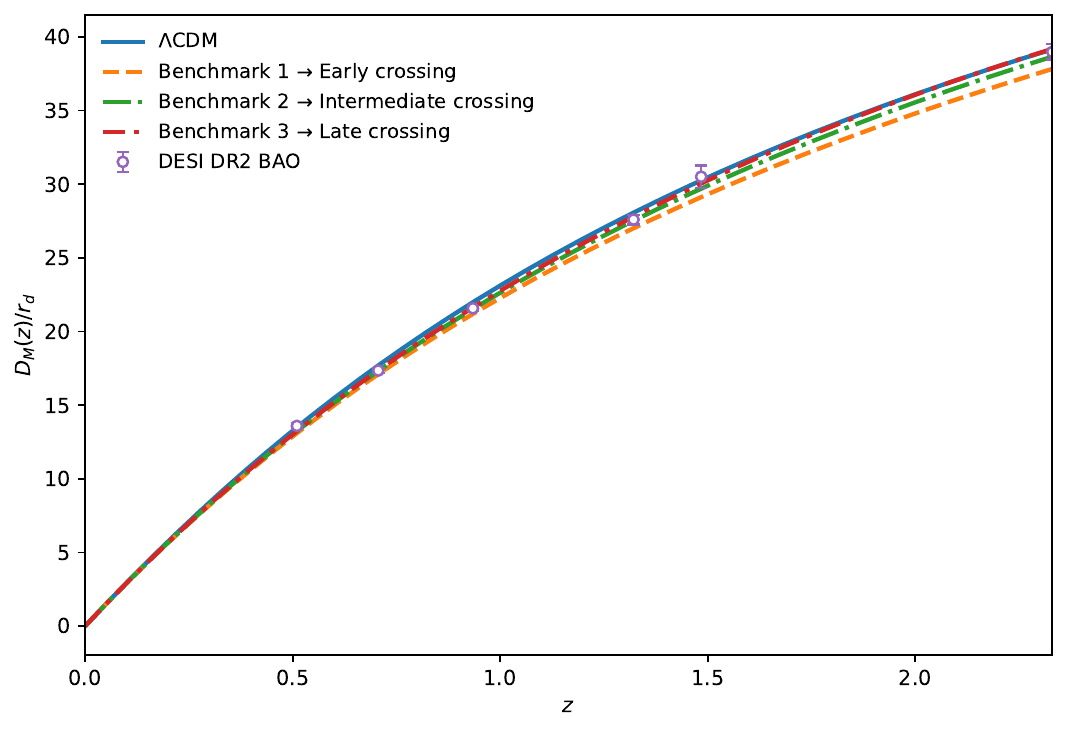} \quad
\includegraphics[width=8.5cm]{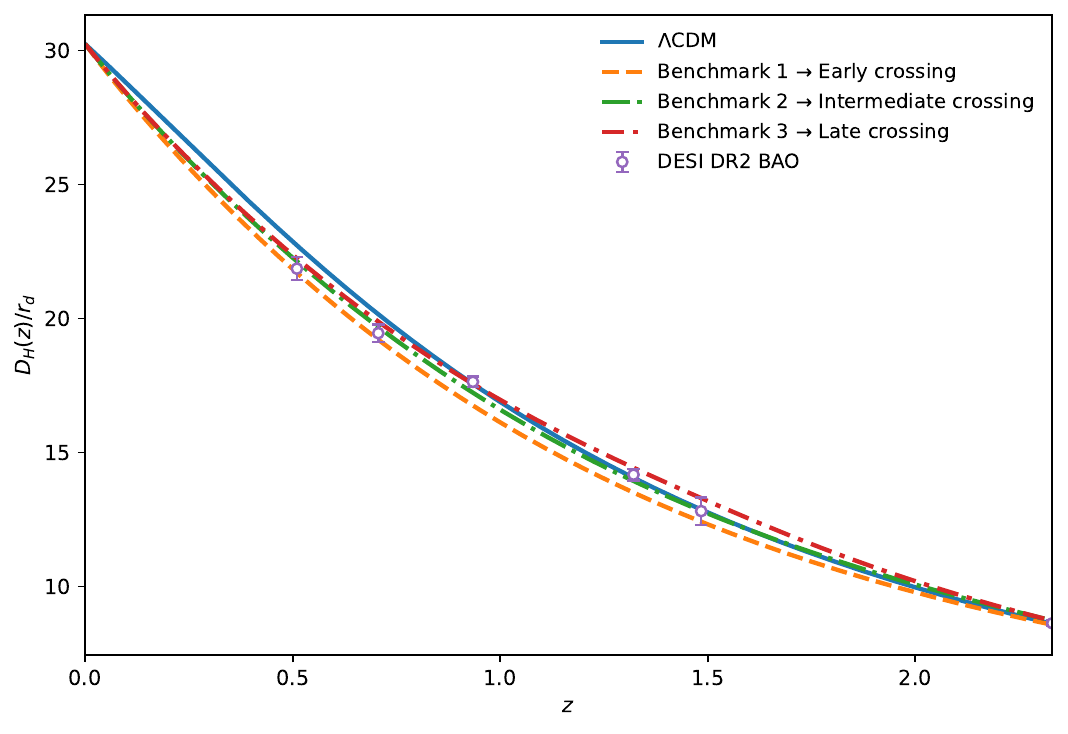} \quad
\includegraphics[width=8.5cm]{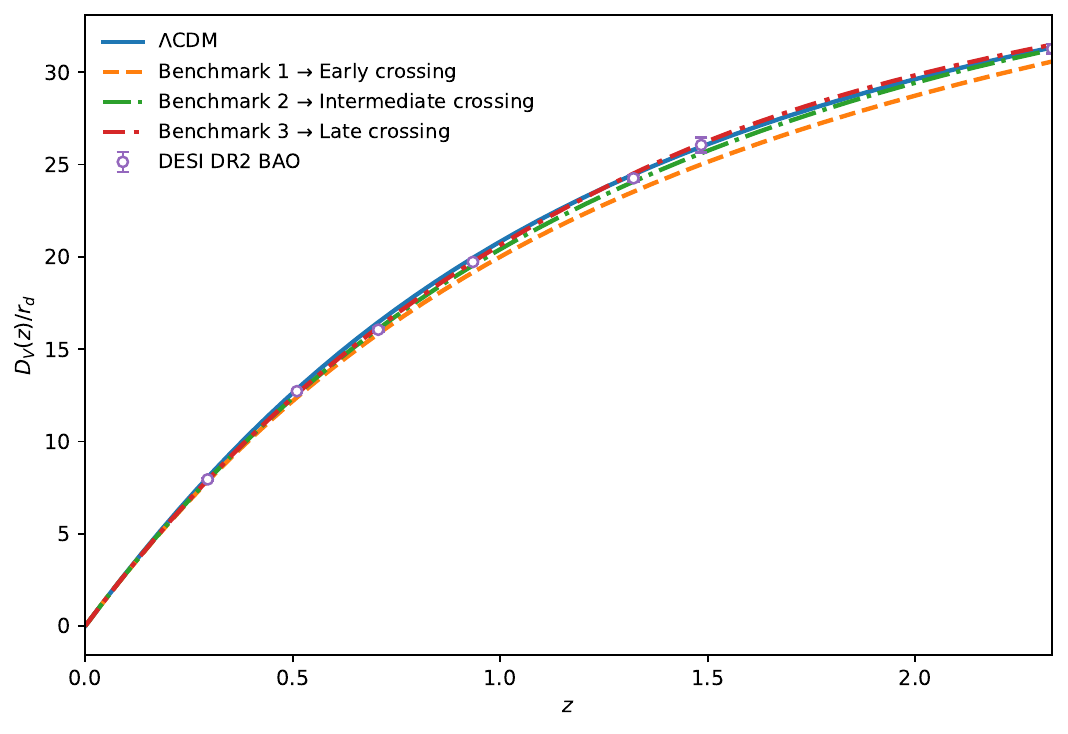}
\caption{ \it The transverse BAO distance ratio $D_M(z)/r_d$ (upper left), the radial BAO distance ratio $D_H(z)/r_d$ (upper right) and the isotropic BAO distance ratio for $D_V(z)/r_d$ for the three quintom benchmarks, compared with $\L$CDM and DESI DR2 BAO measurements. The curves are obtained from the reconstructed $H(z)$, while the DESI points provide a benchmark-level observational comparison of the model geometry.}
\label{DMHVrd}
\end{figure} 
We next propagate the reconstructed expansion history into the distance observables relevant for BAO. For a spatially flat background, the transverse comoving distance is
\be
D_M(z)=c\int_0^z\frac{dz'}{H(z')}
\ee
while the angular-diameter and Hubble distances are given by
\be
D_A(z)=\frac{D_M(z)}{1+z} ~~ {\rm and} ~~ D_H(z)=\frac{c}{H(z)}
\ee
respectively.
We also compute the isotropic BAO distance
\be
D_V(z)=\left[zD_M^2(z)D_H(z)\right]^{1/3}
\ee
The quantities directly constrained by BAO measurements are therefore
\be
D_M(z)/r_d, ~~ D_H(z)/r_d ~~ {\rm and} ~~ D_V(z)/r_d
\ee
where $r_d$ is the sound horizon at the drag epoch. 
\begin{table}[t]
\centering
\begin{tabular}{lcccc}
\hline
Model & $N_{\rm data}$ & $\chi^2_{\rm BAO}$ & $\chi^2_{\rm BAO}/N_{\rm data}$ & $\Delta\chi^2_{\Lambda{\rm CDM}}$ \\
\hline
$\Lambda$CDM & 13 & 28.69 & 2.21 & 0 \\
Benchmark 1: Early crossing & 13 & 88.33 & 6.79 & $+59.63$ \\
Benchmark 2: Intermediate crossing & 13 & 14.93 & 1.15 & $-13.76$ \\
Benchmark 3: Late crossing & 13 & 11.81 & 0.91 & $-16.88$ \\
\hline
\end{tabular}
\caption{
Full-covariance compressed DESI DR2 BAO comparison for the three NPGHU/quintom benchmarks and the fiducial \(\Lambda\)CDM curve used in the plots. The statistic is computed using the official DESI DR2 BAO mean vector and covariance matrix for the compressed observables \(D_V/r_d\), \(D_M/r_d\), and \(D_H/r_d\). The quantity \(\Delta\chi^2_{\Lambda{\rm CDM}}\equiv \chi^2_{\rm BAO}-\chi^2_{\rm BAO,\Lambda{\rm CDM}}\) is shown only as a benchmark-level diagnostic. This table should not be interpreted as a full model-selection analysis, since no scan over the complete NPGHU/quintom initial-condition space is performed.
}
\label{tab:bao_fullcov_chi2}
\end{table}
As we have already mentioned, for the current work we care about the late-time modifications of the dark energy sector therefore we assume that the early thermal history is unchanged. Under this assumption we fix the standard sound-horizon calibration for the benchmark comparison, $r_d = 147.09 ~\rm Mpc$.
In \fig{DMHVrd} we compare the three quintom benchmarks with DESI DR2 BAO distance measurements. The BAO-distance reconstruction shows that the timing of the Quintom-B transition has a visible impact on late-time geometric observables. In particular, the early-crossing benchmark produces the largest departure from the DESI DR2 BAO distance ratios, especially in ($D_M/r_d$) and ($D_V/r_d$), because its modified expansion history affects the distance integral over a longer redshift interval. By contrast, the intermediate- and late-crossing benchmarks remain significantly closer to the DESI DR2 distance points over the redshift range considered. This indicates that our effective quintom model contains phenomenologically viable regions in which the phantom-divide crossing is compatible with the current BAO geometry, while more extreme early-crossing realizations are comparatively disfavored.\\
In this regard and in order to quantify the visual comparison with DESI DR2 BAO data, we also compute the compressed BAO statistic as it is shown in Table~\ref{tab:bao_fullcov_chi2} with
\begin{equation}
\chi^2_{\rm BAO} = \Delta {\cal O}^{T} C^{-1}\Delta {\cal O}, \qquad \Delta{\cal O}_i = {\cal O}^{\rm model}_i- {\cal O}^{\rm DESI}_i,
\end{equation}
where ${\cal O}_i=\{D_V/r_d,D_M/r_d,D_H/r_d\}$ and $C$ is the full DESI DR2 compressed BAO covariance matrix. In this comparison we use the official 13-dimensional DESI BAO mean vector and covariance matrix. We identify the DESI notation $r_s$ with the sound-horizon scale $r_d$ used in the present work.
The full-covariance BAO comparison confirms the qualitative behavior already visible in the distance plots. The early-crossing trajectory gives a substantially larger $\chi^2_{\rm BAO}$ and is therefore disfavored at the benchmark level. By contrast, the intermediate- and late-crossing trajectories give smaller $\chi^2_{\rm BAO}$ values than the fiducial $\L$CDM curve used here, showing that these NPGHU/quintom backgrounds can remain compatible with DESI DR2 BAO distance measurements at the compressed-observable level. We stress that this is still not a full likelihood analysis of the model: a definitive statistical comparison would require scanning the model initial conditions, including the full cosmological data combination and eventually implementing the perturbation sector in a Boltzmann solver which is outside the scope of the current work.
\begin{figure}[t]
\centering
\includegraphics[width=12cm]{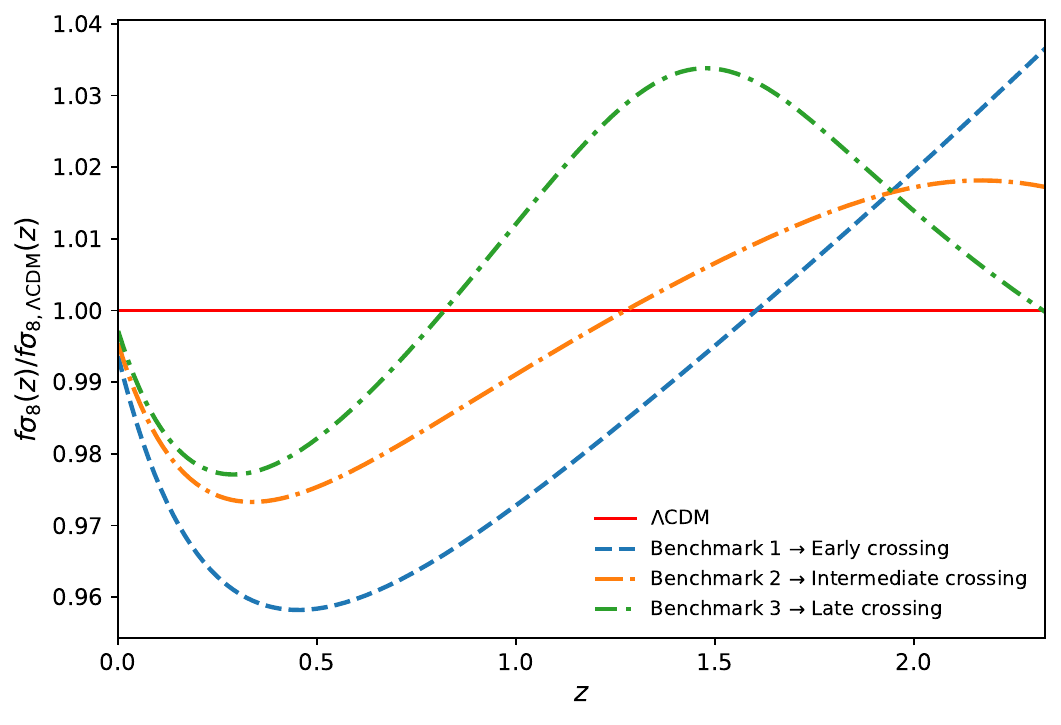}
\caption{ \it The ratio $f\sigma_8(z)/f\sigma_{8, \rm \L CDM}(z)$ for the three quintom benchmarks in the smooth-dark-energy approximation. The solid reference line denotes the $\L$CDM value. The curves show that the benchmark trajectories remain close to the standard growth prediction at the scale-independent linear-growth level.}
\label{fs8fs8LCDM}
\end{figure} 

Finally, we compute a first growth-level observable for the same benchmark trajectories. We work in the smooth-dark-energy approximation, in which dark energy perturbations are neglected on the sub-horizon scales relevant for redshift-space-distortion measurements. The linear matter growth factor $D(N)$ then satisfies
\be
\ddot D+
\left[ 2+\frac{d\ln H}{dN} \right] \dot D - \frac{3}{2}\Omega_m(N)D=0
\ee
with
\be
\Omega_m(N)=
\frac{\Omega_{m0}e^{-3N}}{E^2(N)}
\ee
The logarithmic growth rate is
\be
f(z)=\frac{d\ln D}{d\ln a}=\frac{\dot D}{D}
\ee
and the observable relevant for redshift-space-distortion measurements is
\be
f\sigma_8(z)=
f(z)\sigma_{8,0}\frac{D(z)}{D(0)}
\ee
The resulting curves depicted on \fig{fs8fs8LCDM} remain close to the $\L$CDM growth prediction, indicating that the same benchmark trajectories that pass the background and BAO-distance tests do not show an immediate growth-level pathology in the smooth-dark-energy approximation.
The smooth-growth calculation should be understood in light of the stability analysis of \sect{IFPI}. There we argue that the full scalar perturbation sector (and similarly for the gauge counterpart), including the R-ghost degree of freedom that is subdominant at the background level, admits a perturbatively stable regime. Moreover, we show that the potentially dangerous gravitationally induced vacuum decay of the phantom sector is regulated by the finite localization cut-off $\L$. 
Based on these considerations, the growth analysis performed here uses background trajectories that are theoretically admissible within the effective model. Nevertheless, the calculation remains a scale-independent smooth-dark-energy approximation.
A full perturbation-level likelihood analysis would require implementing the complete scalar, gauge, phantom and R-ghost perturbation system, including possible finite-$\L$ scale-dependent effects, in a Boltzmann solver such as CLASS/CAMB which is left for future work.

Overall, the benchmark analysis shows that the NPGHU-originated effective quintom model can generate viable late-time expansion histories with controlled Quintom-B crossings and non-trivial but moderate deviations from $\L$CDM. The DESI BAO comparison indicates that the timing of the phantom-divide crossing is phenomenologically important: intermediate- and late-crossing trajectories remain closer to the observed BAO distance ratios than the early-crossing case. Finally, the smooth-growth calculation shows no immediate growth-level pathology, while a full perturbation-level likelihood analysis, including scale-dependent finite-cutoff effects, is necessary in order to draw a quantitative conclusion.

\section{Instabilities from perturbations and interactions}\label{IFPI}
The background evolution of our dynamical DE model was analytically and numerically evaluated in the previous section where we showed that representative Quintom-B trajectories can be obtained and can remain compatible with DESI DR2 BAO distances at the benchmark level, without introducing ad-hoc tuning in the scanned initial-condition region. During this process we showed that the spectrum includes two physical fields, two phantom fields and two R-ghosts but only the first two categories were involved in the evaluation of the background EoS since the latter have a subdominant evolution. On the other hand, R-ghosts were crucial in order to maintain a consistent spectrum when comes to the problematic extra pole of the propagators since they counter-act in such a way to alleviate the pole-instability.
However, one should be careful since the system may still develop instabilities which originate either from 1) perturbations around the field background or 2) non-trivial gravitational interactions between physical and phantom fields at quantum level, leading to vacuum decay.
Below we examine both scenarios within our model and we show that neither of these situations leads to an immediate instability within the controlled effective regime considered here.
In order to facilitate our study, in the following we focus once more only on the scalar part of the effective quintom model while similar analysis holds for the gauge counterpart.

\subsection{Expanding around the background}
In contrast to the background evolution where we have shown that the R-ghosts are negligible, when perturbations are considered then all the dof of the spectrum are present. 
In that sense, the system now will include the coupled eom among physical, phantom and R-ghost fields.
In order to study whether instabilities at the perturbation level may drastically affect our system or not we consider fluctuations of both the metric and the scalar fields around their background values. 
In the present work we focus on the linear level and the prescription that we follow suggests that 
\bea\label{deltagdeltaphi12}
g_{\mu\nu} = \bar g_{\mu\nu}(\tau) + \delta g_{\mu\nu}(\tau,x)~, &&~~~ \phi_1(\tau,x) = \varphi_1(\tau) + \delta\phi_1(\tau,x)~,\nn\\
\phi_2(\tau,x) = \varphi_2(\tau) + \d\phi_2(\tau,x) &&~~{\rm and}~~ \chi(\tau,x) = \chi_0(\tau) + \d\chi(\tau,x)
\eea
where $ \bar g_{\mu\nu}(\tau)$ is the background FRW metric of \eq{gmnb}. Since we are working with scalar fields causing perturbations on the spacetime and vice-versa it would be enough to focus only on the scalar perturbations of the metric. In that sense the perturbed metric of \eq{gmnb} becomes
\be\label{metric2}
ds^2 = a^2(\tau) \left( [1+ 2\Psi] d\tau^2 + 2B_{,i} d\tau dx^i - \{ [1 - 2\Phi]\delta_{ij}  -2 E_{,ij} \} dx^idx^j \right)  \,,
\ee
where we follow the standard textbook notation regarding the scalar perturbations 
where $\Psi$ and $\Phi$ are the well known Bardeen potentials (and $\Phi$ here should not be confused with the collective form for the fields that we used in the previous section). 
Of course the above perturbed metric corresponds to the most general form which means that in principle includes non-physical degrees of freedom and needs to be gauge-fixed. In the following we use the conformal-Newtonian gauge ($B = E = 0$) which setts, in the absence of anisotropic stress ($\Psi = \Phi$), that
\be\label{metric3}
ds^2 = a^2(\tau) \left( [1 + 2\Phi] d\tau^2 - [1 - 2\Phi]  dx^2 \right)
\ee
The next step is to put the perturbations back to our effective action which now is given by the full spectrum of \eq{Sscalar}. Then the general form for the eom of a given complex scalar field, $X_i \in \{ \phi_1, \phi_2, \chi \}$ with $\bar X_i \in \{ \phi^*_1, \phi^*_2, \chi^* \}$, at linear level is given by
\bea
\delta X_i'' + 2 {\cal H} \delta X_i' &+& k^2 \delta X_i - 4 \Phi' X_i' + 2 (-1)^{i+1} a^2 \Phi  \frac{\partial V_{\rm scalar}(X, \bar X)}{\partial \bar X_i} \Big |_{X_i = X_{i,0}} \nn\\
  && + (-1)^{i+1} a^2 \sum_{X_j }  \left[ \frac{\partial^2 V_{\rm scalar}(X, \bar X)}{\partial \bar X_i \partial X_j} \d X_j +  \frac{\partial^2 V_{\rm scalar}(X, \bar X)}{\partial \bar X_i \partial \bar X_j} \d \bar X_j \right]_{X_i = X_{i,0}} = 0
\eea
where $X_{i,0} = \{ \varphi_1, \varphi_2, \chi_0 \}$ denotes the associated background field values. In our case the scalar potential is given by
\be
 V_{\rm scalar}(X, \bar X) = - m_{\phi_2}^2  |\phi_2|^2  -  m_{\phi_2}^2  |\chi|^2 + \l_\chi |\chi|^2  |\phi_1 - \phi_2|^2
\ee
therefore the system of equaitons for the linear perturbations $\d X_1 = \d \phi_1$, $\d X_2 = \d \phi_2$ and $\d X_3 = \d \chi$ becomes
\bea
\delta\phi_1'' + 2 {\cal H} \delta\phi_1' + (k^2 + a^2\l_\chi |\chi_0|^2) \delta\phi_1 &-& a^2 \l_\chi |\chi_0|^2 \delta\phi_2 + a^2\l_\chi \D\varphi_{1,2} \D\chi^2 \nn\\
  &&- 4 \Phi' \varphi_1' + 2a^2 \Phi \l_\chi |\chi_0|^2 \D\varphi_{1,2} = 0 \nn \\
\delta\phi_2'' + 2 {\cal H} \delta\phi_2' + (k^2 + a^2 m_{\phi_2}^2 - a^2\l_\chi |\chi_0|^2) \delta\phi_2 &+& a^2 \l_\chi |\chi_0|^2 \delta\phi_1 + a^2\l_\chi \D\varphi_{1,2} \D\chi^2 \nn\\
  &&- 4 \Phi' \varphi_2' + 2a^2 \left[  m_{\phi_2}^2 \phi_2  + \l_\chi |\chi_0|^2 \D\varphi_{1,2} \right] \Phi = 0 \nn\\
  \delta\chi'' + 2 {\cal H} \delta\chi' + (k^2 - a^2 m_{\phi_2}^2 + a^2\l_\chi |\D\varphi_{1,2}|^2) \delta\chi &+& a^2 \l_\chi |\chi_0| \left[  \D\varphi^*_{1,2} (\d\phi_1 - \d\phi_2) +  \D\varphi_{1,2} (\d\phi_1 - \d\phi_2)^* \right]  \nn\\
  &&- 4 \Phi' \chi' + 2a^2 \left[ - m_{\phi_2}^2 \chi_0  + \l_\chi \chi_0 |\D\varphi_{1,2}|^2 \right] \Phi = 0
\eea
where we have defined that 
\be
\D\varphi_{1,2} = \varphi_1 - \varphi_2 ~~~ {\rm and} ~~~ \D \chi^2 = \chi_0^* \d\chi + \chi_0 \d\chi^* \nn
\ee
As usual, assuming that $\Phi$ is relatively small, the source of instability in the above system are the modes that could grow exponentially instead of having an oscillatory behavior. The diagnostic of such a possibility is essentially encoded in the effective mass terms of the modes, which we can read from the above set of equations to be
\bea
(k^2 + a^2\l_\chi |\chi_0|^2) \delta\phi_1 &\equiv & \omega_{\phi_1}^2 \d\phi_1 \\
(k^2 + a^2 m_{\phi_2}^2 - a^2\l_\chi |\chi_0|^2) \delta\phi_2 &\equiv& \omega^2_{\phi_2} \d\phi_2 \\
(k^2 - a^2 m_{\phi_2}^2 + a^2\l_\chi |\D\varphi_{1,2}|^2) \delta\chi &\equiv & \omega^2_\chi \d\chi
\eea
The stability condition then is satisfied when all the frequencies are simultaneously positive, $\omega^2_{X_i} > 0$, a condition that we examine under the two possible signs for $\l_\chi$.
\begin{itemize}
\item If $\l_\chi >0$:\\
Then $\l_\chi \equiv |\l_\chi|$ and $\omega_{\phi_1}^2 > 0$ is automatically satisfied for any mode $k$. On the other hand, positivity for $\omega^2_{\phi_2}$ and $\omega^2_\chi$ is guaranteed for every $k$ when 
\be
m_{\phi_2}^2 > |\l_\chi| |\chi_0|^2 ~~~ {\rm and} ~~~ |\l_\chi| |\varphi_1 - \varphi_2|^2 >  m_{\phi_2}^2 ~~{\rm or}~~    \frac{m_{\phi_2}^2}{|\l_\chi| } > |\chi_0|^2 ~~~ {\rm and} ~~~ |\varphi_1 - \varphi_2|^2 > \frac{ m_{\phi_2}^2}{ |\l_\chi| }\nn
\ee
which collectively suggests the stability range
\be
|\chi_0|^2 < \frac{ m_{\phi_2}^2}{ |\l_\chi| } < |\varphi_1 - \varphi_2|^2 \nn
\ee
Normalizing as usual with $|\varphi_{2, \a}|^2$ every part we get 
\be
|\chi_{0,n}|^2 < \frac{ m_{\phi_2}^2}{ |\l_\chi| |\varphi_{2, \a}|^2 } < |\varphi_{1,n} - \varphi_{2,n}|^2  \nn
\ee
so combining the above with the background evolution of the fields that we evaluated\footnote{Keep in mind that we want to find a bound for $\l_\chi$ that keeps the system safe from instabilities today ($k \approx H_{\rm m, 0}$)
. In that sense the strongest bound is given by the normalized fields today, therefore \eq{lchibound3} is obtained using $|\varphi_{1,n}(N) - \varphi_{2,n}(N)|_{N=0}$ and $|\chi_{0,n}(N)|_{N=0}$. } from \eq{phi1phi2chi0} and substituting $m_{\phi_2}^2 \approx 3 H_{\rm m, 0}^2$ 
we get that 
\be\label{lchibound3}
3.5 \times 10^{-4} \lesssim \frac{ H_{\rm m, 0} }{ \sqrt{|\l_\chi|} |\varphi_{2, \a}| } \lesssim 0.14
\ee
Going one step further by substituting $|\varphi_{2, \a}| \approx 10 H_{\rm m, 0}  $
\be\label{lchibound4b}
 0.5 \lesssim |\l_{\chi}| \lesssim 8 \times 10^5  
\ee

\item If $\l_\chi < 0$:\\
Then $\l_\chi \equiv - |\l_\chi|$ and only $\omega^2_{\phi_2}$ is automatically well defined for every mode while, $\omega^2_{\phi_1}$ and $\omega^2_{\chi}$ constrain the system to have a lower bound on $k$ in order to be stable which is given by
\be
k^2/a^2 > |\l_\chi| |\chi_0|^2 ~~~ {\rm and} ~~~ k^2/a^2 > m_{\phi_2}^2 + |\l_\chi| |\varphi_1 - \varphi_2|^2 
\ee
respectively.
Normalizing once more with $|\varphi_{2, \a}|^2$, evaluating everything on today ($k \approx H_{\rm m,0}$ and $a(N=0) \equiv a(N=0)/a_0 = 1$) and using the same parameters as before we get
\be
\sqrt{|\l_\chi|}  \lesssim  10^{3} \frac{H_{\rm m, 0}}{|\varphi_{2, \a}|} \approx 1  ~~~ {\rm and} ~~~ \frac{H^2_{\rm m, 0}}{|\varphi_{2, \a}|^2} \gtrsim  \frac{H^2_{\rm m, 0}}{|\varphi_{2, \a}|^2} + 0.04 |\l_\chi| \Rightarrow |\l_\chi| < 0
\ee
such that we end up with a contradiction.
\end{itemize}
In summary, according to the above discussion our system never develops an instability at the perturbative level and at the same time the R-ghosts remain negligible for the background evolution, as long as $\l_\chi > 0$ and respects the bound of \eq{lchibound3}.

\subsection{Planck-suppressed gravity-phantom interactions}\label{PSGPI}
In the previous section we showed that for a positive quartic coupling there is a range for its magnitude that keeps the system safe from instabilities at the perturbation level.
Then it is tempting to saturate the lower bound and fix $\l_\chi \equiv 0.5 $ 
according to \eq{lchibound4b} such that we keep the system maximally safe from instabilities without interrupting the background evolution.
However, this is not the end of the story since extra interactions among the physical, the phantom and R-ghost fields will eventually arise due to the presence of gravity. The latter involves the coupling of the various field dof with gravitons which could be an extra source of instabilities but now at the quantum level.

To study the universal gravitational interaction here we exploit once more the scalar part of our effective model. So we consider the action \eq{Sscalar} and expand the space-time metric around the flat space as $g_{\m\n} \simeq \eta_{\m\n} + 2 h_{\m\n}/M_{\rm Pl}$ which at linear order will inherit the system with an interaction part such that 
\be
S_{\rm scalar} \approx \int d^4x \sqrt{-\eta} \left[ |\partial_\m \phi_1|^2 - |\partial_\m \phi_2|^2 + |\partial_\m \chi|^2 +  m_{\phi_2}^2  |\phi_2|^2  +  m_{\phi_2}^2  |\chi|^2 - \l_\chi |\chi|^2  |\phi_1 - \phi_2|^2  + {\cal L}_{\rm int} \right]
\ee
with 
\be
{\cal L}_{\rm int} = -\frac{h^{\m\n}}{M_{\rm Pl}} \left[ T^{\phi_1}_{\m\n} + T^{\phi_2}_{\m\n} + T^{\chi}_{\m\n} + \eta_{\m\n} \l_\chi |\chi|^2  |\phi_1 - \phi_2|^2 \right]
\ee
where $h_{\m\n}$ is the canonically-normalized perturbation which corresponds to the graviton field and $M_{\rm Pl} = 2.4 \times 10^{18}~ \rm GeV$ is the reduced Planck mass.
The stress-energy tensor for the physical field is 
\be
T^{\phi_1}_{\m\n} = 2|\partial_\m \phi_1|^2 - \eta_{\m\n} |\partial_\a \phi_1|^2 
\ee
while for the phantom reads
\be
T^{\phi_2}_{\m\n} = - 2|\partial_\m \phi_2 |^2 -  \eta_{\m\n}\left(- |\partial_\a \phi_2|^2 + m_{\phi_2}^2 |\phi_2|^2\right) 
\ee
and for the R-ghost
\be
T^{\chi}_{\m\n} = 2|\partial_\m \chi|^2 -  \eta_{\m\n}\left(|\partial_\a \chi|^2 + m_{\phi_2}^2 |\chi|^2\right)
\ee
Note that for all the processes (tree- or loop-level) originating from ${\cal L}_{\rm int}$ one should quantize the system. In principle when phantoms are in the spectrum there are two choices: either positive energy states and negative norm or negative energy states and positive norm. The former violates conservation of probability while it keeps a bounded from below Hamiltonian while the latter respects unitarity with a cost of an unbounded negative energy spectrum. However in our case, due to the origin of the model in \fig{Higgsphase2}, above the cut-off (in the Coulomb phase) we have only one scalar dof with correct kinetic sign. Therefore, this fixes the quantization choice to be the unitarity-respecting one and so we are forced to use the same choice after the phase transition for $\phi_1$, $\phi_2$ and $\chi$ as well.
Such a choice is, anyway, the natural and consistent one \cite{Cline:2003gs}.

Based on the above it is of crucial importance to investigate under which conditions the interactions among ghosts and physical particles, such as gravitons, will not lead to an instability. The problematic cases seem to arise from the vacuum decay to physical and non-physical fields via graviton exchange. They mainly come from the gravitational interaction among the field content of the dark energy sector\footnote{Of course the presence of gravitational interactions will couple the DE fields with the SM (and DM \cite{Cline:2023cwm}) which will also lead to vacuum decay problems mainly from processes involving photons. However from the analysis in \cite{Cline:2003gs, Cline:2023cwm} the associated bounds are less strict than the ones that we found here so we will not refer to them in the present work.} ($\phi_1, \phi_2$ and $\chi$)
\bea\label{phi1phi2chiprocess}
\frac{1}{M_{\rm Pl}^2} \left(h^{\m\n} T^{\phi_2}_{\m\n} \right)\left(h^{\r\s} T^{\phi_1}_{\r\s} \right) &\Rightarrow & 0 \to \phi_2 +  \phi^*_2 + \phi_1 +  \phi^*_1 \nn\\
\frac{1}{M_{\rm Pl}^2} \left(h^{\m\n} T^{\chi}_{\m\n} \right)\left(h^{\r\s} T^{\phi_1}_{\r\s} \right) &\Rightarrow & 0 \to \chi +  \chi^* + \phi_1 +  \phi^*_1 \nn \\
\frac{1}{M_{\rm Pl}^2} \left(h^{\m\n} T^{\phi_2}_{\m\n} \right)\left(h^{\r\s} T^{\chi}_{\r\s} \right)&\Rightarrow & 0 \to \phi_2 +  \phi^*_2 + \chi + \chi^* 
\eea
where we have neglected processes arising from the term $\eta_{\m\n} h^{\m\n} \l_\chi  |\chi|^2  |\phi_1 - \phi_2|^2 /M_{\rm Pl} $ since it is ${\cal O}(10^{-5})$ times smaller than the rest due to our $\l_\chi$ choice.
Note that the above processes are independent and lead to different final states therefore, even though $ T^{\phi_2}_{\m\n}$ and $ T^{\chi}_{\m\n}$ come with opposite signs for their kinetic parts, they do not add up in order to cancel each other.
On the other hand, recall that the mass of the phantom and the R-ghost is $m_{\phi_2}^2 \approx H^2_{\rm m, 0}$ while the kinetic part in momentum space will admit momenta much higher than the Hubble rate today.
The associated Feynman rules then become
\bea
M^{\phi_2}_{\m\n} & = & -\frac{i}{M_{\rm Pl}} \left[ p_{1,\m}p_{2,\n} + p_{2,\m}p_{1,\n} - \eta_{\m\n} p_1\cdot p_2 \left(1 - \frac{H^2_{\rm m, 0}}{p_1\cdot p_2} \right) \right] \approx \nn -\frac{i}{M_{\rm Pl}} \left[ p_{1,\m}p_{2,\n} + p_{2,\m}p_{1,\n} - \eta_{\m\n} p_1\cdot p_2  \right] \\
M^{\chi}_{\rho\s} & = & \frac{i}{M_{\rm Pl}} \left[  l_{1,\rho}l_{2,\s} + l_{2, \rho}l_{1,\s} - \eta_{\rho\s} l_1\cdot l_2 \left( 1  - \frac{H^2_{\rm m, 0}}{l_1\cdot l_2} \right) \right] \approx  \frac{i}{M_{\rm Pl}} \left[  l_{1,\rho}l_{2,\s} + l_{2, \rho}l_{1,\s} - \eta_{\rho\s} l_1\cdot l_2 \right] \nn
\eea
According to the above, keeping in mind that the R-ghosts have the correct sign for the kinetic term, the second process in \eq{phi1phi2chiprocess} is kinematically forbidden while the first and third ones will essentially be degenerate in the sense that they will give the same vacuum decay problem. Therefore, we only have to examine one of these processes to see whether the associated decay rate will be catastrophic for the vacuum or not since the phase space of the phantom fields is infinite.
Here we choose to study
\be
\frac{1}{M_{\rm Pl}^2} \left(h^{\m\n} T^{\phi_2}_{\m\n} \right)\left(h^{\r\s} T^{\phi_1}_{\r\s} \right) \Rightarrow  0 \to \phi_2 +  \phi^*_2 + \phi_1 +  \phi^*_1 
\ee
Let us mention that one may wonder about the possible presence of mixed kinetic terms between the physical and phantom fields, in analogy with \cite{Saridakis:2009jq}. A term of the schematic form
\be
\kappa\,\partial_\mu \phi_1\,\partial^\mu\phi_2 + {\rm h.c.} \nn
\ee
would correspond to a non-diagonal field-space metric in the auxiliary two-field description and could lead to interesting phenomenological effects for the phantom crossing. In the leading NPGHU-induced action used in this work, however, the auxiliary-field transformation diagonalizes the higher-derivative scalar sector, leading to the kinetic structure
\be
|\partial_\mu\phi_1|^2-|\partial_\mu\phi_2|^2 \nn
\ee
with no independent mixed kinetic operator retained in the dimension-six approximation. Moreover, the gravitational interaction considered below couples through the stress-energy tensor and generates higher-dimensional derivative interactions rather than a leading two-derivative kinetic mixing between $\phi_1$ and $\phi_2$. Possible kinetic mixing induced by higher-dimensional operators and/or loop effects are therefore left for future work.

Back to our calculation, the way to proceed in our case is to calculate the above decay rate and from that to evaluate the lifetime of the reaction.
If the latter is much longer compared to the lifetime of the observable universe today, then the vacuum will remain stable and the presence of the phantoms will not be problematic. As time is passing and energies are shifting towards the deep IR (Higgs phase) in the far future, localization will be lost and the full 5d construction will be recovered. The latter is ghost-free by definition and the UV-completion in our case (extradimensional, anisotropic and discrete) is free from phantom instabilities.\\
Now, for the above calculation the relevant Feynman rules are given by
\bea
M^{\phi_2}_{\m\n} &\approx & -\frac{i}{M_{\rm Pl}} \left[ p_{1,\m}p_{2,\n} + p_{2,\m}p_{1,\n} - \eta_{\m\n} \left( p_1\cdot p_2 
\right) \right] \nn \\
M^{\phi_1}_{\rho\s} & = & \frac{i}{M_{\rm Pl}} \left[  k_{1,\rho}k_{2,\s} + k_{2, \rho}k_{1,\s} - \eta_{\rho\s} \left( k_1\cdot k_2 \right) \right]
\eea
where $p_i$ corresponds to the phantom momentum and $k_j$ to the physical-field momentum with $i, j = \{1,2\}$.
In addition, we need the graviton propagator for the canonically normalized graviton field
\be
{\rm P}^{\m\n\rho\s} = \frac{\eta^{\m \r } \eta^{\n \s } + \eta^{\m \s } \eta^{\n \r } - \eta^{\m \n } \eta^{\r \s }}{2 s_{q_{1 2}}}
\ee
such that the amplitude of the interaction under consideration becomes
\be\label{calMh}
{\cal M}_{h} \equiv - \frac{ T^{\phi_2}_{\m\n}(p_i) {\rm P}^{\m\n\rho\s} T^{\phi_1}_{\rho\s}(k_j)}{M_{\rm Pl}^2} = \frac{
{ F\left( s_{q_{\a \b}} \times s_{q_{\g \d}}\right) } }{4 M_{\rm Pl}^2 s_{q_{1 2}}} 
\ee
where $F\left( s_{q_{\a \b}} \times s_{q_{\g \d}}\right) \approx s_{q_{1 3}} \times s_{q_{2 4}} + s_{q_{1 4}} \times s_{q_{2 3}} - s_{q_{1 2}} \times s_{q_{3 4}} $ is constructed from the following Lorentz-invariant quantities $s_{q_{\a \b}} = (q_\a + q_\b)^2$ with $q_\a \in \{q_1, q_2, q_3, q_4 \} = \{p_1, p_2, k_1, k_2 \}$. As an indicative example we have that $s_{q_{1 2}} = \left( p_1 + p_2 \right)^2 = \left( k_1 + k_2 \right)^2 = s_{q_{3 4}}$ is the standard s-channel Mandelstam invariant.
In our case $p_i^2 = H^2_{\rm m, 0} \ll \L$ and $k_j^2 = 0 $ such that $s_{q_{\a \b}} \approx 2 q_\a \cdot q_\b$.\\ 
Before we move on with the actual calculation of the decay rate we would like to remind the reader about two crucial characteristics of our model which will drastically affect the above calculation. Both come from the 5d anisotropic-orbifold lattice origin of our effective action and are related to the discussion around \eq{L45UV}.
The first is about the existence of the (finite) cut-off $\L$ on which the model will exhibit a 1st order phase transition. The latter affects only the 5d-projected dof and is purely a consequence of the bulk-boundary (di-)localization indicating that $\L$ is an emergent scale. 
Due to the phase transition the model shifts from the Higgs to the Coulomb phase where the phantom dof and its amplitudes cease to exist.\\
The second characteristic is that since our NPGHU model is by definition constructed on the lattice, then the momenta are naturally restricted to a Brillouin zone in the lattice rest frame (the existence of the orbifold boundaries and the anisotropy in the fifth dimension single out preferred directions breaking explicitly also the 5d Poincaré symmetry).
In that sense, both $\L_{4,\rm UV}$ and $\L$ will provide naturally a preferred rest frame constraining the associated momenta. 
Before the Higgs-Hybrid phase transition towards the IR, in the regime where localization is completely lost and bulk does not affect the boundary, the momenta of the model ($P'_\m$) are confined based on the expected 4d-lattice UV completion under the Brillouin zone 
\be
P'_\m \in (0, \L_{4,\rm UV}]
\ee
Therefore Lorentz Invariance (LI) seems exact up to the UV scale $\L_{4,\rm UV}$ where the spacetime is discretized.
For $\m \lesssim \L$, localization is strong but not exact since the bulk non-trivially affects the boundary by introducing the higher derivative operators along with $\L$. This new emergent scale can be rewritten as $\L \sim F(\g, a_4^f,\b_4,\b_5)/a^f_5$ and in \cite{Irges:2018gra, Irges:2020nap} was shown that the 5d lattice spacing, $a_5^f$, which multiplies the boundary components admits a finite (arbitrarily high) value. On the contrary, $a_4^f$ remains finite. Therefore, $\L$ could naturally be a very small scale which will effectively constrain the 4-momenta ($P_\m $) after the transition, but before localization is completely lost in the deep IR (far future), by 
\be
P_\m \in (0, \L] 
\ee
As a consequence, LI is not fundamental in the NPGHU model and continuous Lorentz boosts are not an exact symmetry of the theory in the presence of a finite cut-off. In other words, the 4d Lorentz invariance is emergent and approximate in the deep IR. In fact, the implicit effect of the bulk dictates the presence of a very low emergent scale in the 4d boundary and the question of how strongly LI is broken below $\L$ is answered, phenomenologically, in the following by demanding the absence of a problematic vacuum decay.

The previous two characteristics implicitly constrain the phase-space of integration and they should be somehow imprinted on the scattering amplitudes. Since localization dynamics is still not understood beyond lattice simulations it is natural to impose its effect directly on the amplitudes themselves. Here we care about the region after the phase transition but still in the vicinity of $\L$.
So we choose, based on cosmological arguments, as the lattice preferred rest frame the cosmological comoving frame where the CMB is (almost) isotropic and we realize the above constraints in a Lorentz-violating way by inserting the following regulator   
\be
{\cal R}_{\rm loc} = \prod_{\m} \Theta( \L - |q_\m| ) 
\ee
on the amplitude level, bounding the physical $4$-momentum to be $|q_\m| \lesssim \L$.
In that sense, the amplitude squared which incorporates the localization effect and we consider here is given by
\be
|{\cal M}^{\rm loc}_{h}|^2 \equiv |{\cal M}_{h}|^2 {\cal R}_{\rm loc} \nn
\ee
Notice that with the above $\Theta$-function we restrict both the usual invariants $s_{q_{1 2} } (s_{q_{3 4} }) $ and the cross invariant terms 
\be
s_{q_{(1,2) (3, 4)}} = \left( p_{(1,2)} + k_{(1,2)} \right)^2 \approx 2 p_{(1,2)} \cdot k_{(1,2)} \nn
\ee
Based on the previous discussions, we are now ready to calculate the vacuum decay rate per unit volume ($\Gamma_0 \equiv \Gamma_{2\phi_22\phi_1}/V$) which will be given by the following formula 
\bea\label{Gamma0}
\Gamma_{0} &=&  \prod^2_{i = 1}\int \frac{d^3 p_i  }{(2\pi)^3 (2 E_{p_i})}  \prod^2_{j = 1} \int \frac{d^3 k_j}{(2\pi)^3 (2 E_{k_j})} (2\pi)^4 \d^{(4)}\left( p_1 + p_2 + k_1 + k_2 \right) |{\cal M}^{\rm loc}_{h}|^2 
\nn\\
 \Gamma_{0} &\equiv & \int d\Pi_{2\phi_2 2\phi_1}  |{\cal M}_{h}|^2 {\cal R}_{\rm loc}
\eea
where 
\be\label{Lips}
d\Pi_{2\phi_2 2\phi_1} = (2\pi)^4 \d^{(4)}\left( p_1 + p_2 + k_1 + k_2 \right) \prod^2_{i = 1}\int \frac{d^3 p_i  }{(2\pi)^3 (2 E_{p_i})}  \prod^2_{j = 1} \int \frac{d^3 k_j}{(2\pi)^3 (2 E_{k_j})} 
\ee
corresponds to the Lorentz-invariant phase space of the two pairs of created particles.
The explicit calculation of the vacuum decay rate \eq{Gamma0} has some tedious algebra while similar calculations have been done recently in \cite{Cline:2023cwm}. Therefore, here we do not present the details of the calculation since it follows the steps performed in \cite{Cline:2023cwm} under the necessary adjustments.
However, it would be useful to schematically present the scaling of this integral within the prescription that we implied above.
For that purpose we deal with each part of \eq{Gamma0} separately under the rescaling $ s_{q_{\a\b}} \equiv \L^2 \hat s_{q_{\a\b}} $, where $\hat s_{q_{\a\b}}$ is dimensionless, which is implicitly enforced by the ${\cal R}_{\rm loc}$.
In that sense the amplitude of \eq{calMh} is rewritten as 
\be\label{calMhr}
{\cal M}_{h} \approx \frac{ s_{q_{12}} s_{q_{12}}}{M_{\rm Pl}^2 s_{q_{12}}} \equiv  \frac{ \L^2 }{M_{\rm Pl}^2} \hat s_{q_{12}} \Rightarrow |{\cal M}_{h}|^2 \approx  \frac{ \L^4 }{M_{\rm Pl}^4} \hat s_{q_{12}}^2 
\ee
For the phase-space analysis we will first bring the associated integrals of \eq{Lips} in their 4-momentum form and insert the unity 
\be
1 = \int \frac{d^4 q}{(2\pi)^4} (2\pi)^4 \d^{(4)}\left( q - p_1 - p_2  \right) \nn
\ee
to rewrite the former collectively as
\bea
d\Pi_{2\phi_2 2\phi_1}
  &=& \int \frac{d^4 q}{(2\pi)^4} \left[(2\pi)^4 \d^{(4)}\left( q - p_1 - p_2  \right) \prod^2_{i = 1} \frac{d^4 p_i}{(2\pi)^3} \d^{(4)}(p^2_i - H^2_{\rm m, 0}) \Theta(- p_i^0) \right] \nn\\
    &&\times \left[(2\pi)^4 \d^{(4)}   \left( q + k_1 + k_2  \right) \prod^2_{j = 1} \frac{d^4 k_j}{(2\pi)^3} \d^{(4)}(k^2_j) \Theta(k_j^0)   \right] {\cal R}_{\rm loc} \nn
\eea
Then the total integral over the phase space is factorized to two sub-phase-space integrals as
\be\label{Lips2}
\int d\Pi_{2\phi_2 2\phi_1} = \int \frac{d^4 q}{(2\pi)^4}  {\cal R}_{\rm loc} \int d \Pi_{\phi_2}( - q \to \phi_2 \phi^*_2) \int d \Pi_{\phi_1}( q \to \phi_1 \phi^*_1)
\ee
and each one of them can be evaluated separately.
In particular, they both will contribute just a dimensionless quantity given by
\bea
d \Pi_{\phi_2}( - q \to \phi_2 \phi^*_2) &\equiv & \frac{\sqrt{1 - \frac{H^2_{\rm m, 0}}{s_{q_{12}}} }}{32\pi^2} d\Omega_{\phi_2} \Rightarrow \int d \Pi_{\phi_2}( - q \to \phi_2 \phi^*_2) \approx \frac{1}{32 \pi^2} \int d\Omega_{\phi_2} \nn\\
d \Pi_{\phi_1}( q \to \phi_1 \phi^*_1) &\equiv & \frac{1}{32\pi^2} d\Omega_{\phi_1} \Rightarrow \int d \Pi_{\phi_2}( - q \to \phi_2 \phi^*_2) = \frac{1}{32 \pi^2} \int d\Omega_{\phi_1} \nn
\eea
where we have used that $\sqrt{1 - H^2_{\rm m, 0}/s_{q_{12}}} \approx 1$ since $s_{q_{12}} \gg H^2_{\rm m, 0}$.
The scaling of the phase-space and its problematic nature due to its infinite boost volume are then completely included in the $q$-integration.
In the following we expose both the scaling and the way that this divergence is regulated here due to the lattice preferred rest frame, by shifting to the rapidity base $(\eta)$.\\
For a fixed invariant mass-squared, $s_{q_{12}} = s_{q_{34}} = q^2$, one could choose a time-like 4-momentum and parameterize its components using $\eta$ by setting
\be
q^0 = \sqrt{s_{q_{12}}} \cosh\eta ~~{\rm and}~~ |q^i| = \sqrt{s_{q_{12}}} \sinh\eta \nn
\ee
which means that the measure of the integral becomes 
\be
d^4 q = \frac{s_{q_{12}}}{2} \sinh^2\eta ~ ds_{q_{12}} d\eta d\Omega_3 \nn
\ee
Then, putting the above into \eq{Lips2} we get 
\be\label{Lips2r}
\int d\Pi_{2\phi_2 2\phi_1} \propto \int ds_{q_{12}} \left[\frac{s_{q_{12}}}{2} \right] \int d\eta \left[ \sinh^2\eta   \right] {\cal R}_{\rm loc} \int d\Omega_{\phi_2} \int d\Omega_{\phi_1}
\ee
where now the regulator is written as
\be
 {\cal R}_{\rm loc} = \Theta( \L - \sqrt{s_{q_{12}}} \cosh\eta ) \prod_{i} \Theta( \L - \sqrt{s_{q_{12}}} \sinh\eta )
\ee
From the time-component it is straightforward to obtain the rapidity bound
\bea
\L  &\ge & \sqrt{s_{q_{12}}} \cosh\eta \Rightarrow \nn\\
 \eta &\le & \cosh^{-1} \left(\frac{\L}{\sqrt{s_{q_{12}}}} \right) \equiv \eta_{\rm max} \nn
\eea
and then the rapidity integral becomes
\be
\int d\eta \left[ \sinh^2\eta   \right] {\cal R}_{\rm loc} \equiv \int_0^{\eta_{\rm max}} d\eta \left[ \sinh^2\eta   \right] = \frac{\sinh 2\eta_{\rm max}}{4} - \frac{\eta_{\rm max}}{2} \approx \frac{1}{2} \frac{\L^2}{s_{q_{12}}}
\ee
where we have kept only the leading-order term in the $\L \gg \sqrt{s_{q_{12}} }$ expansion. 
Based on the previous discussion, the total phase-space now reads
\bea\label{Lips2r2}
\int d\Pi_{2\phi_2 2\phi_1} & \propto & \int_{4 H^2_{\rm m, 0}}^{\L^2} ds_{q_{12}} \left[\frac{s_{q_{12}}}{2} \right] \frac{\L^2}{s_{q_{12}}} \int d\Omega_{\phi_2} \int d\Omega_{\phi_1} \Rightarrow \nn\\
 \int d\Pi_{2\phi_2 2\phi_1} & \propto & \L^4 \int_{\frac{4 H^2_{\rm m, 0}}{\L^2}}^1 d \hat s_{q_{12}} \int d\Omega_{\phi_2} \int d\Omega_{\phi_1} 
\eea
where the rescaling $s_{q_{12}} = \L^2 \hat s_{q_{12}}$ has been used.
Finally, substituting the above along with \eq{calMhr} back to \eq{Gamma0} we get 
\be
\Gamma_0 \sim \frac{\L^8}{M_{\rm Pl}^4}
\ee
and including the numerical factors the decay rate becomes
\be\label{Gamma0rn}
\Gamma_0 \sim 10^{-8}\frac{\L^8}{M_{\rm Pl}^4}
\ee
The associated lifetime of the vacuum decay will be then given by $\tau_{\Gamma_0} = \Gamma_0^{-1/4}$ and
as we have already mentioned, the final step here is to demand that $\tau_{\Gamma_0}$ is much larger than the age of universe today.
The latter is translated, using \eq{Gamma0rn}, to the upper bound
\be\label{Lmax}
\tau_{\Gamma_0}H_{\rm m,0} \gtrsim 1 \Rightarrow \L \lesssim 1~ \rm eV \equiv \L_{\rm max}
\ee
Let us emphasize the assumptions entering the estimate above. The regulator ${\cal R}_{\rm loc}$ is imposed in the preferred lattice frame, which in the present cosmological application is identified with the comoving frame where the CMB is approximately isotropic. Therefore the calculation should not be interpreted as a Lorentz-invariant four-dimensional EFT estimate with an arbitrary frame-dependent cut-off. Rather, the finite NPGHU localization scale defines the physical frame in which the effective four-dimensional description is valid and cuts off the otherwise divergent boost volume of the phantom phase space. The parametric dependence of the result is
\be
|{\cal M}_h|^2\sim \frac{\Lambda^4}{M_{\rm Pl}^4}, ~~ \int d\Pi_{2\phi_2\,2\phi_1} {\cal R}_{\rm loc} \sim \Lambda^4 ~~ \Rightarrow \Gamma_0\sim \frac{\Lambda^8}{M_{\rm Pl}^4} \nn
\ee
up to the numerical factors displayed above. Hence the vacuum-decay constraint is controlled mainly by the hierarchy between the localization scale $\L$ and the upper bound $\L_{\rm max}$. In particular, increasing $\L$ by one order of magnitude enhances the rate by eight orders of magnitude. For the benchmarks considered in this work, $\L\simeq O(10)H_{\rm m,0}$ is many orders of magnitude below $\L_{\rm max}$, so the vacuum is sufficiently long-lived within the controlled effective regime. Finally let us clarify that the detailed benchmark parameters determine the background trajectory $w_q(N)$, but they do not modify the catastrophic vacuum-decay estimate as long as the effective masses remain of order $H_{\rm m,0}$ and the theory is restricted to the same near-transition regime. In other words, the stability conclusion here is primarily controlled by the finite localization cutoff and the preferred-frame assumption, rather than by the detailed shape of a particular benchmark trajectory.

The above indicates that if such a cut-off exists for the bulk-originating fields, then phantoms will not be problematic for our universe.
As we briefly justify in \sect{BoQLM} and also below \eq{mphi2Lambda}, NPGHU will provide a consistent origin for the quintom model as long as today we live in the regime after, but not far from, the Higgs-Hybrid phase transition to keep localization strong. Otherwise there would already be phenomenological signatures that the universe deviates from a 4d descriptions towards a 5d construction which is in contrast to the experiments. 
Therefore and in order to keep a mild tuning regarding the running of couplings and fields \cite{Irges:2019gsb, Irges:2020nap} (neither a very fast nor slow running compered with the Hubble) the cut-off in our framework should be higher but relatively close to the Hubble parameter today. 
In addition, it is also natural to set the initial point of the field-values ($\Phi(N_\a)$) that we used in the previous section on the dawn of the phase transition. Based on these, a physical choice for $\L$ is to be placed when the universe was deep in the matter domination era setting $\L = H(a_\a)$, with the latter defined on $N_\a$ which here we choose to be $N_\a = \ln a_\a/a_0 \approx -1.5 $. In that sense
\be\label{Hcutoff}
\L = H(a_a) \approx H_{\rm m, 0} \left[ \frac{a_0}{a_\a} \right]^{3/2} \Rightarrow \L \approx 10 H_{\rm m, 0} 
\ee
and suggests indeed that $\L \ll \L_{\rm max}$. As a consequence, no catastrophic vacuum decay is experienced by the universe in our model.  
Based on \eq{mphi2Lambda} one price to be paid is that such a choice admits $\{c_6, c_\a\} \approx 10^2 $
. However that number is $\sim {\cal O}(10^{90})$ smaller than the usual choices in quintom models and can be naturally accommodated in our lattice construction as we explained at the end of \sect{FBSGL}.
An other possible issue is that scales related to $\L$, defined as critical wavelengths $\l_{\rm cr} \sim 0.1 H_{\rm m, 0}^{-1} \approx 860 ~ \rm Mpc$ 
and critical time-scales $t_{\rm cr} \sim 0.1 H_{\rm m, 0}^{-1} \approx 2.8 ~ \rm Gyr$, 
may face cosmological instabilities.
Among other effects, the latter may lead the CMB photons to pick up an altered late Integrated Sachs–Wolfe (ISW) signal or produce a scale-dependent modification of the linear growth factor (e.g. distorting the matter power spectrum and the growth of $f\s_8$) at super-cluster regimes. Finally one could worry about whether the instabilities may change the evolution of the density field around the acoustic feature of BAO or not.
We expect that the previous effects will be mild in our case however, one needs to perform a full late-time cosmological analysis in order to concretely define the consequences of the problematic scale band and to explore lower or higher possible values of $\L$. This study is quite involved and we leave it for a future work.  
In total, the outcome of this section is that (at least on particle level) neither the linear perturbations around the background nor the possibility of vacuum decay due to gravitational interactions will lead to an immediate instability of our effective quintom model. 
That is true as long as $\l_\chi $ is positive and respects the bound of \eq{lchibound3} and the cut-off of the bulk-driven phase transition is much smaller than the bound of \eq{Lmax}.
In other words, the particular non-perturbative 5d origin of the action guarantees that on the 4d boundary, where our universe lives, the effective quintom-like behavior of the model remains consistent and renders it a viable dynamical dark energy candidate.


\section{Summary and outlook}\label{Conclusions}
Living in the exciting era of precision cosmology provides us gradually with an ever-increasing number of data which will contribute some novel and non-trivial insights about physics Beyond the Standard Model of both particles and cosmology.
A recent example is that of the data released from the DESI collaboration which are strongly pointing towards the scenario of a dynamical dark energy instead of the established $\L$CDM, favoring in particular a Quintom-B type (crossing of the cosmological boundary towards a quintessence phase) EoS.

In this work, we constructed the most minimal extension of the Standard Model which can naturally act as a UV completion for the original Quintom models, the simplest possibility of DDE with phantom crossing towards a quintessence phase.
In particular, we considered that fundamentally the spacetime is discretized and our universe lives on the 4d boundary of a 5-dimensional anisotropic orbifold lattice with purely a non-Abelian gauge field on its bulk, the so called Non-Perturbative Gauge Higgs Unification (NPGHU) model.
Considering the naive continuum limit and assuming as the simplest case that SM and dark matter are localized on the 4d boundary (decoupled from DE), we entertained the possibility that the components of the 5d bulk gauge field (the scalar $\phi$ and the Abelian field $A_\m$) that survive on the boundary play the role of the dark energy.
The non-trivial phase diagram of the NPGHU model, \fig{Higgsphase2}, inherits the system with a line of emergent low energy cut-offs $\L$, on which a 1st order phase transition takes place. For $\m > \L$ (Coulomb phase) the DE degrees of freedom are that of massless Scalar QED \eq{StmbL}. However at late times, $\m \lesssim \L$ (Higgs phase), we find that the dark energy spectrum is realized by dim-6 derivative operators for the scalar and the gauge fields, \eq{StmsL}.
These higher derivative operators give rise to both physical and phantom scalar and gauge dof, providing essentially the simplest modification of the original Quintom model.
In particular, the consistent basis of the latter should include the scalar and gauge R-ghosts (appeared by fixing the reparameterization freedom of the action  \cite{Irges:2019bzb, Irges:2019gsb, Cembranos:2025dor}) $\chi$ and $B_\m$ respectively.
In that sense, we obtained the complete Lagrangian basis for the late-time effective Quintom action, \eq{SDE3}. 
Based on the latter and showing that for the background evolution the R-ghosts can be neglected, we calculated the background EoS. Obtaining the compact form 
\be
w_q \approx - 1 + 2 \frac{|\dot \varphi_{1,n}|^2 - |\dot \varphi_{2,n}|^2 - e^{-2N}\frac{(\dot A_{2, n})^2}{3} - e^N\frac{m_{A_2}^2}{6 H_{\rm m,0}^2}  A_{2, n}^2 }{|\dot \varphi_{1,n}|^2 - |\dot \varphi_{2,n}|^2 - e^{-2N}\frac{(\dot A_{2, n})^2}{2} - e^{3N} \frac{ m^2_{\phi_2}}{H_{\rm m,0}^2} |\varphi_{2,n}|^2 - e^N\frac{m_{A_2}^2}{2 H_{\rm m,0}^2}  A_{2, n}^2 } \nn 
\ee
we performed a full numerical scan over its free parameters (four initial conditions) to show that our model can naturally provide a Quintom-B type of EoS with negligible fine-tuning.
One of the main novelties of our UV completion is that, below $\L$, naturally provides an effective quintom model with extra gauge phantom dof in contrast to the purely scalar original quintom scenario. In addition, the lattice origin of our model forbids the presence of any polynomial term therefore, its dynamics is strongly constrained.
Actually, under an analytical estimation of the eom we showed explicitly the crucial role of the gauge phantom in obtaining a consistent with the experiments $w_q$ even in the absence of a scalar potential.\\
Of course any quintom-like scenario suffers in principle from classical and quantum instabilities due to the presence of ghosts. However, the UV completion of our model is by definition ghost free and we showed how this information is incorporated to our effective Quintom action.
In particular, at classical level, we studied the behavior of our model under the linear perturbations given the presence of the R-ghosts which couple (at least for the scalar case) the fields of the spectrum. We determined the sign and the magnitude of the associated scalar coupling, $\l_\chi$, which render the effective quintom action classically stable, while similar arguments hold also for the gauge case.
Independently from the presence of extra couplings among the fields, there is always the universal gravitational interaction that couples the phantoms with the DE physical fields and also with the SM/DM spectrum. This leads at quantum level to instabilities since it allows for the spontaneous vacuum decay via the graviton exchange due to the infinite phantom phase space.
The latter is dominantly parameterized in our case by the decay of vacuum to two phantom and two physical scalars, $0 \to \phi_2 +  \phi^*_2 + \phi_1 +  \phi^*_1 $, via graviton exchange.
This rate is indeed infinite and in general severely problematic, however the UV-completion of our model provides a natural way to constrain the momentum phase space.
In particular, since the NPGHU model naturally possesses the finite (low scale) cutoff related to the localization of the 4d brane, $\L \sim F(\g, a_4^f, \b_4,\b_5)/a^f_5$, it suggests that Lorentz Invariance is an IR emergent and approximate symmetry. Based on this, we bounded the Lorentz boosts of the problematic decay rate from above, dramatically reducing the phase space of the reaction.
Then, demanding that the life time of this reaction is higher than the age of our universe we found the upper bound $\L \lesssim 1~ \rm eV$ which renders the vacuum sufficiently long-lived within the effective regime considered here. From phenomenological arguments we showed that the universe today sits in the vicinity of the cut-off towards the Higgs phase which motivates that the localization phase transition happens around the deep matter domination era, $\L \approx {\cal O}(10) H_{\rm m,0}$. The latter is many orders of magnitude smaller that the upper bound above, rendering the vacuum sufficiently long-lived within the effective regime considered here, without introducing catastrophic gravity modifications at experimentally relevant length scales.

The current work is a first step towards the construction of a consistent quintom UV-completion which addresses the usual drawbacks of this class of models and at the same time naturally provides a viable (Quintom-B type) DDE candidate.
Therefore, there are several future directions one could follow in order to extend the phenomenology and the completeness of the model.
Towards this regard some of our future plans are to study: i) a modified cosmological evolution due to the projected fields on the boundary above $\L$ since, the content of the early universe dark energy would be given now by \eq{StmbL}. Such a change of dof above and below $\L$ could account for a viable explanation of the Hubble tension \cite{Lee:2022cyh, Colgain:2025nzf}. Moreover, an interesting possibility is that inflation happened due to $\phi $ and $ A_\m$ non-minimal coupled to gravity considering also the running of $\g(\m)$ such that $\g(H_I) \approx 1$. ii) The scenario of gravitational reheating of the universe with interactions among $\phi, A_\m $ and SM and iii) the case that also DM originates from the bulk (e.g. an $SU(2)\times SU(2)$ bulk gauge group) but SM is still localized on the boundary, allowing for non-trivial interactions between DE and DM.  
Finally, more detailed investigation regarding the non-linear evolution of perturbations in the cosmological background, the sub-horizon scales which may vary the gravitational potential acting as an additional source of the late-time ISW effect, the introduction of scale-dependent modification of the linear growth factor at super-cluster regimes and a stricter bound on $\L$ that may render the previous effects completely negligible are necessary. In that sense, implementing the effective quintom perturbation sector in a Boltzmann solver such as CLASS or CAMB would be the natural next step toward a full likelihood-level comparison with $P(k,z)$, $f\sigma_8(k,z)$, CMB angular spectra and late-time ISW constraints.
Specific modifications along these lines may act as a smoking gun for our NPGHU dark energy model, however due to their involved nature we postpone their studies for a future work as well.

\begin{acknowledgments}
  \noindent
F.K. is thankful to Xinmin Zhang, Jing Ren and Meng-Xiang Lin for useful discussions regarding the quintom phenomenology. Moreover, F.K. is grateful to Jing Ren for the critical discussions on the consistency of the model involving the R-ghosts, on the improvement of the parameter scanning for the EoS and on the stability analysis.
F.K. is also thankful to Nikos Irges for useful discussions regarding the nature of the NPGHU model and for screening the early version of this paper.
Finally  
The work of F.K. is supported in part by the National Natural Science Foundation of China under grant No. 12342502.

\end{acknowledgments}

\section*{Declaration of generative AI and AI-assisted technologies in the manuscript preparation process}

\textit{During the preparation of this work the author used ChatGPT (OpenAI) to assist with the generation of better-looking benchmark-level phenomenological plots based on the auhtor's original Mathematica and Python scripts. After using this tool the author reviewed, verified and edited all material as needed and takes full responsibility for the content of the published article.}

\appendix

\section{From conformal time to the dot basis}\label{dBasis}
In this section we provide an analytic calculation that shifts the basic quantities used in the main text from the conformal time derivatives, $d \tau$, to the dotted ones, $dN = {\cal H} d \tau $.
Let us start with the eom given in \eq{eomdot}. In the general case of a field $\Phi$ with mass $m_\Phi$ and no interactions, the associated eom in FRW space and conformal time reads
\bea\label{Phieom}
\Phi'' + 2 {\cal H} \Phi' + a^2 m_{\Phi}^2 \Phi &=& 0 \Rightarrow \nn\\
\frac{d}{d\tau} \frac{d\Phi}{d\tau} + 2 {\cal H} \frac{d\Phi}{d\tau} + a^2 m_{\Phi}^2 \Phi &=& 0
\eea
where prime denotes derivative with respect to $d\tau$.
Then using that $d \tau = dN/{\cal H}  $ we get 
\bea
{\cal H} \frac{d}{dN} \left( {\cal H} \frac{d\Phi}{dN} \right) + 2 {\cal H}^2  \frac{d\Phi}{dN} + a^2 m_{\Phi}^2 \Phi &=& 0 \Rightarrow \nn\\
{\cal H}^2 \frac{d}{dN} \frac{d\Phi}{dN} + {\cal H} \frac{d {\cal H}}{dN} \frac{d\Phi}{dN} + 2 {\cal H}^2  \frac{d\Phi}{dN} +  a^2 m_{\Phi}^2 \Phi &=& 0 \Rightarrow \nn\\
\ddot \Phi + \left[ 2 + \frac{\dot {\cal H}}{{\cal H}} \right] \dot \Phi + \frac{ a^2 m_{\Phi}^2}{{\cal H}^2} \Phi &=& 0
\eea
while dividing the above with the initial value $|\varphi_2(N_\a)| \equiv |\varphi_{2,\a}| $ we get the normalized and dimensionless eom 
\be\label{ddotPhin}
\ddot \Phi_n + \left[ 2 + \frac{\dot {\cal H}}{{\cal H}} \right] \dot \Phi_n + \frac{ a^2 m_{\Phi}^2}{{\cal H}^2} \Phi_n = 0
\ee
where the subscript $n$ denotes normalized quantities.

Form the previous discussion note that the change from the conformal to the dotted basis includes the total Hubble rate \eq{fFe}. As we have shown in the main text this is broken to the dark energy and matter part with the latter been evaluated in \eq{cHm}.
Therefore, it is important for our analysis to find an explicit form for the comoving DE Hubble rate, ${\cal H}_{\rm DE}$, in the dotted and normalized basis.
For that purpose we use the definition $3 M_{\rm Pl}^2 {\cal H}_{\rm DE} = a^2 \rho_{\rm DE}$ along with the DE background energy density in conformal time \eq{rhoDEsg}, which in the temporal gauge reads
\bea\label{rdep}
\rho_{\rm DE} &=& \frac{|\varphi'_1|^2}{a^2} - \frac{|\varphi'_2|^2}{a^2} + \frac{1}{2 a^4}  (A'_{i,1})^2  - \frac{1}{2 a^4}  (A'_{i,2})^2 - \frac{1}{2 a^2} m_{A_2}^2 A_{i,2}^2 - m_{\phi_2}^2 |\varphi_2|^2 \Rightarrow \nn\\
a^2 \rho_{\rm DE} &=& |\varphi'_1|^2 - |\varphi'_2|^2 + \frac{1}{2 a^2}  (A'_{i,1})^2  - \frac{1}{2 a^2}  (A'_{i,2})^2 - \frac{1}{2} m_{A_2}^2 A_{i,2}^2 - a^2 m_{\phi_2}^2 |\varphi_2|^2
\eea
while keep in mind that the scale factor above is always normalized with its today's value $a_0$.
Now following the same steps that gave us \eq{ddotPhin} we get that  
\bea\label{rdep2}
3 M_{\rm Pl}^2 {\cal H}_{\rm DE}^2 = a^2 \rho_{\rm DE} &=& {\cal H}^2 \frac{d \varphi_{1}}{dN} \frac{d \varphi^*_{1}}{dN}  - {\cal H}^2 \frac{d \varphi_{2}}{dN} \frac{d \varphi^*_{2}}{dN}  + \frac{1}{2 a^2}  \left({\cal H} \frac{dA_{i,1}}{dN} \right)^2  - \frac{1}{2 a^2}  \left({\cal H} \frac{dA_{i,2}}{dN} \right)^2 - \frac{1}{2} m_{A_2}^2 A_{i,2}^2 - a^2 m_{\phi_2}^2 |\varphi_2|^2 \Rightarrow \nn\\
3 M_{\rm Pl}^2{\cal H}_{\rm DE}^2 &=& {\cal H}^2 | \dot \varphi_{1}|^2 - {\cal H}^2 | \dot \varphi_{2}|^2 + {\cal H}^2 \frac{e^{-2N}}{2}  (\dot A_{i,1})^2  - {\cal H}^2\frac{e^{-2N}}{2}  (\dot A_{i,2})^2 - \frac{1}{2} m_{A_2}^2 A_{i,2}^2 - e^{2N} m_{\phi_2}^2 |\varphi_{2}|^2 \Rightarrow \nn\\
{\cal H}_{\rm DE}^2 &=& \frac{{\cal H}^2}{3 M_{\rm Pl}^2} K_{\rm DE} + \frac{V_{\rm DE}}{3 M_{\rm Pl}^2}
\eea
where we have defined for clarity
\be
K_{\rm DE} = | \dot \varphi_{1}|^2 - | \dot \varphi_{2}|^2 + \frac{e^{-2N}}{2}  (\dot A_{i,1})^2  - \frac{e^{-2N}}{2}  (\dot A_{i,2})^2 ~~{\rm and}~~ V_{\rm DE} = - \frac{1}{2} m_{A_2}^2 A_{i,2}^2 - e^{2N} m_{\phi_2}^2 |\varphi_{2}|^2 
\ee
Breaking the total Hubble rate in its parts (${\cal H}^2 = {\cal H}_{\rm DE}^2 + {\cal H}_{\rm m}^2$) and solving for ${\cal H}_{\rm DE}^2$ we obtain
\bea
{\cal H}_{\rm DE}^2 \left[ 1 - \frac{K_{\rm DE}}{3 M_{\rm Pl}^2} \right] &=& \frac{{\cal H}_{\rm m}^2}{3 M_{\rm Pl}^2} K_{\rm DE}  + \frac{V_{\rm DE}}{3 M_{\rm Pl}^2} \Rightarrow \nn\\
{\cal H}_{\rm DE}^2 &=& \frac{{\cal H}_{\rm m}^2}{3 M_{\rm Pl}^2} \frac{K_{\rm DE}}{1 - \frac{K_{\rm DE}}{3 M_{\rm Pl}^2}}  + \frac{V_{\rm DE}}{3 M_{\rm Pl}^2}\frac{1}{1 - \frac{K_{\rm DE}}{3 M_{\rm Pl}^2}} \Rightarrow \nn\\
{\cal H}_{\rm DE}^2 &\approx& \frac{{\cal H}_{\rm m}^2}{3 M_{\rm Pl}^2} K_{\rm DE}  + \frac{V_{\rm DE}}{3 M_{\rm Pl}^2}
\eea
where we kept only the leading in ${\cal O}(1/M_{\rm Pl})$ terms.
The above gives the DE comoving Hubble rate in the dotted basis. So now if we multiply and divide the right hand side of this expression with $|\varphi_{2,\a}|^2$ as previously, use \eq{cHm} and factor out $H_{\rm m,0}^2$, then we end up with 
\be
{\cal H}_{\rm DE}^2  \approx H^2_{\rm m,0} \frac{|\varphi_{2,\a}|^2}{3 M_{\rm Pl}^2} \left[ \left( | \dot \varphi_{1,n}|^2 - | \dot \varphi_{2,n}|^2 + \frac{e^{-2N}}{2} (\dot A_{i,1, n})^2  - \frac{e^{-2N}}{2} (\dot A_{i,2, n})^2 \right)e^{-N} - \frac{m_{A_2}^2}{2 H^2_{\rm m,0}}  A_{i,2, n}^2 - e^{2N} \frac{m_{\phi_2}^2}{H^2_{\rm m,0}} |\varphi_{2,n}|^2  \right]
\ee
which is exactly the dotted and normalized expression for the DE comoving Hubble rate that we used in \eq{cHDEf}.

\section{Choosing the initial field values}\label{CTIFV}
In this section we briefly explain and motivate the choice of initial conditions that we made for the scalar fields in \sect{TEoSEQM}. Recall that our model is originally defined from a 5d anisotropic lattice due to which the bulk-related fields ($\phi$ and $A_\m$) exhibit a phase transition as it is shown in \fig{Higgsphase}. According to that there is an energy cut-off scale, $\m = \L$, above which the spectrum is given by $S^{\rm orb}_{\m>\L}$ in \eq{StmbL} while below this scale by $S^{\rm orb}_{\m < \L}$ of \eq{SDE3}. Let us use only here the subscripts $C$ and $H$ to define the DE scalar in Coulomb and Higgs phase respectively. Following that logic, the dof in each regime should be connected exactly on the cut-off such that the final field-value for $\phi_C$, should match the initial field-value (defined on $N = N_\a$) of $\phi_H$. 
In other words, the above suggests that
\be
\phi_C \big|_{ \m = \L} = \phi_H \big|_{N = N_\a}
\ee
however the latter includes double information according to \eq{phitophi12AtoA12}, so it becomes
\be
\phi_C \big|_{ \m = \L} = \left( \phi_1 - \phi_2 \right) \big|_{N = N_\a}
\ee
Of course the above holds also for the associated background values and allows for the freedom to choose any value\footnote{This is true in the current work however, in a scenario that the orbifold action $S^{\rm orb}_{\m < \L}$ is related with the Early Universe evolution one could imply specific constraints on the field value evolution. Nevertheless this is an interesting perspective, we leave it for a future work.} for the ending point of $\phi_C$. Among the possible choices the most natural one to is consider $\phi_C|_{ \m = \L} = 0$ which corresponds to the most democratic initial values
\be
\phi_1 \big|_{N = N_\a} =  \phi_2 \big|_{N = N_\a}
\ee
which is actually the choice that we used in our analysis of \sect{TEoSEQM}.
On the other hand, similar arguments should also hold for the R-ghost case.
Now, consistency of the theory exactly on the cut-off suggests that 
\be
\chi_C \big|_{ \m = \L} = \chi \big|_{N = N_\a}
\ee
however in contrast to the scalar field case above, the R-ghosts do not exist in the Coulomb phase since there is no extra pole dof (absence of higher derivative operators). Therefore the reparameterization is trivial, $\phi' = \phi + c$ with $c$ a constant, leading to the trivial Jacobian for the path integral measure, ${\cal D}\phi'/{\cal D}\phi = 1$. In other words, $\chi_C$ is identically zero for any $\m >\L$ and that fixes uniquely the initial condition of the R-ghost after the phase transition to
\be
\chi \big|_{N = N_\a} = 0
\ee
justifying our choice in \sect{TEoSEQM} which was necessary to secure the subdominant background contribution of the scalar R-ghost on the EoS.
According to the previous arguments, the set of normalized (with $|\varphi_{2,\a}|$) initial background field values is given by 
\be
\varphi_{1,n, \a} = 1, ~~ \varphi_{2, n, \a} = 1 ~~~ {\rm and} ~~~ \chi_{0, n, \a} = 0
\ee
Similar arguments hold also for the gauge sector of our model providing us with the gauge field initial conditions
\be\label{AnCAn1An2}
A_{\n, 1}\big|_{N = N_\a} = A_{\n, 2}\big|_{N = N_\a} = K_\n ~~~ {\rm and} ~~~ B_{\n}\big|_{N = N_\a} = 0
\ee
assuming that $A_{\n, C}|_{ \m = \L} = 0$.
The constant vector $K_\n$, normalized with $|\varphi_{2,\a}|$, will be one of the model's free parameters. 
Notice that since $A_{\n, 2}$ and $B_\n$ are massive, their spatial components will both have 3 polarization modes in contrast to $A_{\n, 1}$ or to the original gauge field in the Coulomb phase $A_{\n, C}$.
Dividing the spatial parts of the above gauge fields in transverse and longitudinal modes defined as
\bea
A_{i, 2} &=& A^T_{i, 2} + A^L_{i, 2} \frac{k_i}{|k|} \nn\\
A_{i, 1} &\equiv & A^T_{i, 1} \nn\\
B_i &=& B_i^T + B^L_i \frac{k_i}{|k|}
\eea
then the constrain \eq{AnCAn1An2}, imposed due to the non-trivial origin of our model, suggests that 
\bea
A^T_{i, 1}\big|_{N = N_\a} = A^T_{i, 2}\big|_{N = N_\a}\, , ~~~ A^L_{i, 2}\big|_{N = N_\a} = 0 ~~~ {\rm and} ~~~ B_i^T \big|_{N = N_\a} = B^L_i \big|_{N = N_\a} = 0
\eea
Consistency of the model at times larger than $N_\a$, requires that the evolution of $A^L_{i, 2}$ and $B^L_i$ is either zero or negligible. In either case, therefore, $A_{0,2} \propto \dot A^L_{i, 2} \approx 0$ and $B_0 \propto \dot B^L_i \approx 0$ justifying the universal gauge choice $A_0 = 0$ and the presence of only the transverse modes in our calculations at the main text.

\bibliographystyle{JHEP}
\bibliography{References}

\appendix

\end{document}